\documentclass[fleqn,usenatbib,onecolumn]{mnras}

\usepackage{newtxtext,newtxmath}
\usepackage[T1]{fontenc}

\usepackage{pdflscape}

\DeclareRobustCommand{\VAN}[3]{#2}
\let\VANthebibliography\thebibliography
\def\thebibliography{\DeclareRobustCommand{\VAN}[3]{##3}\VANthebibliography}

\usepackage{graphicx}	% Including figure files
\usepackage{amsmath}	% Advanced maths commands
\usepackage{subcaption} % ADDED PACKAGE: for using subfigures
\usepackage{orcidlink}  % In order for this to work, must include the oridlink.sty file in the project folder!

\newcommand{\dd}{\mathrm{d}}

\newcommand{\psrpoppy}{\texttt{P}{\footnotesize\texttt{SR}}\texttt{P}{\footnotesize\texttt{OP}}\texttt{P}{\footnotesize\texttt{Y}} }
\newcommand{\psrpoppycaption}{\texttt{P}{\scriptsize\texttt{SR}}\texttt{P}{\scriptsize\texttt{OP}}\texttt{P}{\scriptsize\texttt{Y}} }

\newcommand{\figtype}{.png}

\defcitealias{mcgrath2021}{MC21}
\defcitealias{DOrazio2021}{DL21}

\title[Measuring $H_0$ in Pulsar Timing]{Measuring the Hubble Constant with Double Gravitational Wave Sources in Pulsar Timing}

\author[McGrath, D'Orazio, \& Creighton]{
Casey McGrath \textsuperscript{\orcidlink{0000-0002-6155-3501}},$^{1}$ $^{2}$ $^{3}$ \thanks{casey.d.mcgrath@nasa.gov}
Daniel J. D'Orazio \textsuperscript{\orcidlink{0000-0002-1271-6247}},$^{4}$\thanks{daniel.dorazio@nbi.ku.dk}
and Jolien Creighton \textsuperscript{\orcidlink{0000-0003-3600-2406}},$^{5}$\thanks{jolien@uwm.edu}
\\
$^{1}$Center for Space Sciences and Technology, University of Maryland, Baltimore County, Baltimore, MD 21250, USA \\
$^{2}$Gravitational Astrophysics Lab, NASA/GSFC, Greenbelt, MD 20771, USA \\
$^{3}$Center for Research and Exploration in Space Science and Technology II, NASA/GSFC, Greenbelt, MD 20771, USA \\
$^{4}$Niels Bohr International Academy, Niels Bohr Institute, Blegdamsvej 17, DK-2100 Copenhagen, Denmark \\
$^{5}$Center for Gravitation, Cosmology and Astrophysics, Department of Physics, University of Wisconsin-Milwaukee, P.O. Box 413, Milwaukee, WI 53201, USA
}

\date{Accepted 2022 September 09. Received 2022 September 09; in original form 2022 August 12}

\pubyear{2022}

\begin{document}
\label{firstpage}
\pagerange{\pageref{firstpage}--\pageref{lastpage}}
\maketitle

\begin{abstract}
Pulsar timing arrays (PTAs) are searching for gravitational waves from supermassive black hole binaries (SMBHBs). Here we show how future PTAs could use a detection of gravitational waves from individually resolved SMBHB sources to produce a purely gravitational wave-based measurement of the Hubble constant.  This is achieved by measuring two separate \textit{distances} to the same source from the gravitational wave signal in the timing residual: the luminosity distance $D_L$ through frequency evolution effects, and the parallax distance $D_\mathrm{par}$ through wavefront curvature (Fresnel) effects.  We present a generalized timing residual model including these effects in an expanding universe. Of these two distances, $D_\mathrm{par}$ is challenging to measure due to the pulsar distance wrapping problem, a degeneracy in the Earth-pulsar distance and gravitational wave source parameters that requires highly precise, sub-parsec level, pulsar distance measurements to overcome.  However, in this paper we demonstrate that combining the knowledge of \textit{two} SMBHB sources in the timing residual largely removes the wrapping cycle degeneracy.  Two sources simultaneously calibrate the PTA by identifying the distances to the pulsars, which is useful in its own right, and allow recovery of the source luminosity and parallax distances which results in a measurement of the Hubble constant.  We find that, with optimistic PTAs in the era of the Square Kilometre Array, two fortuitous SMBHB sources within a few hundred Mpc could be used to measure the Hubble constant with a relative uncertainty on the order of 10 per cent.
\end{abstract}


%%%%%%%%%%%%%%%%%%%%%%%%%%%%%%%%%%%%%%%%%%%%%%%%%%
%%%%%%%%%%%%%%%%% BODY OF PAPER %%%%%%%%%%%%%%%%%%

    \section{Introduction}\label{sec:intro}

Pulsar timing arrays (PTAs) are currently searching for gravitational waves with frequencies $\mathcal{O}$(1 to 100 nHz) produced by supermassive black hole binaries (SMBHBs) at the centers of coalescing galaxies. These experiments time highly regular millisecond pulsars across our Galaxy and look for irregularities with the arrival times of the pulsar's pulses.  A continuous gravitational wave source is expected to cause these arrival times to periodically drift in and out of synchronization with a reference clock.  While the first gravitational wave detection using this experimental method may be a stochastic background of unresolved sources \citep[see, however,][]{Kelley+2018}, these experiments are expected to become more sensitive to the point where loud individual sources will be resolved and parameter estimation will be able to extract the source parameters from the data.

Of notable interest is the ability of a gravitational wave-based experiment to measure cosmological parameters such as the Hubble constant $H_0$.  These new gravitational wave-based measurements are important in helping us to resolve the current tension between experimental measurements of $H_0$ \citep[][and references therein]{gw170817_h0, gw170817_h0_updated, Feeney_standardsirens}.  In this study we are motivated by the recent work of~\citet{DOrazio2021} and~\citet{mcgrath2021} (hereafter \citetalias{DOrazio2021} and \citetalias{mcgrath2021}).  \citetalias{DOrazio2021} demonstrated how a measurement of the Hubble constant could be made from a purely gravitational wave-based method, by measuring the source's luminosity distance $D_L$ and it's comoving distance $D_c$.  In this approach $D_L$ is recovered from frequency evolution in the timing residual signal, and $D_c$ comes from probing the curvature of the wavefront across the Earth-pulsar baseline of the PTA experiment.  In this paper we show that more generally, the distance measured from the curvature of the wavefront is the ``parallax distance'' $D_\mathrm{par}$, which is equivalent to $D_c$ in a flat universe.  \citetalias{mcgrath2021} generalized the current gravitational wave timing residual models by classifying these wavefront curvature effects into the ``Fresnel regime,'' and they studied how well this new distance parameter could be recovered for different PTA constraints and source parameters. Both of these studies were strongly motivated by and synthesized the previous work of \citet{DF_main_paper} and \citet{CC_main_paper}.

For comparison, the ``standard sirens'' approach to measuring $H_0$ is a hybrid technique requiring a source observable via both gravitational wave and electromagnetic messengers \citep{Schutz_1986, holz_standardsirens}. Here the luminosity distance $D_L$ is measured via gravitational waves from a chirping source, while the redshift $z$ is measured from the host galaxy through electromagnetic observations.  It has been showed that future PTAs may also be able to make measurements of $H_0$ through this technique~\citep{pta_standardsiren}.  Combining these measurements of $D_L$ and $z$ allows an inference of $H_0$ to be made.  As an example of a purely gravitational wave-based measurement of $H_0$, gravitational wave signals from binary neutron star or neutron star black hole mergers could be used to obtain both measurements of the source's luminosity distance $D_L$ and redshift $z$~\citep[][]{messenger_gwH0, ghosh_2022, Shiralilou_2022}.  This would require well constrained knowledge of the neutron star equation of state from numerous detections, which may be possible with future Einstein Telescope and Cosmic Explorer-era gravitational wave observatories.  The approach presented in this paper is also purely gravitational wave-based, but here we trade the redshift measurement for a \textit{second} distance measurement made from the Fresnel wavefront curvature effects, and we do not require any intrinsic knowledge of rest-frame source properties.

In this study we take the models developed in \citetalias{mcgrath2021} and further generalize them to a cosmologically expanding universe.  We apply the same Bayesian framework and methods described in that study, in order to predict how well future PTA experiments may be able to measure the Hubble constant.  When detecting a single SMBHB source, we find that in order to recover the parallax distance parameter (and hence measure $H_0$), we require highly accurate measurements of the distances to the pulsars in our array.  But crucially, we find that when detecting two SMBHB sources simultaneously, no prior knowledge on the pulsar distances is required in order to recover $D_\mathrm{par}$ and $H_0$.  Multiple simultaneous continuous wave SMBHBs have not been widely investigated within PTA research~\citep[some studies include][]{babak_multipleSMBHBsI,babak_multipleSMBHBsII,qian_2022}, therefore this work helps us to motivate a particular insight that multiple sources can provide over just a single source.  The results presented in this work also demonstrate how the same methods used to measure $H_0$ do so through improved measurements of the pulsar distances, which is an important result in its own right.  In this paper we focus on the $H_0$ measurement, but an additional paper (\textcolor{blue}{McGrath, D'Orazio, \& Creighton, in preparation}) will provide greater details on the pulsar distance recovery.

This paper is outlined as follows.  Section~\ref{sec:timing residual theory} presents the theoretical background and generalization of our timing residual models, which shows how the Hubble constant enters the pulsar timing model.  Section~\ref{sec:methods} explains the methods used to estimate system parameters and $H_0$ from mock observations. Section~\ref{sec: wrapping problem solution} presents the primary obstacle to a practical $H_0$ measurement, the pulsar distance wrapping problem, and our solution. Section~\ref{sec:measurement of H0} presents the main results, simulated measurements of the Hubble constant for different mock PTAs. Finally, conclusions and future directions are presented in Section~\ref{sec: conclusions}.

    \section{Theoretical Background}\label{sec:timing residual theory}

%---------------------------------------------------------------------------------------------------
    \subsection{Fresnel and Frequency Evolving Regimes}\label{subsec:timing residual theory}
\begin{figure}
        \centering
        \includegraphics[width=0.5\linewidth]{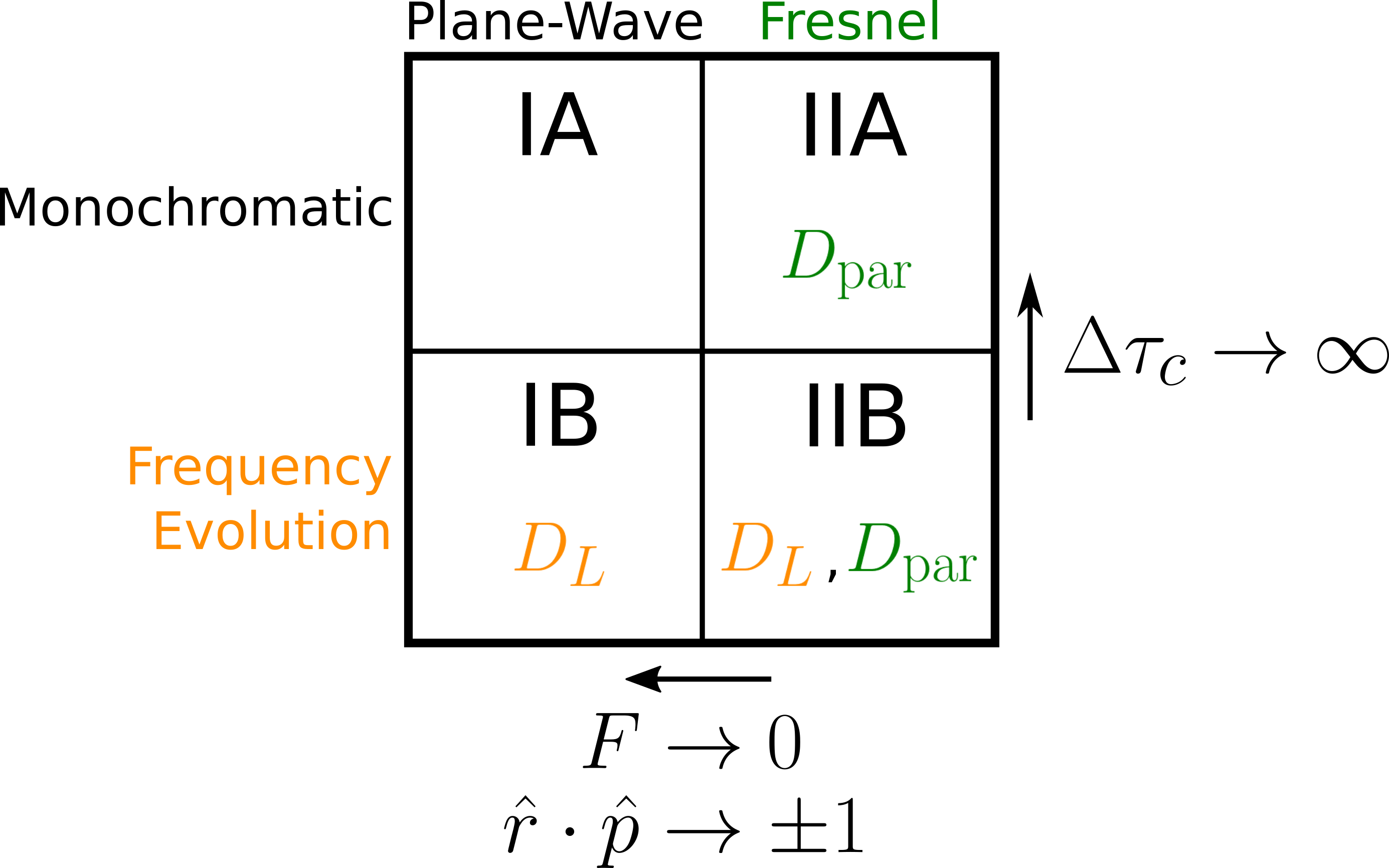}
        % --> located in:  Personal Laptop - /Research/My_Research_Papers/figures & sketches/sketches/4 Model Regimes/4_model_regimes-cosmo_v2.eps
    \caption{A classification summary of the four gravitational wave regimes, increasing in generality from left-to-right and top-to-bottom.  The frequency evolution regime reduces to the monochromatic regime in the large coalescence time limit ($\Delta \tau_c \rightarrow \infty$), and the Fresnel regime reduces to plane-wave regime in either the small Fresnel number limit ($F \rightarrow 0$) or the natural plane-wave limit ($\hat{r}\cdot\hat{p} \rightarrow \pm 1$).  Importantly, in a cosmologically expanding universe the frequency evolution regimes allow for the direct measurement of the source luminosity distance $D_L$, and the Fresnel regimes allow for the direct measurement of the source parallax distance $D_\mathrm{par}$.  In the fully general IIB model, both of these distances can be independently measured, and therefore through equation~\ref{eqn: Hubble constant} a value of $H_0$ can be inferred.}
        \label{fig: 4 model regimes}
    \end{figure}

    To detect and extract information from gravitational waves using PTAs, one must model the pulse-time-of-arrival deviations induced by a gravitational wave passing through the array.  Such gravitational wave-induced timing residual models can be classified based on two physical qualities/assumptions of the model: the shape of the incoming gravitational wavefront, and its frequency evolution.  Following \citetalias{mcgrath2021}, Figure~\ref{fig: 4 model regimes} categorizes four distinct model regimes.  Models that assume no wavefront curvature over the Earth-pulsar baseline make the ``plane-wave'' assumption (I), while models which account for the first-order curvature terms are classified as ``Fresnel'' models (II).  Then, if the frequency of the gravitational wave is assumed constant over the thousands of years it takes a photon to travel from the pulsar to the Earth, the model is ``monochromatic'' (A); otherwise it is in a ``frequency evolving'' (B) regime.
    
    The result is four distinct timing residual models, increasing in generality from left-to-right, and top-to-bottom.  While simpler models are mathematically and computationally simpler and less expensive to calculate, they may produce inaccurate predictions of the timing residual if the source is producing gravitational waves which fall into a more complicated regime.  The appropriate limits wherein one model approaches another are governed by the coalescence time $\Delta \tau_c$ (equation~\ref{eqn: frequency evolution (B)})  of the source, the Fresnel number $F$ (equation~\ref{eqn: Fresnel number}), and the relative source-pulsar-Earth orientation ($\hat{r}\cdot\hat{p}$; see equation~\ref{eqn:source/pulsar basis vectors}).  Therefore considering the relative size of these quantities is crucial in deciding when one model's prediction will fail in comparison to another.
    
    All four models are provided and studied explicitly in \citetalias{mcgrath2021}, which critically assumed a flat, static universe.  However, \citetalias{DOrazio2021} show that in a cosmologically expanding universe, two distinct distances appear in the gravitational wave-induced pulsar timing residual model -- the luminosity distance $D_L$ and the comoving coordinate distance $D_c$.  Accounting for frequency evolution effects (row B, Figure \ref{fig: 4 model regimes}) in the timing model allows the direct measurement of $D_L$, while accounting for Fresnel effects (column II, Figure \ref{fig: 4 model regimes}) allows the direct measurement of the parallax distance $D_\mathrm{par}$, which reduces to $D_c$ in a flat universe.  The goal of this paper is to remove the assumption that the universe is static on cosmological scales, re-derive the formulae of \citetalias{mcgrath2021} in a cosmologically expanding universe, and study implications for recovering $H_0$.  We provide a general list of assumptions made in this work in Appendix~\ref{app: assumptions}.

%---------------------------------------------------------------------------------------------------
    \subsection{Incorporating Cosmological Expansion and the Hubble Constant}\label{subsec: generalizing the flat, static model}
    
    In a flat static universe we have eight gravitational wave source parameters for a circular binary SMBHB system in the Newtonian regime, $\{R, \theta, \phi, \iota, \psi, \theta_0, \mathcal{M}, \omega_0 \}$.  These are the Earth-source comoving coordinate distance (for a flat static universe), sky angles, orientation Euler angles (inclination and polarization), initial phase (which also can be interpreted as the third orientation Euler angle), system chirp mass $\mathcal{M}\equiv \left(m_1 m_2\right)^{3/5} / \left(m_1 + m_2\right)^{1/5}$ (for component masses $m_1$ and $m_2$), and initial (angular) orbital frequency.  The pulsar parameters are $\{L, \theta_p, \phi_p \}$, which are the Earth-pulsar distance, and pulsar sky angles.
    
    In our cosmological framework we divide space-time into two types of frames: a ``global cosmological frame'' where the background metric is Friedmann–Lema\^{i}tre–Robertson–Walker (FLRW), and ``local frames'' where the background metric is Minkowski (see Appendix \ref{app: assumptions}, assumption~\ref{as: metric transition}).  The gravitational waves (as described by the metric perturbation below) are generated in the local source frame, propagate over the cosmological frame, and reach the Milky Way Galaxy where we assume they are in the local observer frame.  Therefore the cosmological effects of an expanding universe only need to be considered during the gravitational wave's propagation between the local source and observer frames.
    
    We begin by assuming that the FLRW metric describes our global space-time background between the Earth and our gravitational wave source:
    \begin{align}
        \dd s^2 &= -c^2 \dd t^2 + a^2(t)\left[ \frac{\dd r^2}{1-kr^2} + r^2 \dd \Omega^2 \right], \nonumber \\
        &= a^2(\eta) \left[ -c^2 \dd \eta^2 + \dd \chi^2 + r(\chi)^2 \dd \Omega^2 \right] 
    \label{eqn: FLRW metric}
    \end{align}
    where $\dd \Omega^2 \equiv \dd \theta^2 + \sin^2(\theta)\dd \phi^2$, $a$ is the universe scale factor, and $k$ is the space-time curvature constant. The coordinates $(t, r, \theta, \phi)$ are the comoving coordinates, while the version shown in the second line is made through the change to conformal time $\dd \eta \equiv \frac{\dd t}{a}$ (i.e. $\eta \equiv \int^t \frac{\dd t'}{a}$) and the spatial coordinate transformation:
    \begin{equation}
        r(\chi) \equiv \begin{cases}
                    \frac{1}{\sqrt{k}}\sin\left(\sqrt{k}\chi\right) &, \quad k > 0 \\
                    \chi &, \quad k = 0 \\
                    \frac{1}{\sqrt{|k|}}\sinh\left(\sqrt{|k|}\chi\right) &, \quad k < 0 .
                  \end{cases}
    \label{eqn: r(chi) definition}
    \end{equation}
    Here we are using the convention that the curvature parameter carries units of $\left[\mathrm{length}^{-2}\right]$, where $k>0$, $k=0$, or $k<0$ for closed, flat, and open spatial geometries, respectively.  This implies that our time-dependent scale factor $a$ is unitless.  We also choose to normalize $a(t_0) \equiv 1$ at the present day time $t_0$, but we will often write it explicitly in our derivations here.  Therefore, the Gaussian curvature of a spatial slice of our universe at the present time is $k$.
    
    In this framework, without an independent measurement of the source's redshift (either from electromagnetic observations, or as we will explain, from effects of the curvature of the wavefront itself), the redshift parameter $z$ cannot be disentangled from the source rest frame $R$ (now the transverse comoving distance; see Appendix~\ref{app: cosmological distances}), $\mathcal{M}$, and $\omega_0$ parameters \citep[see for example,][]{maggiore_2008}.  The result is that we can only observe the luminosity distance $D_L = (1+z)R$, a redshifted system chirp mass $\mathcal{M}_\mathrm{obs} \equiv (1+z) \mathcal{M}$, and a redshifted orbital frequency parameter $\omega_{0,\mathrm{obs}} = \frac{\omega_0}{(1+z)}$.
    
    As the gravitational wave propagates through the FLRW universe, the wavefront is traced by the retarded time. Therefore in order to understand time delay effects of the wave arriving at our pulsar compared to the time delay effects of the wave arriving at the Earth, we must calculate the expression for the retarded time of the wave at a field position $\vec{x}$ for a source at the location $\vec{x}'$ \citep[see for example,][]{greens_functions_1993}.  Crucially, if we make the assumption that the $a$ does not evolve appreciably over the time it takes a photon to travel from our pulsar to the Earth (see assumption~\ref{as: scale factor evolution}), then $\eta_\mathrm{ret} - \eta \approx \frac{1}{a(t_0)} \left(t_\mathrm{ret} - t\right)$ and we can write:
    \begin{align}
        &\eta_\mathrm{ret} = \eta - \frac{\left|\vec{x}' - \vec{x}\right|}{c} \nonumber \\
        & \therefore \quad t_\mathrm{ret} \approx t - a(t_0)\frac{\left|\vec{x}' - \vec{x}\right|}{c} .
    \label{eqn: tret - eta and t versions}
    \end{align}
    Next we must calculate $\left|\vec{x}' - \vec{x}\right|$ in the three possible spatial geometries of our universe.  The details of this are shown in Appendix~\ref{app: tret calculation}, and the result is that regardless of the background curvature of the spatial slices of our cosmology, the generalized expression for the retarded time evaluated with the field position set to the pulsar's location is:
    \begin{equation}
        t_\mathrm{ret}\left(t, \vec{x}=L\hat{p}\right) \ \ \approx \ \ t \ \ - \underbrace{\frac{D_c}{c}}_{\mathrm{``Far} \ \mathrm{Field"}} \  + \ \ \underbrace{\left(\hat{r}\cdot\hat{p}\right)a(t_0)\frac{L}{c}}_{\mathrm{``Plane-Wave"}} \ \ - \ \ \underbrace{\frac{1}{2}\left(1-\left(\hat{r}\cdot\hat{p}\right)^2\right)a^2(t_0)\frac{L}{c}\frac{L}{D_\mathrm{par}}}_{\mathrm{``Fresnel"}} \ \ + \ \ \ldots 
    \label{eqn: tret - pulsar term}
    \end{equation}
    The Fresnel term gives the first-order curvature of the physical wavefront, and introduces an additional time delay to the arrival time of the wavefront as predicted by the plane-wave approximation.  A useful proxy for determining when this term in the expansion will contribute a significant correction to the plane-wave approximation is the Fresnel number (see \citetalias{mcgrath2021} and \citetalias{DOrazio2021} for more discussion of this quantity):
    \begin{align}
        F &\equiv \frac{L^2}{\lambda_\mathrm{gw,obs} D_\mathrm{par}} \approx 0.0003 \left(\frac{\omega_{0,\mathrm{obs}}}{1\text{ nHz}}\right) \left(\frac{L}{1\text{ kpc}}\right)^2 \left(\frac{100\text{ Mpc}}{D_\mathrm{par}}\right) , \label{eqn: Fresnel number} \\
        &\mathrm{and}\quad \lambda_\mathrm{gw,obs} \equiv \frac{\pi c}{\omega_{0,\mathrm{obs}}} \approx (30.5 \text{ pc}) \left(\frac{1\text{ nHz}}{\omega_{0,\mathrm{obs}}}\right) . \label{eqn: gw wavelength}
    \end{align}

    There are two notable results that come out of the expression in equation~\ref{eqn: tret - pulsar term}.  First, we find that \textit{two} cosmological distance measurements to the same source appear in the ``Fresnel'' term in the expansion -- the line-of-sight comoving distance $D_c$, and the parallax distance $D_\mathrm{par}$.  In practice, however, we can only measure $D_\mathrm{par}$ from our timing residual models.  This is because in addition to $\frac{D_c}{c}$ appearing as the first term in the expression for the retarded time measured at the pulsar, it also appears in the expression for the retarded time measured at the Earth (for the Earth, just set $L=0$ in equation~\ref{eqn: tret - pulsar term}).  Therefore, when the rest of our timing residual model is worked out (see for example the IIB model in equations~\ref{eqn: Res(t) fr freq evo} and~\ref{eqn: Res(t) fresnel freq evo phase E and P}), we can simply choose the fiducial time $t_{0,\mathrm{obs}} = -\frac{D_c}{c}$ and all dependence on $D_c$ vanishes (this point is discussed in \citetalias{mcgrath2021}).  Second, is that in principle we could also choose to include the cosmological curvature constant $k$ as a parameter to attempt to directly measure it from the gravitational wave signal.  Equation~\ref{eqn: Dpar - R - DM - DC - DL relationship} provides the connection between $D_\mathrm{par}$ and $D_L$, and therefore would introduce both $z$ and $k$ as additional model parameters.  However, since the Fresnel corrections are already smaller order corrections and $k$ is currently understood to be very close to zero, we will restrict our attention in this work to a geometrically flat universe and assume $k=0$.

    Therefore, working under the assumption of a geometrically flat universe, the relationships given in Appendix~\ref{app: cosmological distances} provide us the connection to the Hubble constant:
    \begin{equation}
        k = 0 \quad \longrightarrow \quad \begin{cases}
            D_L = (1+z) D_\mathrm{par} , \\[4pt]
            H_0 = \frac{c}{R}\int^z_0 \frac{dz'}{E(z')} = \frac{c}{D_\mathrm{par}}\int^{D_L/D_\mathrm{par}-1}_0 \frac{\dd z'}{E(z')} .
        \end{cases}
    \label{eqn: Hubble constant}
    \end{equation}
    By procuring a measurement of both the distances $D_L$ and $D_\mathrm{par}$ from our pulsar timing model (see Section~\ref{subsubsec: model IIB}), we can directly measure the source's redshift.  By then assuming values of the cosmological density parameters which appear in the Hubble function $E$ (see equation~\ref{eqn: dimensionless Hubble function}), or by using the small redshift approximation (see the discussion around equation~\ref{eqn: low redshift R approximation}), we can directly measure the Hubble constant $H_0$. In this work we simulate a flat $\Lambda$CDM universe using the density parameters values $\Omega_r = 0$, $\Omega_M = 0.3$, $\Omega_\Lambda = 0.7$, and $\Omega_k = 0$, and a Hubble constant value of $H_0 = 70 \ \mathrm{km} \ \mathrm{s}^{-1} \ \mathrm{Mpc}^{-1}$.

%---------------------------------------------------------------------------------------------------
    \subsection{Constructing the Gravitational Wave Timing Residual}\label{subsec: constructing the timing residual}    
    
    We now derive the response of a PTA to the passing of a gravitational wave in the most general regime (IIB of Figure \ref{fig: 4 model regimes}), in an expanding universe.  Throughout, the notation and conventions used in this paper mirror those in \citetalias{mcgrath2021}.  The main pedantic distinction is the ``obs'' subscript on many of the quantities, simply to remind the reader that unlike in \citetalias{mcgrath2021}, here there is a difference between source frame quantities and ``observer'' frame quantities.  
    
    The metric perturbation produced by a circular binary system can be written as~\citep{maggiore_2008,creighton_anderson_2011}:
    \begin{equation}
        \begin{cases}
        h_{+}(t_\mathrm{obs}) &\equiv -h(t_\mathrm{obs}) \cos\big(2\Theta(t_\mathrm{obs})\big), \\
        h_{\times}(t_\mathrm{obs}) &\equiv -h(t_\mathrm{obs}) \sin\big(2\Theta(t_\mathrm{obs})\big), \\
        h(t_\mathrm{obs}) &\equiv \frac{4(G\mathcal{M}_\mathrm{obs})^{5/3}}{c^4 D_L}\omega(t_\mathrm{obs})^{2/3}, \\
        h_{0,\mathrm{obs}} &\equiv \frac{4(G\mathcal{M}_\mathrm{obs})^{5/3}}{c^4 D_L}\omega_{0,\mathrm{obs}}^{2/3} ,
      \end{cases} \label{eqn: h+x(t) & h(t)}
    \end{equation}
    for `plus' and `cross' polarizations, and where $t_\mathrm{obs}$ denotes the observer's clock time.  The angular phase and frequency functions $\Theta(t_\mathrm{obs})$ and $\omega(t_\mathrm{obs})$ are defined by the monochromatic regime (A) or the frequency evolving regime (B), depending on which model we desire:
    \begin{align}
        \underset{(A)}{\text{Monochromatism}} \hspace{0.5cm} & \begin{cases}\begin{tabular}{l l}
            $\omega(t_\mathrm{obs})$ & $= \omega_{0,\mathrm{obs}}$ , \\ 
            $\Theta(t_\mathrm{obs})$ & $= \theta_0 + \omega_{0,\mathrm{obs}} (t_\mathrm{obs} - t_{0,\mathrm{obs}})$ , 
            \end{tabular}\end{cases} \label{eqn: monochromatism (A)} \\[4pt]
        \underset{(B)}{\text{Frequency Evolution}} \quad & \begin{cases} \begin{tabular}{l l l r}
            $\omega(t_\mathrm{obs})$ & $= \omega_{0,\mathrm{obs}}\left[1 - \frac{t_\mathrm{obs}-t_{0,\mathrm{obs}}}{\Delta\tau_{c,\mathrm{obs}}} \right]^{-3/8}$ , & & \\
            $\Theta(t_\mathrm{obs})$ & $= \theta_0 + \theta_c\left[ 1 - \left(\frac{\omega(t_\mathrm{obs})}{\omega_{0,\mathrm{obs}}}\right)^{-5/3} \right]$ , & & \\
            $\Delta\tau_{c,\mathrm{obs}}$ & $\equiv \frac{5}{256}\left(\frac{c^3}{G\mathcal{M}_\mathrm{obs}}\right)^{5/3} \frac{1}{\omega_{0,\mathrm{obs}}^{8/3}}$ & $= (1+z)\Delta\tau_c$ , & \\
            $\theta_c$ & $=  \frac{1}{32}\left(\frac{c^3}{G\mathcal{M}_\mathrm{obs}\omega_{0,\mathrm{obs}}}\right)^{5/3}$ & $\equiv \frac{8}{5}\Delta\tau_{c,\mathrm{obs}} \omega_{0,\mathrm{obs}}$ & $= \frac{8}{5}\Delta\tau_c \omega_0$ ,
            \end{tabular}\end{cases} \label{eqn: frequency evolution (B)}
    \end{align}
    where $t_{0,\mathrm{obs}}$ here (and below) denotes the fiducial time for the model.  The frequency evolution regime is governed by the physically significant quantities $\Delta \tau_{c,\mathrm{obs}}$ which is the ``observed coalescence time,'' and $\theta_c$ which is the ``coalescence angle'' (the total angle swept out in the source's orbital plane before the system coalesces).  A more in-depth discussion of these quantities and the physical assumptions in these regimes is given in \citetalias{mcgrath2021}.
    
    The strength of the gravitational wave as it reaches the path that the pulsar photons take between the pulsar and the Earth also depends on the relative Earth-pulsar-source geometrical alignment.  With the Earth at the center of the coordinate system, the source frame orientation vectors and the Earth-to-pulsar vector are defined by:
    \begin{equation}
        \begin{cases}
            \begin{tabular}{c l c r l}
                $\hat{r}$ &$\equiv \big[\sin(\theta)\cos(\phi),$ &$\sin(\theta)\sin(\phi),$ &$\cos(\theta) \big]$ \ , & (Earth-to-gravitational wave source unit vector) \\
                $\hat{\theta}$ &$\equiv \big[\cos(\theta)\cos(\phi),$ &$\cos(\theta)\sin(\phi),$ &$\sin(\theta) \big]$ \ , & (transverse plane basis vector) \\
                $\hat{\phi}$ &$\equiv \big[-\sin(\phi),$ &$\cos(\phi),$ &$0 \big]$ \ , & (transverse plane basis vector) \\
                $\hat{p}$ &$\equiv \big[\sin(\theta_p)\cos(\phi_p),$ &$\sin(\theta_p)\sin(\phi_p),$ & $\cos(\theta_p) \big]$ \ , & (Earth-to-pulsar unit vector)
            \end{tabular}
            \end{cases} \label{eqn:source/pulsar basis vectors} 
    \end{equation}
    from which we define the following generalized polarization tensors:
    \begin{align}
        \begin{cases}
            e^{\hat{r}+}_{ij} &\equiv \hat{\theta}_i\hat{\theta}_j - \hat{\phi}_i\hat{\phi}_j, \\
            e^{\hat{r}\times}_{ij} &\equiv \hat{\phi}_i\hat{\theta}_j + \hat{\theta}_i\hat{\phi}_j, \\
            E^{\hat{r}+}_{ij} &\equiv \frac{1}{2}\left(1 + \cos^2(\iota)\right) \left[\cos(2\psi)e^{\hat{r}+}_{ij} + \sin(2\psi)e^{\hat{r}\times}_{ij} \right], \\
            E^{\hat{r}\times}_{ij} &\equiv \cos(\iota) \left[-\sin(2\psi)e^{\hat{r}+}_{ij} + \cos(2\psi)e^{\hat{r}\times}_{ij} \right].
        \end{cases} \label{eqn: generalized polarization tensors}
    \end{align}
    This notation groups all of the geometrical orientation and location angles into the definition of the polarization tensor, in order to keep them from mixing the plus and cross metric perturbations.  With these we then define the following antenna patterns which decide the detector's sensitivity to the gravitational wave based on the Earth-pulsar-source geometrical alignment:
    \begin{align}
        \begin{cases}
            \begin{tabular}{l l l}
                $f^{+}$ &$\equiv \frac{\hat{p}^i\hat{p}^j e^{\hat{r}+}_{ij}}{\left(1-\hat{r}\cdot\hat{p}\right)}$ &$= \frac{\hat{p}^i\hat{p}^j\hat{\theta}_i\hat{\theta}_j - \hat{p}^i\hat{p}^j\hat{\phi}_i\hat{\phi}_j}{\left(1-\hat{r}\cdot\hat{p}\right)} \quad = \frac{\left(\hat{p}\cdot\hat{\theta}\right)^2 - \left(\hat{p}\cdot\hat{\phi}\right)^2}{\left(1-\hat{r}\cdot\hat{p}\right)}$ , \\[3pt]
                $f^{\times}$ &$\equiv \frac{\hat{p}^i\hat{p}^j e^{\hat{r}\times}_{ij}}{\left(1-\hat{r}\cdot\hat{p}\right)}$ &$= \frac{\hat{p}^i\hat{p}^j\hat{\phi}_i\hat{\theta}_j + \hat{p}^i\hat{p}^j\hat{\theta}_i\hat{\phi}_j}{\left(1-\hat{r}\cdot\hat{p}\right)} \quad = \frac{2\left(\hat{p}\cdot\hat{\theta}\right)\left(\hat{p}\cdot\hat{\phi}\right)}{\left(1-\hat{r}\cdot\hat{p}\right)}$ , \\[3pt]
                $F^+$ &$\equiv \frac{\hat{p}^i\hat{p}^j E^{\hat{r}+}_{ij}}{\left(1-\hat{r}\cdot\hat{p}\right)}$ &$= \frac{1}{2}\left(1+\cos^2(\iota)\right) \left[ \cos(2\psi) f^+ + \sin(2\psi) f^\times \right]$ , \\[3pt]
                $F^\times$ &$\equiv \frac{\hat{p}^i\hat{p}^j E^{\hat{r}\times}_{ij}}{\left(1-\hat{r}\cdot\hat{p}\right)}$ &$= \cos(\iota) \left[ -\sin(2\psi) f^+ + \cos(2\psi) f^\times \right]$ .
            \end{tabular}
        \end{cases} \label{eqn: antenna patterns}
    \end{align}
    The final gravitational waveform in the transverse-traceless gauge along the $\hat{r}$-axis can now be expressed as:
    \begin{equation}
        h^{TT\hat{r}}_{ij} = E^{\hat{r}+}_{ij}h_{+} + E^{\hat{r}\times}_{ij}h_{\times} = E^{\hat{r}\textsc{A}}_{ij}h_{\textsc{A}}  \qquad \mathrm{for}\quad \textsc{A} \in [+, \times].  \label{eqn: general metric perturbation}
    \end{equation}
        
    In the local observer frame the gravitational waves will affect the timing of local pulsars.  The derivation at this point remains unchanged to what was shown in~\citetalias{mcgrath2021} (see, for example, Section A2 of that paper), and we again compute the effect that the gravitational waves have on the timed signals of these pulsars.  The gravitational wave-induced fractional shift of the pulsar's period $T$ is:
    \begin{equation}
        \frac{\Delta T^\mathrm{GW}}{T}(t_\mathrm{obs}) \approx \frac{1}{2}\hat{p}^i\hat{p}^j E^{\hat{r}\textsc{A}}_{ij} \int\limits^{t_\mathrm{obs}}_{t_\mathrm{obs} - \frac{L}{c}} \frac{\partial h_{\textsc{A}}\Big(t_\mathrm{obs,ret}\left(t'_\mathrm{obs},\vec{x}\right)\Big)}{\partial t'_\mathrm{obs}}\Bigg\rvert_{\vec{x}=\vec{x}_{0}\left(t'_\mathrm{obs}\right)} \dd t'_\mathrm{obs} ,
    \label{eqn: Delta T / T}
    \end{equation}
    where $t_\mathrm{obs}$ is the time a pulsar's photon is observed arriving at Earth, $t_\mathrm{obs,ret}$ is the retarded time of the gravitational wave, and $\vec{x}_{0}(t_\mathrm{obs})$ is the spatial path of the photon between the pulsar and the Earth.  Finally the gravitational wave-induced timing residual is the integrated fractional period shift due to the gravitational wave over the observation time.  Conceptually this is the difference between the observed and expected time-of-arrival of a pulsar's pulse~\citep{creighton_anderson_2011, schneider_2015, maggiore_2018}:
    \begin{equation}
        \mathrm{Res}^\mathrm{GW}(t_\mathrm{obs}) = \int\frac{\Delta T^\mathrm{GW}}{T}\left(t'_\mathrm{obs}\right) \dd t'_\mathrm{obs} = \int\frac{T_\mathrm{obs} \left(t'_\mathrm{obs}\right) - T}{T} \dd t'_\mathrm{obs}  = \ \mathrm{Obs}(t_\mathrm{obs}) - \mathrm{Exp}(t_\mathrm{obs}) . \label{eqn:timing residual}
    \end{equation}

%-----------------------------------------------------------------------------------------------------
    \subsubsection{The Fresnel, Frequency Evolution Model (Regime IIB)}\label{subsubsec: model IIB}

    Section 3 of \citetalias{mcgrath2021} details the four gravitational wave-induced timing residual model regimes IA-IIB, characterized by frequency evolution and curvature effects (see Figure~\ref{fig: 4 model regimes}).  Generalizing to a flat cosmologically expanding universe does not change the derivation behind those four models, but it does change the interpretation of some of the model parameters.
    
    Specifically, cosmological redshift causes the luminosity distance $D_L$ to become the parameter that appears in the amplitude of the metric perturbation (equation~\ref{eqn: h+x(t) & h(t)}), while similarly, we now recognize that the chirp mass and orbital frequency parameters are measured in the observer frame of reference, and are no longer equivalent to their values in the source frame (although the source frame values can be obtained if the source redshift $z$ is recovered through parameter estimation and equation~\ref{eqn: Hubble constant}).
    Importantly, the parallax distance $D_\mathrm{par}$ (i.e. the transverse comoving distance $D_M$, or comoving distance, $D_c$, in our flat universe; see equation~\ref{eqn: Dpar - R - DM - DC - DL relationship}) now enters via the retarded time calculated in a flat expanding universe (equation \ref{eqn: tret - pulsar term}).
    
    Hence, the most general IIB model becomes:
    \begin{align}
        \mathrm{Res}^\mathrm{GW}(t_\mathrm{obs}) &= \frac{F^\textsc{A}}{4} \left[ \frac{h_\textsc{A}\Big(\omega_{0E},\Theta_E-\frac{\pi}{4}\Big)}{\omega_{0E}} - \frac{h_\textsc{A}\Big(\overline{\omega}_{0P},\overline{\Theta}_P-\frac{\pi}{4}\Big)}{\overline{\omega}_{0P}} \right]   \qquad \mathrm{for}\quad \textsc{A} \in [+, \times], \label{eqn: Res(t) fr freq evo} \\[4pt]
        &\underset{\left(t_{0,\mathrm{obs}} \ = \ -\frac{D_c}{c}\right)}{\mathrm{where}}\quad \begin{cases}
            \omega_{0E} &\equiv \omega_{0,\mathrm{obs}} , \\
            \overline{\omega}_{0P} &\equiv \omega_{0,\mathrm{obs}}\left[1 + \frac{\left(1-\hat{r}\cdot\hat{p}\right)\frac{L}{c} + \frac{1}{2}\left(1-\left(\hat{r}\cdot\hat{p}\right)^2\right)\frac{L}{c}\frac{L}{D_\mathrm{par}}}{\Delta\tau_{c,\mathrm{obs}}}  \right]^{-3/8} , \\
            \theta_{0E} &= \theta_0 , \\
            \overline{\theta}_{0P} &= \theta_0 + \theta_c\left(1-\left[1+ \frac{\left(1-\hat{r}\cdot\hat{p}\right)\frac{L}{c} + \frac{1}{2}\left(1-\left(\hat{r}\cdot\hat{p}\right)^2\right)\frac{L}{c}\frac{L}{D_\mathrm{par}}}{\Delta\tau_{c,\mathrm{obs}}}\right]^{5/8}\right) , \\
            \Theta_E & = \theta_{0E} + \omega_{0E}t_\mathrm{obs} , \\
            \overline{\Theta}_P & = \overline{\theta}_{0P} + \overline{\omega}_{0P}t_\mathrm{obs} ,
            \end{cases}\label{eqn: Res(t) fresnel freq evo phase E and P}
    \end{align}    
    along with equations~\ref{eqn: h+x(t) & h(t)}, \ref{eqn: frequency evolution (B)}, \ref{eqn:source/pulsar basis vectors}, \ref{eqn: generalized polarization tensors}, and~\ref{eqn: antenna patterns}.  Written this way the timing residual is the sum of an ``Earth term'' (subscripted with an ``E'') and a ``pulsar term'' (subscripted with a ``P'').  Note that the overbar notation here is used for the same distinguishing purpose as in \citetalias{mcgrath2021}, to remind the reader that this model has not been analytically derived, but rather proposed.  \citetalias{mcgrath2021} explains why this model is physically motivated.

    \section{Methods}\label{sec:methods}

%---------------------------------------------------------------------------------------------------
    \subsection{Model Parametrization}\label{subsec:model parametrization}

    In this cosmological framework, the gravitational wave timing residual in the full IIB regime is governed by the source parameters: $\vec{s} = \{D_L, D_\mathrm{par}, \theta, \phi, \iota, \psi, \theta_0, \mathcal{M}_\mathrm{obs}, \omega_{0,\mathrm{obs}} \}$.  Following the explanation given in \citetalias{mcgrath2021} we swap the $D_L$ parameter in our model with the ``observed Earth term timing residual amplitude'' parameter, defined as:
    \begin{equation}
        A_{E,\mathrm{obs}} \equiv \frac{h_{0,\mathrm{obs}}}{4 \omega_{0,\mathrm{obs}}} = \frac{\left(G\mathcal{M}_\mathrm{obs}\right)^{5/3}}{c^4 D_L \omega_{0,\mathrm{obs}}^{1/3}} \approx (140 \text{ ns}) \left(\frac{\mathcal{M}_\mathrm{obs}}{10^9 M_\odot}\right)^{5/3} \left(\frac{100\text{ Mpc}}{D_L}\right) \left(\frac{1\text{ nHz}}{\omega_{0,\mathrm{obs}}}\right)^{1/3} .
    \label{eqn: earth term amplitude}
    \end{equation}
    
    Because present day capabilities on measuring the distances to pulsars often cannot constrain them to better than the order of 100~pc~\citep{NG_11yr_data, pulsar_parallax2019}, which is larger than our gravitational wave wavelengths (see equation~\ref{eqn: gw wavelength}), we include all of the pulsar distances as free parameters in our model~\citep[][\hspace{-4.6pt}; \citetalias{mcgrath2021}]{CC_main_paper, GWastro_Lee2011}.  Therefore we can divide the model parameters into source parameters and pulsar distance parameters $\vec{X} = \left[\vec{s}, \vec{L}\right]$.  

    To measure $H_0$, one  measures $D_\mathrm{par}$, and also $A_{E,\mathrm{obs}}$, $\mathcal{M}_\mathrm{obs}$, $\omega_{0,\mathrm{obs}}$, which together give $D_L$ via equation~\ref{eqn: earth term amplitude}. Then combining $D_L$ and $D_\mathrm{par}$ in equation~\ref{eqn: Hubble constant} gives $H_0$.  Note that this does require knowledge of the density parameters that go into $H_0$ (equation~\ref{eqn: expansion rate}).  In practice we take a simplified approach by using the small redshift approximation to replace the parallax distance parameter $D_\mathrm{par}$ with $H_0$. This does not qualitatively change the calculation, but allows computational simplicity for this proof-of-principle study and is relevant for the source distances for which Fresnel effects are prominent~(\citetalias{DOrazio2021, mcgrath2021}). For a flat $k=0$ universe and for $z \ll 1$, the Hubble Law is $z \approx \frac{H_0}{c} D_L$, so we can combine equations~\ref{eqn: Dpar - R - DM - DC - DL relationship} and~\ref{eqn: earth term amplitude} to write:
    \begin{align}
        D_\mathrm{par} &\approx \left[ \frac{c^4 A_{E,\mathrm{obs}} \omega_{0,\mathrm{obs}}^{1/3}}{\left(G\mathcal{M}_\mathrm{obs}\right)^{5/3}} + \frac{H_0}{c}\right]^{-1}  \qquad\qquad\qquad (\text{for } k = 0, \ z \ll 1), 
    \label{eqn: low redshift R approximation}
    \end{align}
    Note that if one were to use the equally valid small redshift approximation $z \approx \frac{H_0}{c} D_\mathrm{par}$, then our result for $H_0$ would differ by $\mathcal{O}(z^2)$ ($\lesssim$ a few percent for the fiducial cases in this study).
    
    This approach is particularly useful when working with the two source problem described in Section~\ref{subsec:breaking the wrapping cycle degeneracy}, as equation~\ref{eqn: low redshift R approximation} will reduce the dimensionality of the model by one ($D_\mathrm{par,1}$ and $D_\mathrm{par,2}$ are both replaced by $H_0$), therefore giving us the direct \textit{joint} posterior on $H_0$ from the two sources.  In the full calculation, one must make an estimate of the joint posterior on $H_0$ (for example, with a kernel density estimate) using the measured parameters from both sources.  The low-redshift approach builds into our model the additional knowledge that $H_0$ is a constant irrespective of the source we are detecting.  But again the approximation is only valid for $z \ll 1$.

    As a final note, unless otherwise stated, for the studies presented in this work we chose to inject the following default values for the gravitational wave source parameters:  $D_\mathrm{par} = 100$ Mpc, $D_L = 102.35$ Mpc, $\mathcal{M}_\mathrm{obs} = 10^9 \ \mathrm{M}_\odot$, $\omega_{0,\mathrm{obs}} = 30$ nHz, and angular parameters $\{\theta, \ \phi, \ \iota, \ \psi, \ \theta_0\} = \{\frac{3\pi}{7}, \ \frac{5\pi}{3}, \ \frac{2\pi}{5}, \ \frac{\pi}{7}, \ 1\}$.

%---------------------------------------------------------------------------------------------------
    \subsection{Likelihood and Priors}\label{subsec:likelihood and priors}

    Following~\citetalias{mcgrath2021}, for this work we assume that the timing residual data is the sum of the underlying gravitational wave-induced residual plus some random noise, that is $\overrightarrow{\mathrm{Res}} = \overrightarrow{\mathrm{Res}}^\mathrm{GW} + \vec{N}$.  For simplicity the noise we model is white noise, and each data point collected has some uncertainty uncorrelated between observations/pulsars, that is the timing covariance matrix $\bmath{\Sigma} = \mathrm{diag}\left(\sigma_1^2, \sigma_2^2, \ldots, \sigma_d^2\right)$.  Therefore the likelihood function we propose and use in this work is:
    \begin{align}
        \mathcal{L}\left(\overrightarrow{\mathrm{Res}}\mid\vec{X}\right) &= \frac{1}{\sqrt{(2\pi)^d \mathrm{det}(\bmath{\Sigma})}}\exp{\left[-\frac{1}{2}\bigg( \overrightarrow{\mathrm{Res}} - \overrightarrow{\mathrm{Res}}^\mathrm{GW}\left(\vec{X}\right) \bigg)^T \bmath{\Sigma}^{-1} \bigg( \overrightarrow{\mathrm{Res}} - \overrightarrow{\mathrm{Res}}^\mathrm{GW}\left(\vec{X}\right) \bigg) \right]} , \nonumber \\
        &= \frac{1}{\sqrt{(2\pi)^d}\sigma^d} \prod^d_{a=1} \exp{\left[-\frac{1}{2}\frac{1}{\sigma_a^2}\left( \mathrm{Res}_a - \mathrm{Res}^\mathrm{GW}_a\left(\vec{X}\right) \right)^2 \right]} .
    \label{eqn: likelihood function}
    \end{align}
    Here $\overrightarrow{\mathrm{Res}}$ is the $d$-dimensional timing residual data (with dimension equal to the number of pulsars times the number of observations per pulsar, indexed by $a$), $\overrightarrow{\mathrm{Res}}^\mathrm{GW}$ is our residual model (for the IIB model, see Section~\ref{subsubsec: model IIB}) for every data point, and the model parameters are contained in the vector $\vec{X}$.  Log-parameters are used for the non-angular parameters $\left\{\log_{10}\left(A_{E,\mathrm{obs}}/A_{E,\mathrm{obs},\mathrm{true}}\right), \log_{10}\left(D_\mathrm{par}/D_\mathrm{par,true}\right), \log_{10}\left(\mathcal{M}_\mathrm{obs}/\mathcal{M}_\mathrm{obs,\mathrm{true}}\right),\right.$
    $\left.\log_{10}\left(\omega_{0,\mathrm{obs}}/\omega_{0,\mathrm{obs},\mathrm{true}}\right) \right\}$, as well as $\log_{10}\left(H_0/H_{0,\mathrm{true}}\right)$. For our priors, we required all of the physical parameters be non-negative ($A_{E,\mathrm{obs}}$, $\mathcal{M}_\mathrm{obs}$, $\omega_{0,\mathrm{obs}}$, $D_L$, $D_\mathrm{par}$, $L$, $H_0$), and that $D_\mathrm{par} \leq D_L$ (which assumes $z > 0$ in equation~\ref{eqn: Dpar - R - DM - DC - DL relationship}).  For the angular parameters we placed the general boundaries: $0 < \theta, \ \iota, \ \theta_0 < \pi$, and $0 < \phi, \ \psi < 2\pi$.

%---------------------------------------------------------------------------------------------------
    \subsection{Mock Observations}\label{subsec:fiducial pta}
    
    In order to create the mock timing residual data for this study, we generate different mock PTAs, namely ``fiducial PTAs'' and ``Square Kilometre Array (SKA)-era PTAs.''  The details for constructing these arrays follow below, but each serves a distinct purpose in this study.  While the fiducial PTA is physically motivated, it is a simpler construction, and is primarily used throughout this work to look for various trends and scaling laws in our predicted measurement capabilities.  On the other hand, the various SKA PTAs are meant to represent more realistic hypothetical future timing arrays, and are therefore used to simulate more realistic results.
    
    Note that for simplicity, in this work we do not add noise $\vec{N}$ on top of our gravitational wave-induced timing residuals.  Therefore the simulated timing residuals are purely the gravitational wave timing residual component, which is why in the MCMC results shown (Figures~\ref{fig: error envelopes}, \ref{fig: 2 Sources - MCMC target}, and~\ref{fig: error envelope progression}) the posteriors are centered on the true injected parameters.  Finally, for all of our simulations in this paper, we use an observation time of 10 years, and a timing cadence of 30 observations per year.

%---------------------------------------------------------------------------------------------------
        \subsubsection{A Fiducial PTA}

        To gauge dependence of parameter recovery on pulsar number and timing precision we choose a simple fiducial PTA.  The fiducial PTA is realistically motivated by using a simple Milky Way Galaxy structure model to create a density distribution of pulsars $n_\mathrm{psr}$ \citep{schneider_2015} to sample from,
        \begin{equation}
            n_\mathrm{psr} \propto \exp\left(-\frac{r}{h_R}\right)\Bigg[\exp\left(-\frac{|z|}{h_\mathrm{thin}}\right) + 0.02\exp\left(-\frac{|z|}{h_\mathrm{thick}}\right) \Bigg] \times \exp\left(-\frac{1}{2}\left[\left(\frac{x}{R}\right)^2 + \left(\frac{y}{R}\right)^2 + \left(\frac{z}{R}\right)^2\right]\right)H(r-r_\mathrm{min}) ,
        \label{eqn: fiducial PTA}
        \end{equation}
        where the center of the coordinate system is at the Earth's position, the Galactic center is at $x_\mathrm{MW} \approx 8$ kpc, and $r \equiv \sqrt{(x-x_\mathrm{MW})^2 + y^2}$.  Here $h_R$ is the scale length of the Galactic disc ($h_R \approx 3.5$ kpc), $h_\mathrm{thin}$ is the scale height of the thin disc ($h_\mathrm{thin} \approx 0.325$ kpc), and $h_\mathrm{thick}$ is the scale height of the thick disc ($h_\mathrm{thick} \sim 1.5$ kpc).  The first term in this expression creates the distribution of stars in the Milky Way, while the second term simply places preference on stars within a Gaussian ball centered on Earth.  The motivation for including this second term is simply the idea that we will likely be more sensitive to timing pulsars within some scale distance $R$ from the Earth.  We also placed a minimum distance $r_\mathrm{min}$ on this sphere (hence the Heaviside function).  For our fiducial PTA we chose $R=5$ kpc and $r_\mathrm{min} = 0.1$ kpc.  An example of 1000 pulsars generated from this distribution is shown in Figure~\ref{fig: PTAs}.

%---------------------------------------------------------------------------------------------------
        \subsubsection{Simulated SKA-era PTAs}
        \label{S:SKAPTA}
        In addition to the fiducial PTA, we use the \psrpoppy package \citep{PsrPopPy:2014} to generate a population of millisecond pulsars (MSPs) that could be discovered and used for timing in the SKA era.  We begin by generating a population of $3\times10^4$ MSPs in the galaxy \citep[motivated by][]{SmitsSKA+2009} and simulate an SKA-like survey which detects $8893$ of these with signal-to-noise (SNR) above 9. We separate the detected MSPs into 20 radial bins out to the furthest detected pulsars at $\sim18$~kpc (with the closest at $\sim0.2$~kpc).  To simulate which MSPs will be best for high-precision timing, we sort the MSPs by SNR in each radial bin and retain the top $p\%$ for the final PTA. In the outer three bins which contain $<10$ MSPs we take only the highest SNR pulsar, each of which have SNRs $\geq 20$. Depending on the value of $p$, the final PTAs contain between $\sim40$ and 1800 pulsars (See Appendix \ref{app: Pulsar Populations} for further details). 
        
        Because our SKA simulation allows us to estimate detection SNRs and spin periods for the simulated SKA-era population of MSPs, we can additionally estimate a timing uncertainty for the $a^{\mathrm{th}}$ MSP in the array. To do so, we assume that the pulse time-of-arrival uncertainty is proportional to the radiometer noise  \citep[e.g.][]{Verbiest_timing_noise:2018},
        \begin{equation}
            \sigma_a \propto \frac{ P_a}{\sqrt{\Delta T_a }  \mathrm{SNR}_a} ,
        \end{equation}
        for integration time $\Delta T$, pulse period $P$, and detection signal-to-noise SNR. While this represents only part of the noise budget, it allows a study where pulsars have a more realistic spread of timing uncertainties than in the fiducial PTAs.
        We then consider two scenarios, one where the integration time is constant for all pulsars ($\Delta T_a =\mathrm{cst.}$), and one where the observation strategy is intelligently chosen to boost the signal from lower SNR MSPs ($\Delta T_a \propto \mathrm{SNR}^{-1}_a$), 
        \begin{equation}
            \sigma_a = \sigma_{\min}  \left( \frac{P_a}{1 \mathrm{ms}} \right)
                \begin{cases}
        			 \left(\frac{\mathrm{SNR}_a}{\mathrm{SNR}_{\max}} \right)^{-1} , & \Delta T_a =\mathrm{cst.} \\
                     \left(\frac{\mathrm{SNR}_a}{\mathrm{SNR}_{\max}} \right)^{-1/2} , & \Delta T_a \propto \mathrm{SNR}^{-1}_a,
        		 \end{cases}
        		 \label{eqn:sigmai}
        \end{equation}
        where we choose a characteristic, best-case timing uncertainty $\sigma_{\mathrm{min}} = \{1, 10\}$~ns and enforce $\sigma_a \geq \sigma_{\mathrm{min}}$.  While somewhat arbitrary, we assume an $\mathrm{SNR}_{\max}=10^4$, corresponding to an SNR threshold above which pulsars with $1$~ms pulse periods can be timed to $\sigma_{\mathrm{min}}$. Mock array pulsar distributions and timing uncertainties are shown in Appendix \ref{app: Pulsar Populations}.

%---------------------------------------------------------------------------------------------------
    \subsection{Fisher Matrix and MCMC}\label{subsec:fisher and mcmc}

    To recover best-fit source and pulsar parameter values we carry out both Fisher matrix and Markov Chain Monte Carlo (MCMC) analyses which compare our model and mock timing residuals through the likelihood (equation \ref{eqn: likelihood function}). 
    
    A Fisher matrix is a useful tool for quick parameter estimation.  Calculating the inverse of the Fisher matrix gives the estimated parameter covariance matrix $\mathbfss{F} = \mathbfss{C}^{-1}$ for the experiment.  Therefore finding a model's inverse Fisher matrix can help to tell us which parameters are covariant with each other, and roughly how well we might expect to recover each model parameter given our experimental set-up.  Computationally, this is efficient and can quickly allow us to test many different experiments.  Fisher-based surveys are useful for quickly searching large parts of parameter space for trying to understand what types of sources would result in good parameter estimation.  But it does require that one assume a more robust search (such as with an MCMC analysis) could successfully identify the \textit{true} mode from any potential secondary modes.  This limitation is due to the fact that the Fisher matrix is only meant to approximate the shape of the posterior near the maximum likelihood.  If for example the posterior is multimodal (see Section~\ref{sec: wrapping problem solution}), then the Fisher matrix will not capture this behavior.  For a more extensive (and computationally expensive) targeted study of the posterior, we use MCMC.
    
    From our likelihood equation~\ref{eqn: likelihood function} and the definition of the Fisher matrix as $\textsf{F}_{ij} \equiv \Big \langle \left( \frac{\partial}{\partial X_i}\ln{\mathcal{L}\left(\overrightarrow{\mathrm{Res}}\mid\vec{X}_\mathrm{true}\right)}\right) \left(\frac{\partial}{\partial X_j}\ln{\mathcal{L}\left(\overrightarrow{\mathrm{Res}}\mid\vec{X}_\mathrm{true}\right)}\right)  \Big\rangle$ we have:
    \begin{equation}
        \textsf{F}_{ij} = \sum_a \frac{1}{\sigma_a^2}\left(\frac{\partial \mathrm{Res}^\mathrm{GW}_a\left(\vec{X}_\mathrm{true}\right)}{\partial X_i}\right) \left(\frac{\partial \mathrm{Res}^\mathrm{GW}_a\left(\vec{X}_\mathrm{true}\right)}{\partial X_j}\right) .
    \label{eqn: Fisher matrix for residuals}
    \end{equation}
    In some instances, namely when using the Fresnel models, inverting the Fisher matrix cannot be done accurately because inclusion of the pulsar distance parameters introduces too much uncertainty and covariance amongst the other parameters (this typically happens with a single gravitational wave source). In these instances we add uncorrelated Gaussian pulsar distance priors to the Fisher matrix \citep[Wittman, unpublished notes\footnote{Wittman D., no date, Fisher Matrix for Beginners, UC Davis, \url{http://wittman.physics.ucdavis.edu/Fisher-matrix-guide.pdf.}};][]{gwfast_fisher}.  See Section 5 of \citetalias{mcgrath2021} for more details on adding these priors, and for a discussion of why this problem is more prominent in the Fresnel regimes.

    Since all parameters in a Fisher matrix approximation are normally distributed, we often use the coefficient of variation (CV), the predicted fractional error on a given parameter, as a useful way of quantifying the measurability of a given parameter when performing a Fisher matrix analysis:
    \begin{equation}
        \mathrm{CV}_x \equiv \frac{\text{Standard Deviation of } x}{\text{Expectation Value of } x} = \begin{cases} \frac{\sigma}{\mu} ,  &\quad\text{Normal Distribution in $x$}  \\[8pt]
        \sqrt{e^{\sigma^2 \ln(10)^2} - 1} .  &\quad\text{Log}_{10}\text{-Normal Distribution in $x$}
        \end{cases}
    \label{eqn: CV}
    \end{equation}
    Here $\mu$ and $\sigma$ are the distributions' parameters.  If the parameter $x$ is normally distributed then $\mu$ and $\sigma$ are also the distribution's mean and standard deviation, and if the parameter is log-normally distributed then CV only depends on the log-normal $\sigma$ parameter.  A resulting CV of the order of unity or larger suggests the parameter would not be measureable given the source and experiment, while smaller CVs suggest better parameter recovery.

    We use the PYTHON \texttt{emcee} package~\citep{emcee} to carry out MCMC sampling of the posterior.  We use differential evolution jump proposals with the number of walkers set between 5 and 7 times the number of parameter dimensions, and the results shown in this paper ran between $350~000$ to $500~000$ iterations.  We target our searches by initializing the MCMC walkers in very small Gaussian ``balls'' about the true injected parameters, as well as around nearby theoretical secondary modes of interest.  For the covariance matrix of these Gaussian balls we use the inverse of the Fisher matrix prediction, typically multiplied by some overall scale factor.  This choice helps the MCMC simulations more quickly find and explore the local distributions around the true parameters.
    
    While our MCMC simulations help provide proof-of-principle for the main ideas in this paper, future work should explore using different MCMC samplers to perform consistency checks on the work presented here, such as with parallel tempering or nested samplers~\citep{samajdar_2022}.  Further investigation of custom jump proposals should also be implemented to try and improve the exploration of walkers from one posterior mode to another.

    \section{A Two-Source Solution to the Pulsar Distance Wrapping Problem}\label{sec: wrapping problem solution}

The pulsar distance wrapping problem creates a challenge for parameter estimation in this work.  As explained in detail in \citetalias{mcgrath2021}, due to the way the pulsar distance $L$ affects the phase of the pulsar term (equation~\ref{eqn: Res(t) fresnel freq evo phase E and P}), increasing or decreasing the pulsar distance by specific $\Delta L$ values will cause the phase to wrap around the interval $\left[0, 2\pi\right)$, resulting in the same timing residual for multiple pulsar distances.

However, we show here that the wrapping problem is prominent primarily when there is only \textit{one} gravitational wave source in the signal.  
When there are two or more sources simultaneously creating residuals in every pulsar, the degeneracy of the wrapping cycle is broken.  
This has a significant effect on our ability to recover the parallax distance $D_\mathrm{par}$ and subsequently the Hubble constant, largely because breaking the wrapping degeneracy can strongly constrain the pulsar distances.

%---------------------------------------------------------------------------------------------------
    \subsection{Classifying the Error Envelope (Single Source)}\label{subsec: classifying the error envelope}
    
    More formally, in the monochromatic regimes IA and IIA, the wrapping cycle degeneracy arises because the timing residual is identical for $L \pm \Delta L_n$:  
    \begin{equation}
    \Delta L_n \equiv \begin{cases}
        n \frac{\lambda_\mathrm{gw}}{\left(1-\hat{r}\cdot\hat{p}\right)} ,  &\quad\text{(Plane-Wave, monochromatic)} \\[10pt]
        -\left(\frac{D_\mathrm{par}}{\left(1+\hat{r}\cdot\hat{p}\right)} + L\right) + \sqrt{ \left(\frac{D_\mathrm{par}}{\left(1+\hat{r}\cdot\hat{p}\right)} + L\right)^2 + n\frac{2\lambda_\mathrm{gw}D_\mathrm{par}}{\left(1-\left(\hat{r}\cdot\hat{p}\right)^2\right)}} , &\quad\text{(\textit{Heuristic} Fresnel, monochromatic)}
    \end{cases}
    \label{eqn: wrapping cycle}
    \end{equation}
    for $n \in \mathbb{Z}$.\footnote{See~\citetalias{mcgrath2021} and note that in equation~\ref{eqn: wrapping cycle} we replace $R \rightarrow D_\mathrm{par}$ to fit with our generalization described in Section~\ref{sec:timing residual theory}.}  Hence, the wrapping cycle distance $\Delta L$ is a scaled version of the gravitational wave wavelength $\lambda_\mathrm{gw}$, which includes geometric factors dependent on the relative positions of the pulsar and the source.  Most notably, the more aligned a source and pulsar are on the sky (within 90\degr, \ $0 < \hat{r}\cdot\hat{p} < 1$), the larger the wrapping cycle becomes, and vice versa as the source becomes more anti-aligned (more than 90\degr, \ $-1 < \hat{r}\cdot\hat{p} < 0$).
    
\begin{figure}
    \centering
      \begin{subfigure}[b]{0.48\linewidth}
      \centering
      \textbf{1 Source IA Model} \\
      (Plane-Wave, Monochromatic)
        \includegraphics[width=1\linewidth]{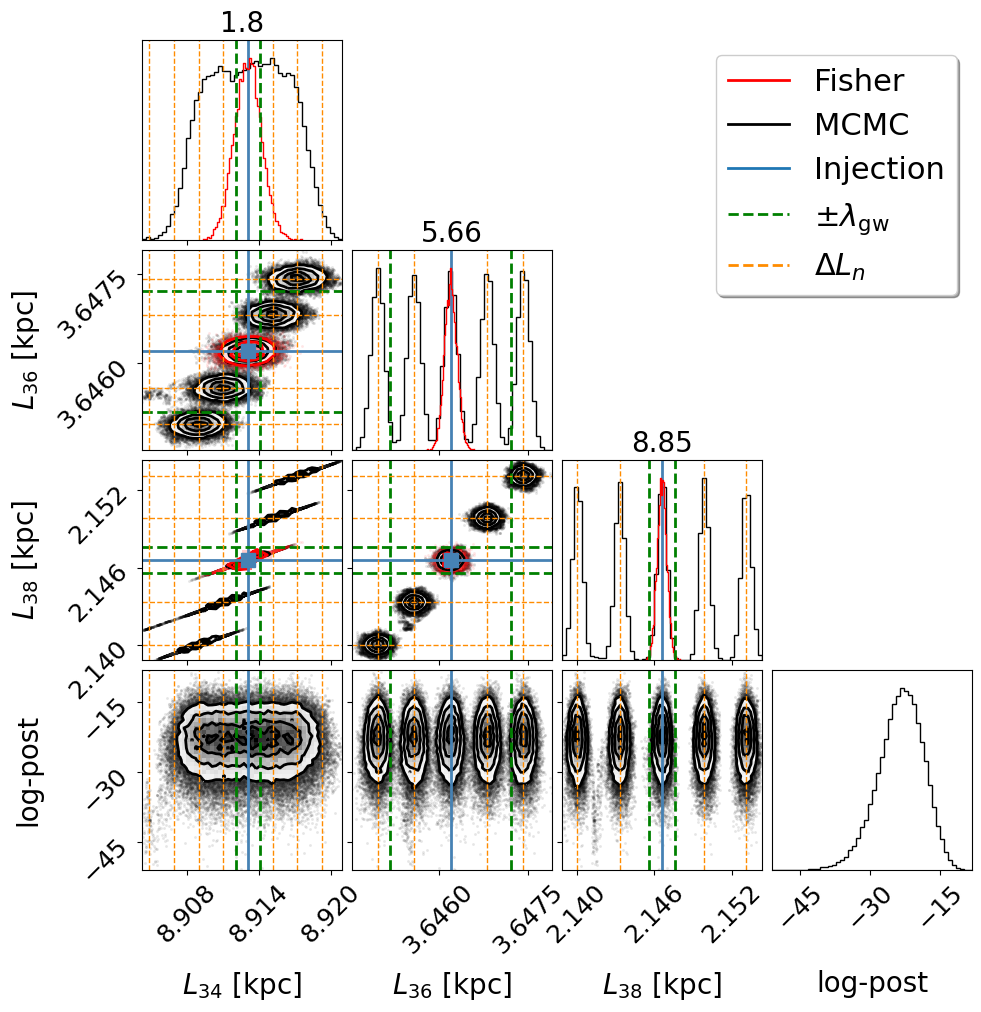}
        % --> located in:  cosmo_study/IA/envelope_runs_allparams/data/figures/run_9_L_badautocorr.jpg
      \end{subfigure}
      \hfill
      \begin{subfigure}[b]{0.48\linewidth}
      \centering
      \textbf{1 Source IB Model} \\
      (Plane-Wave, Frequency Evolution)
      \includegraphics[width=1\linewidth]{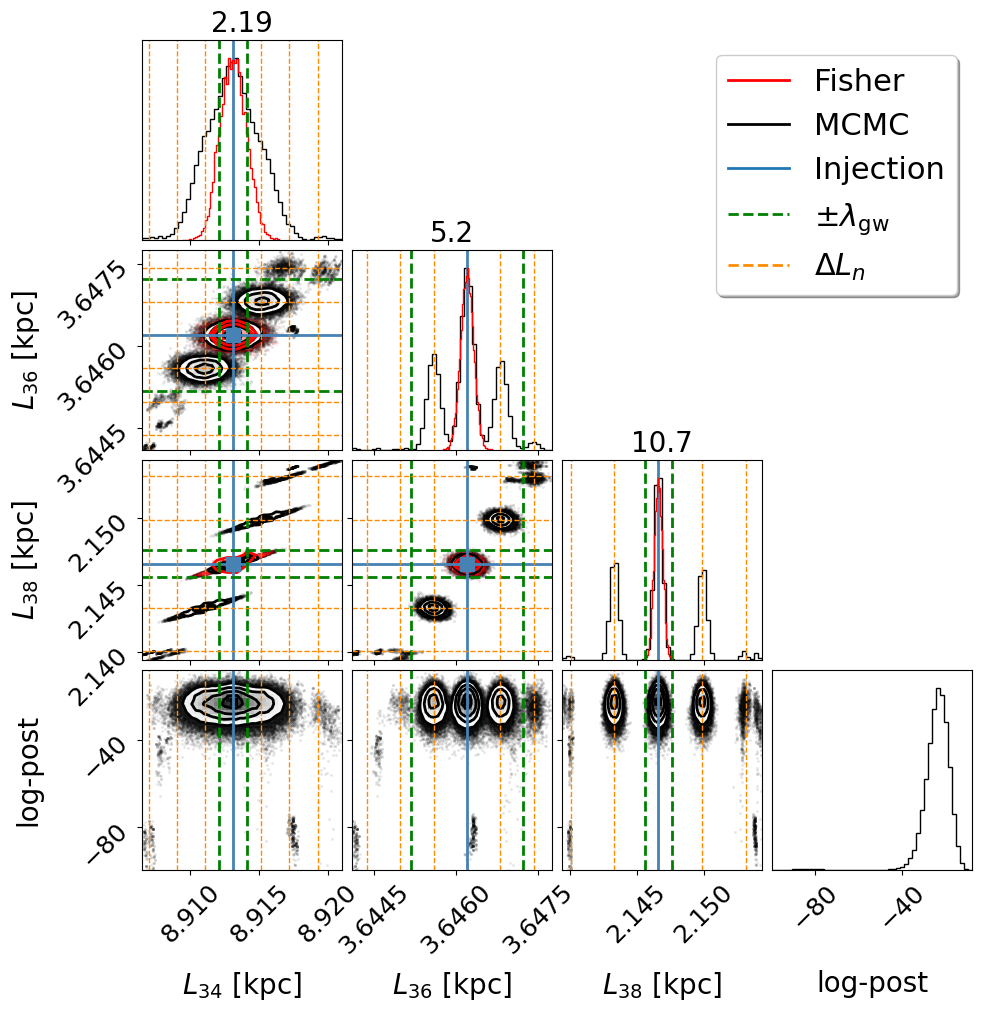}
        % --> located in:  cosmo_study/IB/envelope_runs_allparams/data/figures/run_16_L_cleaned-100.png
      \end{subfigure}
    \caption{
    Corner plots demonstrating the pulsar distance error envelope from an MCMC simulation of a PTA with a single gravitational wave source (default parameters). The results are compared using the (non-Fresnel regime) IA versus IB models (\citetalias{mcgrath2021}) for the exact same PTA and source.  The full PTA contained 40 pulsars from the fiducial PTA (see Figure~\ref{fig: error envelope progression}), with a uniform timing uncertainty of $\sigma = 0.2$ ns.  This corner plot singles out three pulsars and displays the modal overlap criterion ratios from equation~\ref{eqn: error envelope criteria} above each 1D histogram.  The bottom row shows the (un-normalized) log-posterior values.
    \textbf{Left-hand~panel:}~For the monochromatic IA model, each mode receives the same support in the likelihood function at every wrapping cycle.
    \textbf{Right-hand~panel:}~For the IB model, we see how frequency evolution helps us to weaken the wrapping cycle degeneracy by reducing support in the likelihood at secondary wrapping cycles modes.  Note the samples in this panel have been cleaned, such that all samples with log-posterior $\leq -100$ have been removed from the data set (see the discussion of cleaning in Section~\ref{subsec:mcmc results}).
    }
    \label{fig: error envelopes}
    \end{figure}

    Consider the problem now in terms of parameter estimation using the likelihood function equation~\ref{eqn: likelihood function}.  If we work entirely in one of the monochromatic regimes IA or IIA, meaning we use one of these models to both calculate the timing residuals of our injected source $\overrightarrow{\mathrm{Res}}$ \textit{and} recover the parameters $\overrightarrow{\mathrm{Res}}\left(\vec{X}\right)$, then the likelihood function is perfectly multimodal at every $L \pm \Delta L_n$ for every pulsar in the PTA (see the left-hand panel of Figure~\ref{fig: error envelopes}).  Therefore in either of these regimes, we cannot identify the true pulsar distance with this likelihood since all modes have equal probability.

    In the frequency evolution regimes IB and IIB this wrapping cycle does not formally exist as the frequencies are now time-dependent.  While this technically breaks the degeneracy, there still exists some support at $L \pm \Delta L_n$.  That is, the IB and IIB likelihood functions are still multimodal at each pulsar's true distance modulo its wrapping cycle distance (see the right-hand panel of Figure~\ref{fig: error envelopes}).  However, as moving out a wrapping cycle no longer returns the same exact timing residual as the timing residual at the true pulsar distance, these modes become less and less probable for every additional cycle away from the true distance.
    
    Additionally, every mode will have some width.  If these modes are close enough to each other (relative to the size of their widths), then the modes will blend together.  \citeauthor{CC_main_paper} identified this blending of the pulsar distance wrapping modes in their model as an ``error envelope'' which forms about the true pulsar distance.  The effect is that the true distance will be further buried within the uncertainty surrounding the secondary modes, making it much harder to measure in practice (see for example the top-most 1D histogram in both panels of Figure~\ref{fig: error envelopes}).  However, \citeauthor{CC_main_paper} did not fully classify the properties of the error envelope.  Namely the geometric factors in the wrapping cycle (equation~\ref{eqn: wrapping cycle}) are of crucial importance in determining the separation between the true mode and the secondary modes. 
    
    \begin{figure}
    \centering
      \begin{subfigure}[c]{0.48\linewidth}
      \centering
        \includegraphics[width=1\linewidth]{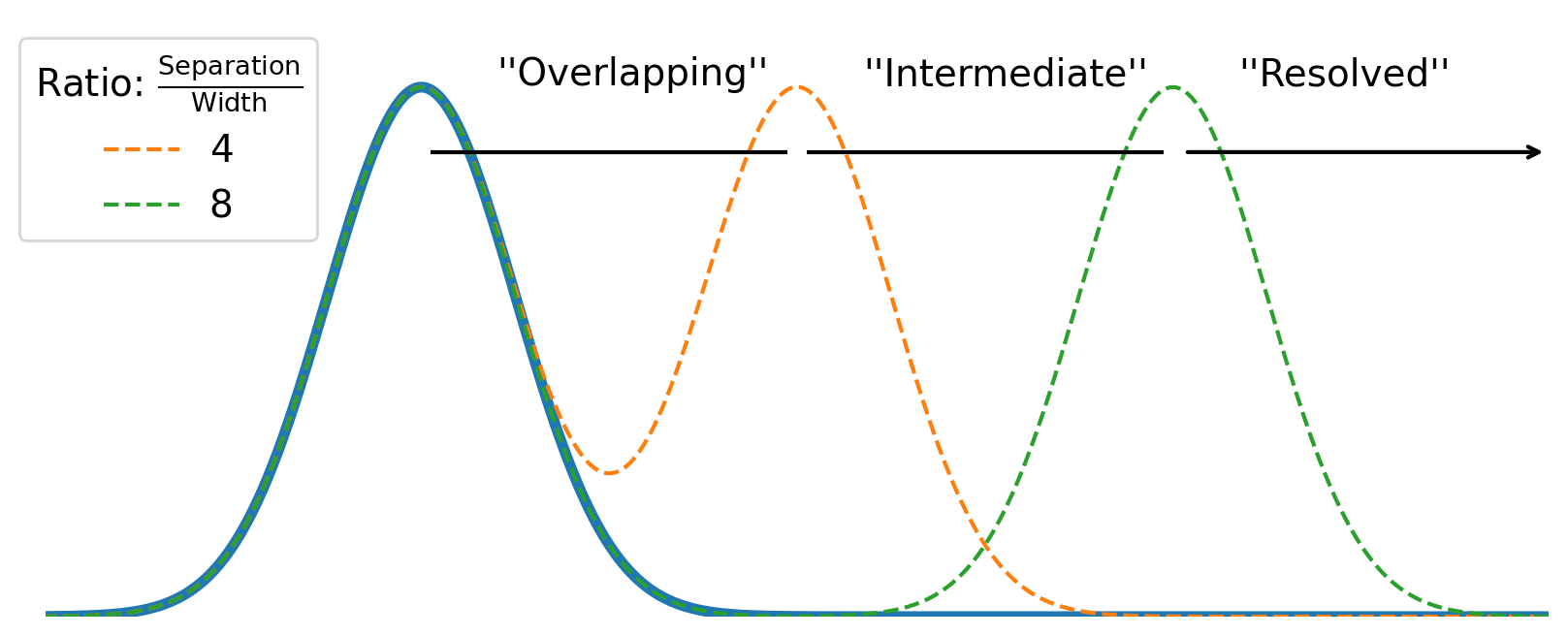}
        % --> located in:  Personal Laptop - Research/study_cosmo/Overlapping Gaussians Visualization.ipynb
      \end{subfigure}
      \hfill
      \begin{subfigure}[c]{0.48\linewidth}
      \centering
        \includegraphics[width=1\linewidth]{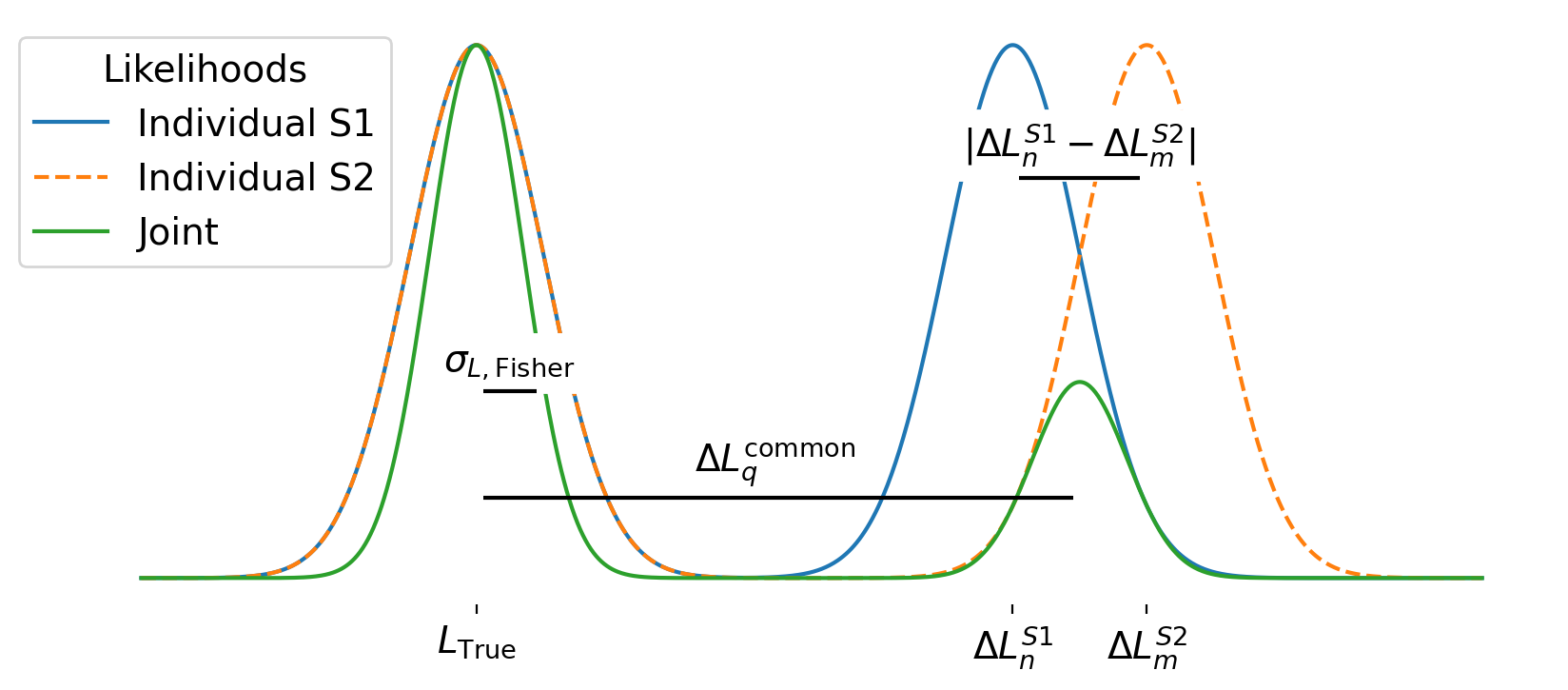}
        % --> located in:  Personal Laptop - Research/study_cosmo/Overlapping Gaussians Visualization.ipynb
      \end{subfigure}
    \caption{Here we visualize the modal overlap criterion used to analyse the pulsar wrapping cycle degeneracy (equations \ref{eqn: error envelope criteria}-\ref{eqn: 2 Source Common error envelope criteria}).
    \textbf{Left-hand panel:} The ``modal overlap'' criterion between the primary and secondary modes in the pulsar distance wrapping problem is defined as the ratio of the separation of the primary and secondary modes ($\Delta L_1$ for a single source, and $\Delta L^\mathrm{common}_1 = |\Delta L_n^\mathrm{S1} + \Delta L_m^\mathrm{S2}| / 2$ for two sources) to the width of the modes ($\sigma_{L,\mathrm{Fisher}})$.  We classify values of this ratio $<4$ as ``overlapping'', between $4$ and $8$ as ``intermediate'', and $>8$ as ``resolved.''
    \textbf{Right-hand panel:} \textit{Individual} likelihoods for two sources S1 and S2 will agree on the location of the primary mode in the pulsar distance, but as long as the sources are located on different parts of the sky and/or have different frequencies, they will have different sets of pulsar distance wrapping cycles (equation~\ref{eqn: wrapping cycle}).  Therefore the \textit{joint} likelihood for both sources will find support at the distances where there is some common modal overlap between $\Delta L_n^\mathrm{S1}$ and $\Delta L_m^\mathrm{S2}$.
    }
    \label{fig: 2 Source Common Wrapping Cycle}
    \end{figure}
    
    Therefore we construct a useful way of classifying the amount of ``modal overlap'' between the primary (thick blue line) and secondary modes (dashed orange lines).  Overlap between the first and second mode will depend on two factors: the width of each mode, and the separation between modes.  If each mode is a Gaussian of approximately the same width, then the ratio of mode separation to mode width provides a numerical quantification of the amount of overlap (see the left-hand panel of Figure~\ref{fig: 2 Source Common Wrapping Cycle}).  We define such a modal overlap criterion by calculating the ratio of the first wrapping cycle distance $\Delta L_1$ (i.e. the distance between our primary and secondary modes) to the approximated Fisher width $\sigma_{L,\mathrm{Fisher}}$ of each mode:
    \begin{equation}
        \frac{\Delta L_1}{\sigma_{L,\mathrm{Fisher}}} \sim \begin{cases}
        < 4 \ ,    &\text{Overlapping}  \\
        4 - 8 \ ,  &\text{Intermediate}  \\
        > 8 \ .    &\text{Resolved} 
        \end{cases}
    \label{eqn: error envelope criteria}
    \end{equation}
    Here $\sigma_{L,\mathrm{Fisher}}$ is the uncertainty of a given pulsar distance as predicted by the Fisher matrix analysis (that is, the standard deviation of a pulsar distance $L$ calculated from the inverse Fisher matrix, $\mathbfss{C} = \mathbfss{F}^{-1}$).  Conceptually, the values we choose here are motivated such that a secondary mode ``overlapping'' with the primary mode will have it's peak within $4\sigma_{L,\mathrm{Fisher}}$ of the true value, and ``resolved'' when it's peak is greater than $8\sigma_{L,\mathrm{Fisher}}$ from the true value.  Figure~\ref{fig: error envelopes} shows an example of three pulsars from a simulation with overlapping, intermediate, and resolved modes clearly visible.  A complete version is given in Figure~\ref{fig: error envelope progression}.
    
    Note that in general the Fisher uncertainty predictions are only valid around the \textit{true} mode, and therefore cannot accurately forecast the multimodal behavior of the pulsar distances.  However, in practice we observe that the secondary and primary modes have approximately the same width.  Therefore, we use $\sigma_{L,\mathrm{Fisher}}$ as our proxy in defining equation~\ref{eqn: error envelope criteria}.  This effectively assumes that the behavior of the likelihood function near each wrapping cycle will look the same as the behavior of the likelihood function near the true mode.
    
    In practice, this specific modal overlap criterion also works best if no prior knowledge is placed on the pulsar distances.  This is partly because $\sigma_{L,\mathrm{Fisher}}$ would be affected by a prior (see discussion in Section~\ref{subsec:fisher and mcmc}), while $\Delta L_1$ would not.  In our MCMC simulations, we find that when a very constricting prior is placed on the pulsar distances (such as a prior of the order of the wrapping cycle distance $\sim \mathcal{O}\left(\Delta L_1\right)$), then the location of the secondary modes in the posterior can shift away from $\Delta L_1$ and closer to the true mode.  This results in a change in the numerical ranges for which we find the overlapping, intermediate, and resolved modes in equation~\ref{eqn: error envelope criteria}.
    
    One other important caveat to note is that this criterion also works best if there are no secondary modes in the other source parameters.  Consider equation~\ref{eqn: wrapping cycle} for the wrapping cycle modes.  In the case of the plane-wave, monochromatic wrapping cycle distance, $\Delta L_n = \Delta L_n\left(\theta,\phi,\omega_{0,\mathrm{obs}}\right)$.  If the likelihood has secondary modes in the frequency parameter $\omega_{0,\mathrm{obs}}$, then those secondary mode solutions can produce their own wrapping cycle modes, which we observed in several MCMC tests.  Specifically, for some high mass systems $\mathcal{M}_\mathrm{obs} > 10^9 \ \mathrm{M}_\odot$, secondary modes (of lower probability) form in the $\mathcal{M}_\mathrm{obs}$ and $\omega_{0,\mathrm{obs}}$ parameters, which then change the locations of the observed secondary modes in pulsar distances (meaning our criterion becomes less accurate in this case).  In such cases, these new modes may be the result of stronger frequency evolution effects within the model, and should be kept in mind when performing these types of analyses.
    
    Hence, equation~\ref{eqn: error envelope criteria} provides a new and more general criterion for quantifying the wrapping problem.  While this criterion relies on the Fisher predicted uncertainty $\sigma_{L,\mathrm{Fisher}}$, which does not offer an immediately intuitive connection to the system parameter values, equation~\ref{eqn: error envelope criteria} can still be calculated for all pulsars in a PTA without requiring a full MCMC simulation.  Therefore we can quickly predict for any pulsar in our PTA whether or not that pulsar will have an error envelope.  The more pulsars in the PTA which are predicted to have resolved modes, the more likely it will be that we can obtain a measurement of the desired parallax distance $D_\mathrm{par}$ and hence the Hubble constant.

%---------------------------------------------------------------------------------------------------
    \subsection{Breaking the Wrapping Cycle Degeneracy With Two Sources}\label{subsec:breaking the wrapping cycle degeneracy}

    Now consider \textit{two} gravitational wave sources leaving their combined signal in the timing residuals of every pulsar.  With a single source, every pulsar has a specific wrapping cycle distance (equation~\ref{eqn: wrapping cycle}) which primarily depends on the frequency of the source and the angular sky separation between the source and the pulsar.  But with two sources, every pulsar will have \textit{two} wrapping cycle distances.  The key idea is that these two sets of wrapping cycle distances will not be the same as long as the two sources differ in either frequency and/or angular sky position. Therefore, the \textit{joint} likelihood for both sources should only contain secondary modes in an individual pulsar's distance at common multiples of \textit{both} the wrapping cycle distances.  In principle adding even more sources would further break the degeneracy, because secondary modes should then only form in the joint likelihood at common multiples of \textit{all} sources.  We restrict our attention to two SMBHB sources in this work, and leave it open to future work to consider more sources.
    
    Consider the right-hand panel of Figure~\ref{fig: 2 Source Common Wrapping Cycle}.  The joint likelihood should find some support at  common distances where the two separate wrapping cycle distance uncertainty modes overlap.  The amount of support at those common distances will depend on how strong the modal overlap is.  And if there is strong modal overlap at a common distance, we can then check if that common mode will overlap with the primary mode to create an error envelope (as we did in equation~\ref{eqn: error envelope criteria}).  Therefore we can define two new criteria.  The first criterion predicts the ``common mode overlap strength,'' and the second criterion classifies the ``common modal overlap'' with the primary mode.
    
    For the common mode overlap strength criterion we choose to write:
    \begin{equation}
        \frac{\left|\Delta L_n^\mathrm{S1} - \Delta L_m^\mathrm{S2}\right|}{\sigma_{L,\mathrm{Fisher}}}  \sim \begin{cases}
        0 - 2 \ ,   &\text{``strong'' support} \\
        2 - 4 \ ,   &\text{``weak'' support} \\
        \end{cases} ,
    \label{eqn: 2 source mode condition}
    \end{equation}
    where $\Delta L_n^\mathrm{S1}$ and $\Delta L_m^\mathrm{S2}$ are the $n$th and $m$th wrapping cycle distances of source 1 and source 2, respectively.  As an example, if this quantity equaled zero then the $n$th and $m$th modes would perfectly overlap, hence there would be strong support for a mode here.  At a value of two, the point in between the $n$th and $m$th modes (the average distance) would be $1\sigma_{L,\mathrm{Fisher}}$ from each peak, and at a value of four, the point in between the $n$th and $m$th modes would be $2\sigma_{L,\mathrm{Fisher}}$ from each peak.  We then expect that the common wrapping cycle distance is at approximately the average distance between these individual modes, that is $\Delta L^\mathrm{common}_q \equiv \frac{\Delta L_n^\mathrm{S1} + \Delta L_m^\mathrm{S2}}{2}$, for the $q$th common wrapping cycle (between the $n$th and $m$th wrapping cycles of sources 1 and 2).  Note that when calculating positive or negative common wrapping cycles, both $\Delta L^\mathrm{S1}_n$ and $\Delta L^\mathrm{S2}_m$ should be the same sign.
    
    Note that for the purpose of our studies, we will only be searching for common modes where we think there is strong support, as we define in the criterion in equation~\ref{eqn: 2 source mode condition}.  The reason for this is that ``weak'' common modes will not have nearly as much support in the likelihood function.  So even if the modal overlap between a weak common mode and the primary mode is such that it would produce an error envelope (using equation~\ref{eqn: 2 Source Common error envelope criteria}), we expect that this envelope would not contribute significantly to widening the uncertainty about the true pulsar distance.
    
    For the common modal overlap criterion we write the following expression in analogy to equation~\ref{eqn: error envelope criteria} (see also the left-hand panel of Figure~\ref{fig: 2 Source Common Wrapping Cycle}):
    \begin{equation}
        \frac{\Delta L^\mathrm{common}_1}{\sigma_{L,\mathrm{Fisher}}} 
        \sim \begin{cases}
        < 4 \ ,    &\text{Overlapping}  \\
        4 - 8 \ ,  &\text{Intermediate}  \\
        > 8 \ .    &\text{Resolved}
        \end{cases}
    \label{eqn: 2 Source Common error envelope criteria}
    \end{equation}
    This condition checks if the uncertainty around the 1st common wrapping cycle will be close enough to the uncertainty around the true distance as to blend those uncertainties into a greater envelope.  Once again, in order to make this prediction we are assuming as a proxy that the width of the true mode predicted by the Fisher matrix for the joint source likelihood also approximates the secondary modes at the individual source wrapping cycles.
    
    The key point is that with more than one source, it is more likely that common modes in the pulsar distance will form further away from the true distance mode.  This means that the uncertainties surrounding these common modes will be less likely to blend with the uncertainty surrounding the true distance to form an error envelope, hence resulting in better recovery of the pulsar distances.  Furthermore, recall that in the regimes IB and IIB, frequency evolution alone nominally breaks the wrapping cycle degeneracy (Section~\ref{subsec: classifying the error envelope}).  Therefore two sources with significant frequency evolution will further constrain the pulsar distances. All of this coupled with prior knowledge on our pulsar distances thanks to electromagnetic observations provide three separate means of localizing the pulsars in our PTA.  In application to our Hubble constant measurement, better pulsar distance measurements mean we can better recover the source parallax distance $D_\mathrm{par}$ from the Fresnel effects, which is what we show in Section~\ref{sec:measurement of H0}.

%---------------------------------------------------------------------------------------------------
    \subsection{Targeted MCMC Results for Two Sources}\label{subsec:mcmc results}
    
    Next we run a targeted MCMC simulation to test our discussion from Section~\ref{sec: wrapping problem solution} of how two sources break the wrapping cycle degeneracy.  For two gravitational wave sources, our likelihood function (equation~\ref{eqn: likelihood function}) now has $\overrightarrow{\mathrm{Res}}^\mathrm{GW} = \overrightarrow{\mathrm{Res}}^\mathrm{GW}_1 + \overrightarrow{\mathrm{Res}}^\mathrm{GW}_2$ for sources ``1'' and ``2.'' Decomposing this \textit{joint} likelihood function, we can write it as the product of the likelihood function of just source 1, the likelihood of just source 2, and their cross terms.  Fortunately adding a second source does not double our parameter space, since both sources will share the pulsar distance parameters $\vec{L}$.  We simply double the number of source parameters, so $\vec{X} = \vec{X}_1 \cup \vec{X}_2 = \left[\vec{s}_1, \vec{s}_2, \vec{L}\right]$.  Furthermore, using the small redshift approximation (equation~\ref{eqn: low redshift R approximation}) to replace the parallax distance $D_\mathrm{par,1}$ and $D_\mathrm{par,2}$ parameters with $H_0$ reduces the model parameter dimensionality by one and gives us the joint posterior recovery of $H_0$.
    
    In order to test the ideas in Section~\ref{subsec:breaking the wrapping cycle degeneracy}, our methodological approach was to inject balls of walkers in specific areas of parameter space in order to strategically initialize them (as mentioned in Section~\ref{subsec:fisher and mcmc}).  All walkers were initialized near the true injected source parameters, but the pulsar distance parameters were initialized distinctly.  One ball of walkers was placed around the true mode pulsar distance parameter values ($L_\mathrm{true}$); two balls of walkers were placed at the source one wrapping cycles $L_\mathrm{true} + \Delta L^\mathrm{S1}_{-1,+1}$; two balls of walkers were placed at the source two wrapping cycles $L_\mathrm{true} + \Delta L^\mathrm{S2}_{-1,+1}$; and finally, two balls of walkers were placed at the first predicted \textit{strong} (such that equation~\ref{eqn: 2 source mode condition} was $\leq 2$) common wrapping cycles $L_\mathrm{true} + \Delta L^\mathrm{common}_{-1,+1}$.
    
    This was done because we wanted to see if indeed the walkers placed near the original wrapping cycles of the individual sources would no longer find any meaningful support in posterior, as they did in Figure~\ref{fig: error envelopes}.  The walkers near the new predicted common wrapping cycles were placed there in order to see if any support in the posterior would be found at those locations.  However, depending on how far away those common modes are from the true mode we did not necessarily expect to find meaningful support their either.  This is because we know that when frequency evolution is included in the model, secondary modes further away from the true mode become less probable (see again, the right-hand panel of Figure~\ref{fig: error envelopes}).
    
\begin{figure}
        \centering
        \textbf{2 Source IIB Model} \\
        (Fresnel, Frequency Evolution) \\       
          \begin{subfigure}[b]{0.48\linewidth}
          \centering
          \textit{``Uncleaned'' Samples }
            \includegraphics[width=1\linewidth]{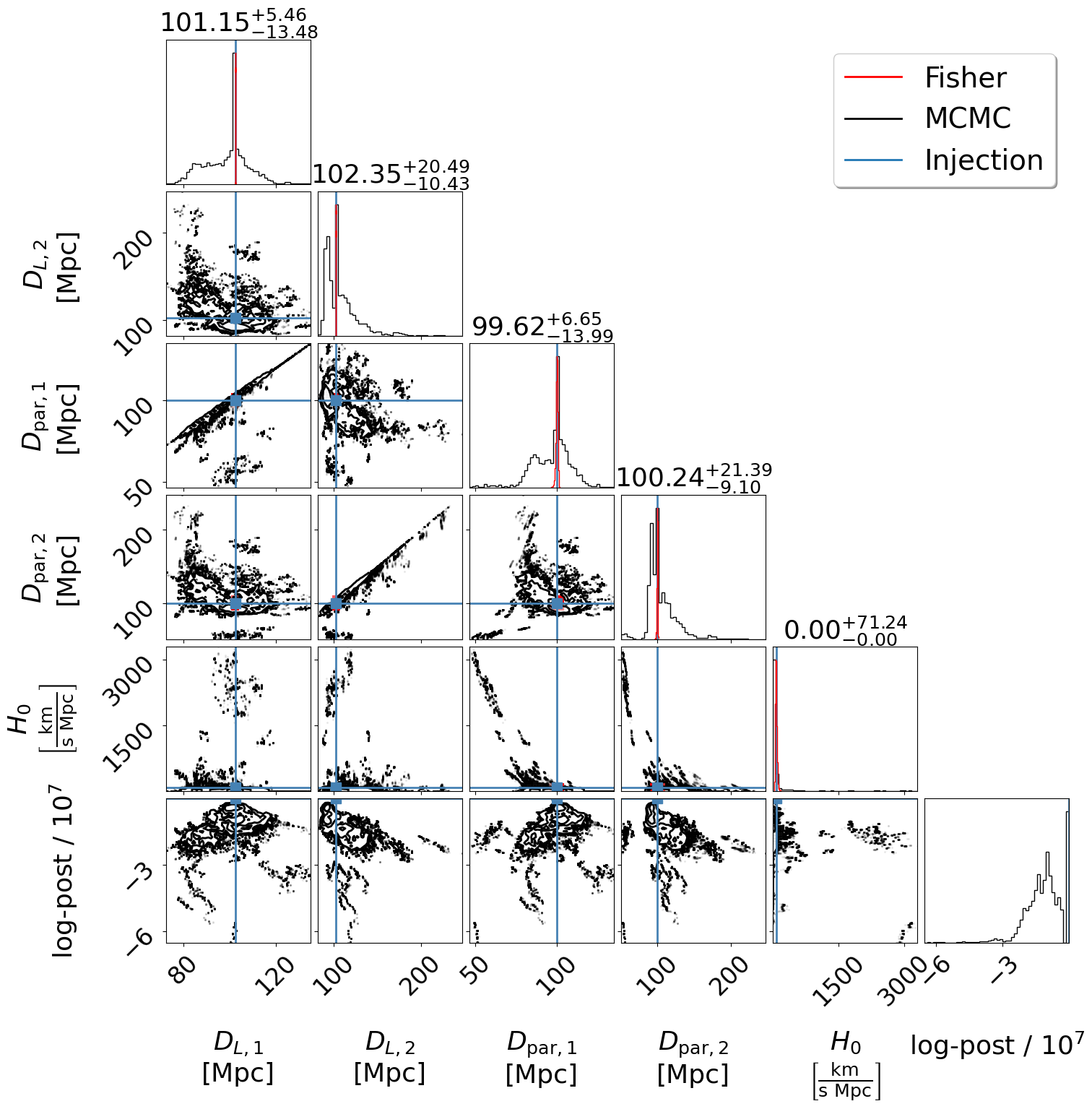}
        % --> located in:  Data15/mcgrathc/cosmo_study/IIB/2Sources_H0param_v4/data/figures/
          \end{subfigure}
          \hfill
          \begin{subfigure}[b]{0.48\linewidth}
          \centering
          \textit{``Cleaned'' Samples}
          \includegraphics[width=1\linewidth]{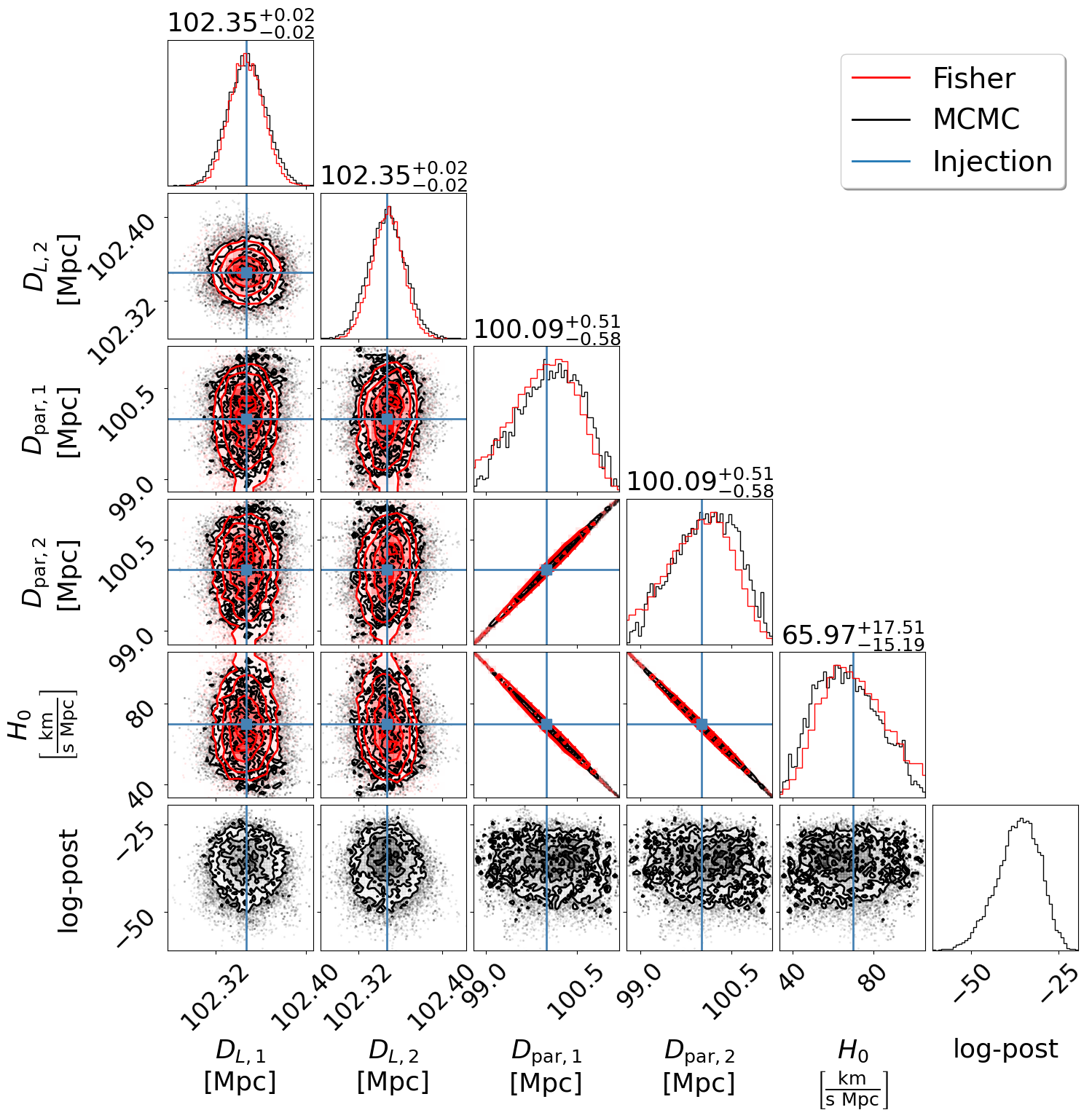}
        % --> located in:  Data15/mcgrathc/cosmo_study/IIB/2Sources_H0param_v4/data/figures/
          \end{subfigure}
        \caption{A targeted MCMC simulation of the recovery of $H_0$ for two sources (using the IIB model).  For simplicity, both sources are identical \textit{except} for being located on different parts of the sky.  Source 1 is at the default location (and matches the single source in Figure~\ref{fig: error envelopes}), while source 2 has sky angles $\theta_2 = \frac{\pi}{3}$ rad and $\phi_2 = \frac{\pi}{5}$ rad.  Here 40 pulsars from our fiducial PTA are timed with $\sigma = 0.2$ ns uncertainty (same as in Figure~\ref{fig: error envelopes}).  Both panels are of the same simulation, showing just the recovery of the two sources' distances ($D_L$ and $D_\mathrm{par}$), and the \textit{joint} recovery of $H_0$.  The bottom row shows the (un-normalized) log-posterior values.  For each pulsar, if we look at the first common mode where the common mode overlap strength criterion equation~\ref{eqn: 2 source mode condition} is $\leq 2$, then for this given source and PTA our common modal overlap criterion predicts that there will be 0 overlapping, 0 intermediate, and 40 resolved pulsar error envelopes.  For this reason, our prediction before even running the MCMC is that we should not find any strong support at secondary modes, making this is an ideal candidate for the $H_0$ measurement, which for this set of source and PTA parameters has a predicted $\mathrm{CV}_{H_0} = 0.257$.  Indeed, this is what we see here -- unlike in Figure~\ref{fig: error envelopes}, our walkers did not seem to find any high probability secondary modes.
        \textbf{Left-hand panel:}  For the full simulation we initialized multiple ``balls'' of walkers as described in Section~\ref{subsec:mcmc results}.  The bottom row shows many orders of magnitude difference in the values of the log-posterior for the walkers centered on the true mode versus the walkers away from the true mode.  If we ran the MCMC simulation long enough, we would expect that these low-probability walkers would eventually find and settle on the true mode.
        \textbf{Right-hand panel:}  This is the same corner plot, but now with all values of the log-posterior $\leq -2\times10^6$ removed from the data set.  This ``cleaned'' data reveals the sharply peaked true values in parameter space, which closely agree with the Fisher predictions.  
        }
    \label{fig: 2 Sources - MCMC target}
    \end{figure}

    The result of a targeted MCMC search is shown in Figure~\ref{fig: 2 Sources - MCMC target}, which is identical to the set-up in Figure~\ref{fig: error envelopes}, but now with the inclusion of a second SMBHB source (and using the full 2 Source IIB model).  For this example, our common modal overlap criterion predicts that all 40 pulsars would be resolved (again given that for each pulsar we look at the first common mode where the common mode overlap strength criterion equation~\ref{eqn: 2 source mode condition} is $\leq 2$).  Compare this to the single source IB model results in Figure~\ref{fig: error envelope progression}, where only 10 pulsars were resolved.  Therefore the addition of a second source changed this prediction such that the remaining 30 pulsars should be resolvable.  Additionally, the predicted $\mathrm{CV}_\mathrm{H0} = 0.257$.  Therefore our expectation was that we would find no meaningful support in the posterior at the secondary modes as compared to the primary mode, resulting in a measurement of the Hubble constant.  The left-hand panel of Figure~\ref{fig: 2 Sources - MCMC target} shows that the posterior has sharp peaks at the true parameter values with broad bases surrounding them.
    
    But if we consider the bottom row of the left-hand panel, which shows the (un-normalized) log-posterior values of these MCMC results, we also see that the samples span a large range of parameter space with very low probability.  Therefore if we try ``cleaning'' this data by simply removing all samples with log-posterior values below a certain threshold (in this case, we cleaned all values for which the log-posterior $\leq -2\times10^6$), then we obtain the results shown in the right-hand panel.  This reveals that walkers near the true mode very closely traced the Fisher matrix prediction, walkers away from the true mode had very low-probability support, and that there were no clearly defined modal peaks -- which starkly contrasts the results found in Figure~\ref{fig: error envelopes}.  Therefore these results support our conclusion about how two sources can help break the wrapping cycle degeneracy.
    
    It is important to point out that we have shown proof-of-principle in the ability to disentangle secondary posterior modes from the primary mode.  The caveat, as we see here is that the true mode is sharply peaked, and therefore may be difficult to find in a blind search over parameter space.  Other techniques may be needed to find these true modes, such as parallel tempering MCMC, nested sampling, or different jump proposals to improve sampling efficiency.  We leave this open to further investigation in future studies.

    \section{Measurement of the Hubble Constant}\label{sec:measurement of H0}

%---------------------------------------------------------------------------------------------------
    \subsection{Comparing \texorpdfstring{$H_0$}{H0} Measurements With One versus Two Gravitational Wave Sources}\label{subsec:1v2 sources H0 recovery}
    
\begin{figure}
        \centering
        \includegraphics[width=0.75\linewidth]{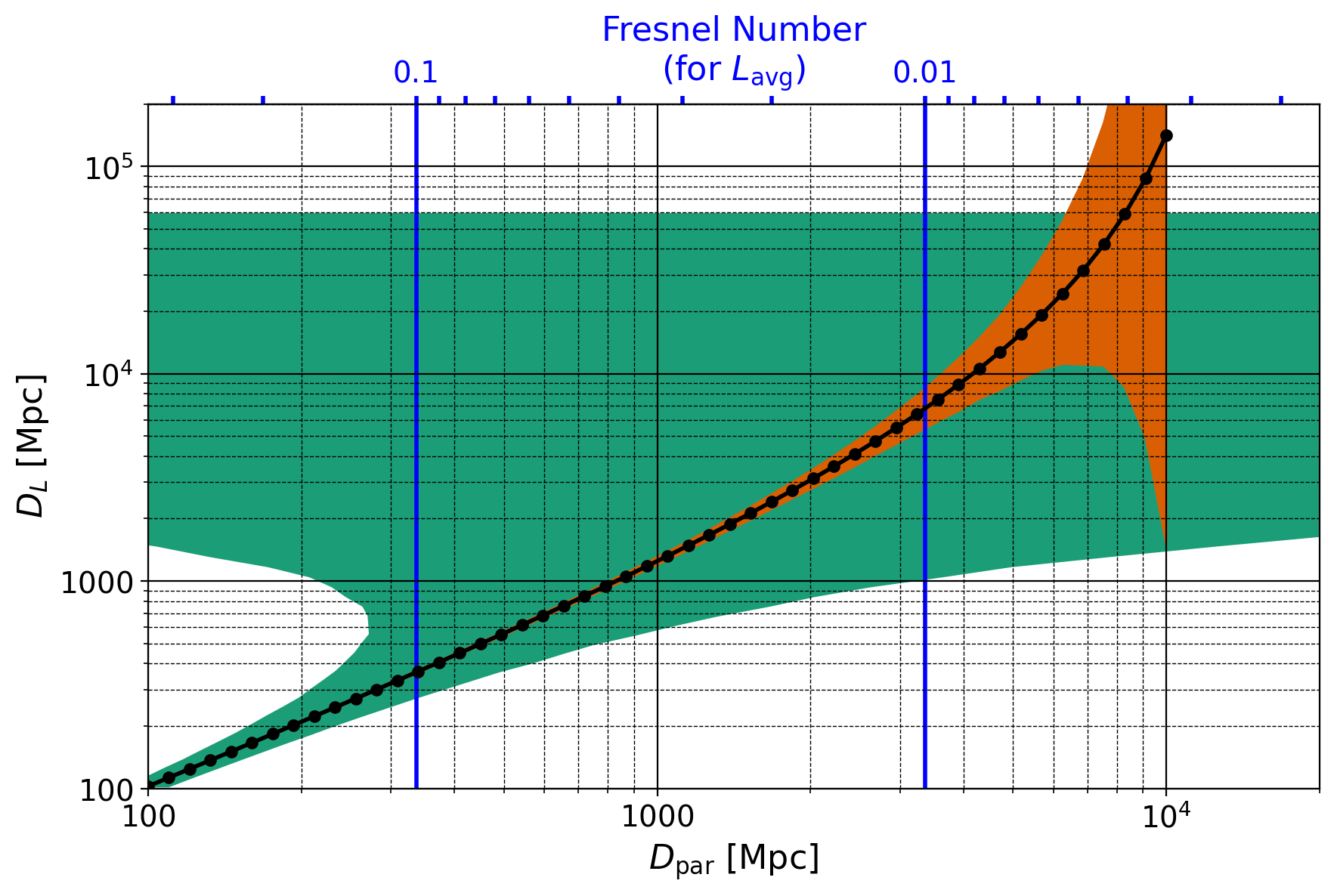}
        % --> located in:  cosmo_study/IIB/Fsurvey_1S/R-DL_paper/data/figures/RvDL_1.eps
    \caption{An example comparison of the uncertainties on the $D_L$ and $D_\mathrm{par}$ parameters using a Fisher analysis for a \textit{single} gravitational wave source (using the IIB model and default parameters).  For this simulation, we set uniform uncertainties $\sigma = 10$ ns and $\sigma_L = 0.1 \lambda_\mathrm{gw}$.  This PTA contains 100 pulsars from our fiducial PTA.  The orange region shows the uncertainty range on the luminosity distance, while the green region shows the uncertainty range on the parallax distance.  For reference, the Fresnel number (equation~\ref{eqn: Fresnel number}) for a pulsar at the average pulsar distance $L_\mathrm{avg}$ for this particular array is indicated on the top axis.  Generally, the luminosity distance is easier to measure from the frequency evolution, as compared to the parallax distance from the Fresnel corrections, which for this example, is measurable for $D_\mathrm{par}\lesssim 400$~Mpc.}
        \label{fig: 1 Source: R vs DL measurement}
    \end{figure}

    \citetalias{mcgrath2021} and \citetalias{DOrazio2021} found that with a single gravitational wave source we must know a priori the pulsar distances to sub-gravitational wavelength uncertainties in order to avoid the pulsar distance wrapping problem.  Only then can $D_\mathrm{par}$ be recovered with sufficient accuracy to measure the Hubble constant through equation~\ref{eqn: Hubble constant} (exact) or equation~\ref{eqn: low redshift R approximation} (approximate).  An example showing the recovery of $D_L$ compared to $D_\mathrm{par}$ for a single source and assuming pulsar distance uncertainties of $\sigma_L = 0.1 \lambda_\mathrm{gw}$ is shown in Figure~\ref{fig: 1 Source: R vs DL measurement}.  However, this level of precision in the pulsar distance measurement is far beyond current capabilities for all but the most nearby pulsars, making this a very challenging measurement restricted to gravitational wave sources within $\leq 100$~Mpc (\citetalias{DOrazio2021} and \citetalias{mcgrath2021}).
    
\begin{figure}
        \centering
        \begin{subfigure}[b]{1.0\linewidth}
          \centering
            \includegraphics[width=0.8\linewidth]{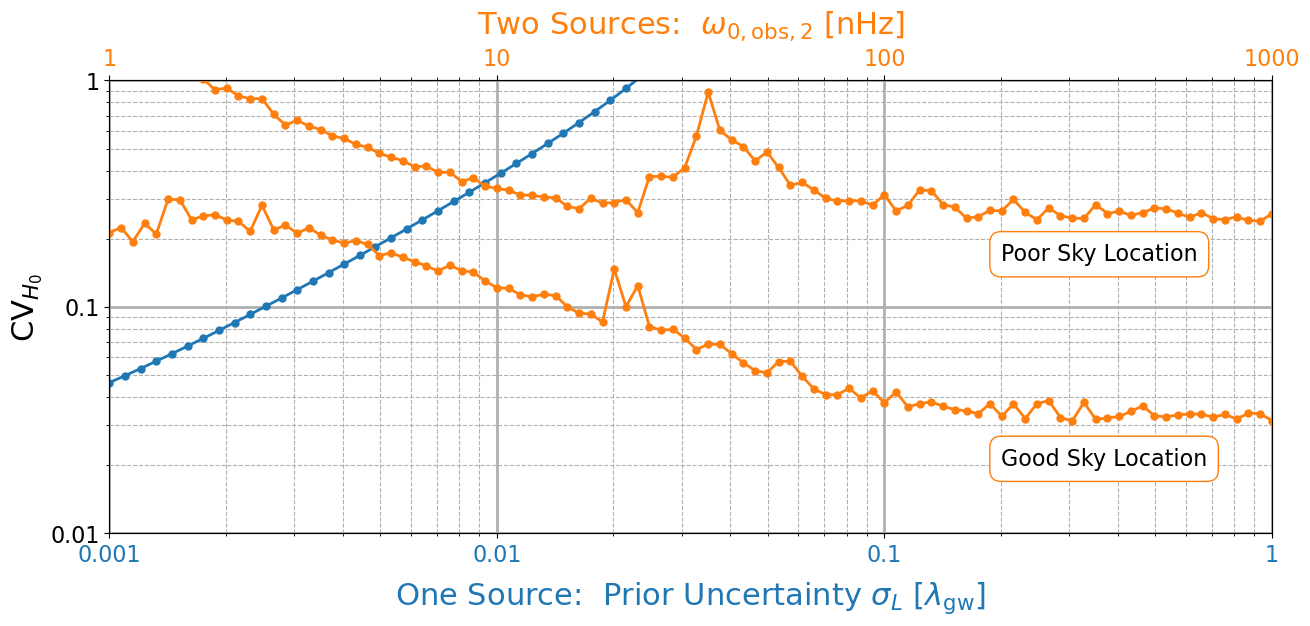}
        % --> located in:  cosmo_study/IIB/Fsurveys_2SH0param/omega02-H0_paper/data/figures/CVH0_1v2S.eps
          \end{subfigure} \\
          \begin{subfigure}[b]{1.0\linewidth}
          \centering
          \includegraphics[width=0.8\linewidth]{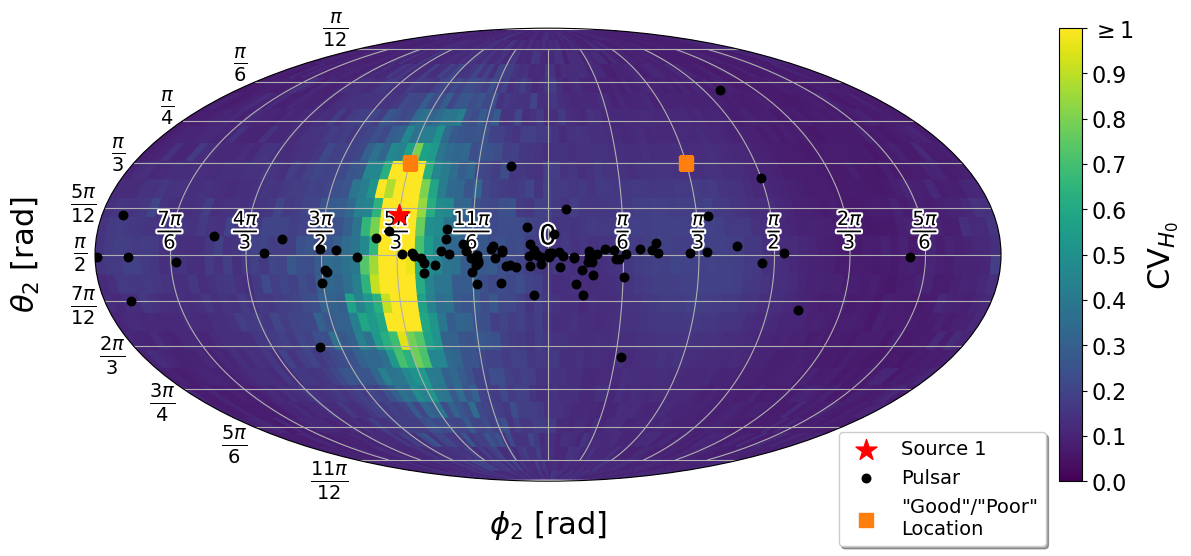}
        % --> located in:  cosmo_study/IIB/Fsurveys_2SH0param/sky_survey_paper/data/figures/doublesourcesky_2.eps
          \end{subfigure}
        \caption{Fisher analysis predictions of the recovery of $H_0$ (using the IIB model).  Both sources have $\mathcal{M}_\mathrm{obs} = 2\times10^9 \ \mathrm{M}_\odot$, default parameters otherwise, and the PTA contains 100 pulsars from the fiducial PTA with $\sigma = 1$ ns timing uncertainty.
        \textbf{Top panel:}  A comparison of the recovery of the Hubble constant using one vs. two gravitational wave sources.  For one source, the recovery of $H_0$ is plotted as a function of the required pulsar distance prior knowledge, set uniformly on \textit{all} pulsars in the PTA.  For two sources, \textit{no prior knowledge} on the pulsar distances is assumed. Instead, recovery of $H_0$ is shown as a function of the second source's frequency.  Both changing the second source's frequency as well as its sky position relative to the first source breaks the pulsar distance wrapping cycle degeneracy, so here we show two cases: one where the second source is in a ``good'' sky position ($\theta_2 = \frac{\pi}{3}$ rad, $\phi_2 = \frac{2\pi}{3}$ rad), and one where it is in a ``poor'' sky position ($\theta_2 = \frac{\pi}{3}$ rad, $\phi_2 = \frac{5\pi}{3}$ rad).  Both of these locations are marked on the sky plot in the bottom panel -- these two positions lie on the same latitude, but on different parts of the sky.
        \textbf{Bottom panel:} $H_0$ recovery with two sources, as a function of the second source's sky position.  Here we see a degenerate region centered on the first source (red star) and reflected about the galactic plane.  In the limit that the two sources become perfectly aligned the recovery of the Hubble constant becomes very difficult, even when the two sources have different frequencies.
    }
    \label{fig: 1v2 Source H0 recovery}
    \end{figure}

    The top panel of Figure~\ref{fig: 1v2 Source H0 recovery} shows a direct comparison of measuring $H_0$ using one versus two gravitational wave sources.  With one gravitational wave source, when performing a Fisher matrix analysis the only way to accurately invert the Fisher matrix requires that we first add in pulsar distance priors which would constrain the uncertainty on the pulsar distances down to the order of the wrapping cycle (see Section~\ref{subsec:fisher and mcmc}).  Otherwise, the Fisher matrix is ill-conditioned, due to the strong covariances introduced from the pulsar distances and parallax distance parameters.  So for a single source, the blue line in the top panel of Figure~\ref{fig: 1v2 Source H0 recovery} shows the recovery of $H_0$ as a function of the pulsar distance prior, which is a constraint applied to \textit{all} of the pulsars in the PTA uniformly. We see that $H_0$ can only be recovered in this example for distance priors smaller than a few per cent of the gravitational wave wavelength.
    
    With two gravitational wave sources, we can achieve the same level of accuracy in $H_0$ recovery without pulsar distance prior knowledge. The orange lines in the top panel of Figure~\ref{fig: 1v2 Source H0 recovery} demonstrate this for both a ``good'' and ``poor'' sky location. Simply having two sources at favorable sky separations and frequencies results in recovery of the Hubble constant better than a single source with sub-gravitational wavelength prior knowledge. Even unfavorable source sky positions still result in better recovery than with a single source.
    
    The bottom panel of Figure~\ref{fig: 1v2 Source H0 recovery} demonstrates $H_0$ recovery for all sky realizations of the relative position of the second source, assuming the first source is located at the position of the red star.  In the limit that the two sources become perfectly aligned with each other on the sky, the recovery of $H_0$ becomes effectively impossible.  This is a rather interesting result, because from our previous discussions in Sections~\ref{subsec: classifying the error envelope} and~\ref{subsec:breaking the wrapping cycle degeneracy} we originally predicted that as long as the source frequency of both sources were different (even for the same sky position), the wrapping cycle degeneracy should break.  Mathematically this still happens, but when testing various scenarios we found that in the case of perfect source sky alignment the model parameters for our PTA and sources needed to be highly favorable, making it very unlikely that such a circumstance would occur naturally.
    
    Hence, we see that the Fisher matrix analysis demonstrates an important result, namely that two sources remove the need for prohibitively precise pulsar distance measurements a priori in order to measure the Hubble constant.  Our one source simulation only obtained values of $\mathrm{CV}_{H_0} \sim 0.1$ when all pulsars distances were known with precision $\sigma_L < 0.01 \lambda_\mathrm{gw}$, which is far beyond our current capabilities.  However, two sources without any prior pulsar distance knowledge resulted in $H_0$ measurements ranging from a percent to 10's of percent, dependent on the relative sky positions of the two sources.
    
\begin{figure}
        \centering
          \begin{subfigure}[b]{0.48\linewidth}
          \centering
            \includegraphics[width=1\linewidth]{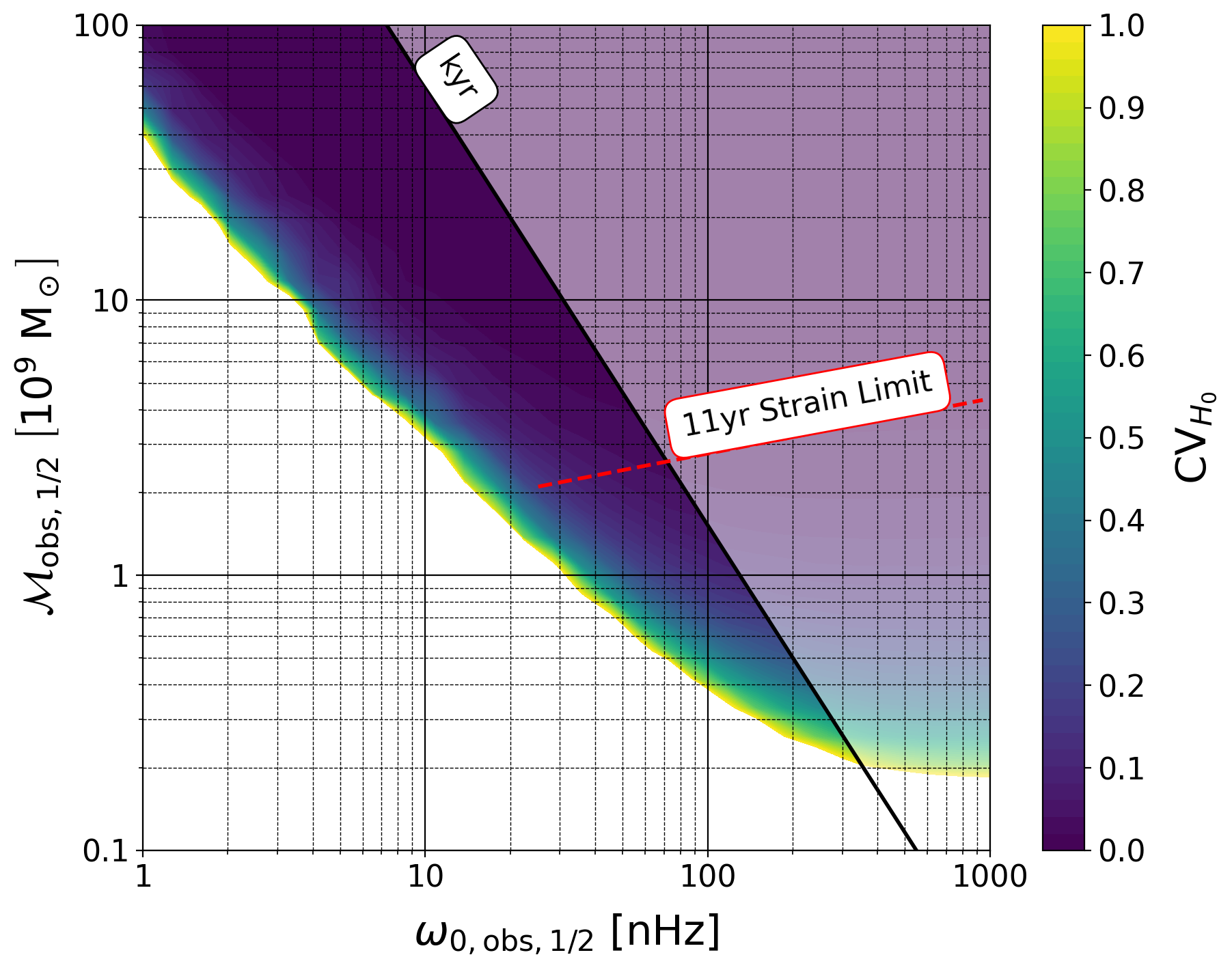}
          \end{subfigure}
        % --> located in:  cosmo_study/IIB/Fsurveys_2SH0param/M12-omega012-H0_paper/data/figures/M12omega012H0.png
          \hfill
          \begin{subfigure}[b]{0.48\linewidth}
          \centering
          \includegraphics[width=1\linewidth]{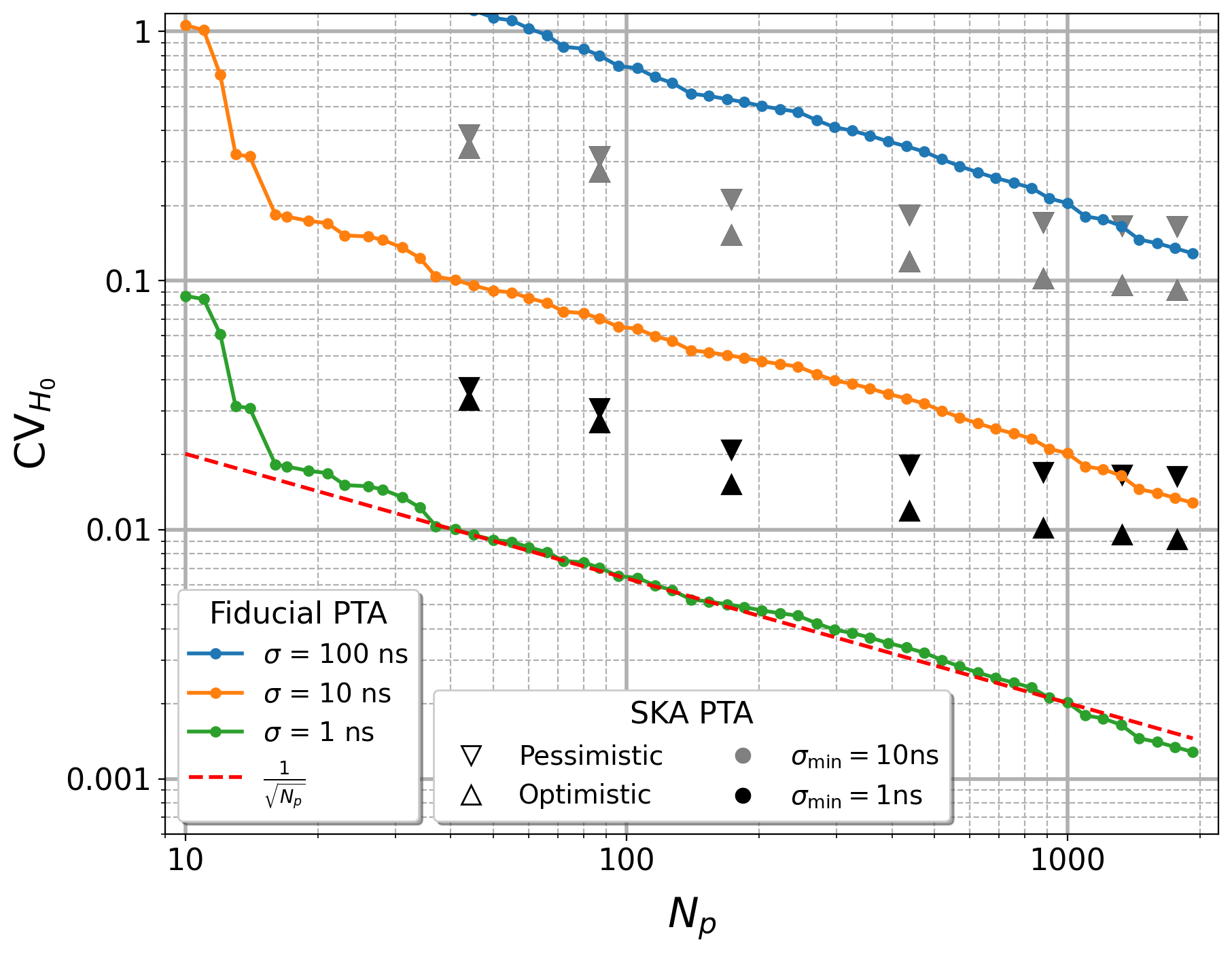}
          \end{subfigure}
        % --> located in:  cosmo_study/IIB/Fsurveys_2SH0param/Np-H0_paper/data/figures/CVH0vsNp_100Mpc-SKA_20220809.png
        \caption{Fisher analysis predictions for the recovery of $H_0$ (using the IIB model) with two sources, as a function of the observed chirp mass and orbital frequency of the source, and as a function PTA timing uncertainty and number of pulsars. For simplicity, both sources are assumed to be identical  (signified by the ``$1/2$'' notation) \textit{except} for being located on different parts of the sky.  Source 1 is at the default location, while source 2 has sky angles $\theta_2 = \frac{\pi}{3}$ rad and $\phi_2 = \frac{\pi}{5}$ rad.
        \textbf{Left-hand panel:}  Here 100 pulsars from our fiducial PTA are timed with $\sigma = 1$ ns uncertainty.  Hypothetical sources with $\Delta \tau_c < 1$ kyr have been shaded out as these represent highly chirping systems (see the general discussion of assumptions in \citetalias{mcgrath2021}).  Additionally the results of \citet{NG_11yr_cw} are used to include a rough estimate of NANOGrav's 11 yr continuous wave strain upper limit, $h_{0,\mathrm{11yr}} \approx 10^{-6} \frac{\omega_0}{\pi (1 \ \mathrm{Hz})}$, as reference.  
        \textbf{Right-hand panel:}  Here both sources have $\mathcal{M}_\mathrm{obs} = 10^{10} \ \mathrm{M}_\odot$.  From our fiducial PTA which has constant timing uncertainty, we see that the recovery of $H_0$ scales as $\mathrm{CV}_{H_0} \propto \sigma/\sqrt{N_p}$.  For variable timing uncertainty, the SKA PTA no longer achieves the same improvement by adding more pulsars, but rather now scales as $\mathrm{CV}_{H_0} \propto \sigma_\mathrm{min}$ at large $N_p$.
        }
    \label{fig: 2 Sources - H0 recovery}
    \end{figure}

    Figure~\ref{fig: 2 Sources - H0 recovery} shows Fisher analysis surveys of the $H_0$ measurability in terms of intrinsic source chirp mass and frequency, as well as PTA characteristics like timing accuracy and number of pulsars.  Not surprisingly, we measure $H_0$ best from sources with high chirp mass and frequency, since these produce strong frequency evolution effects in the signal. For reference, we include a rough estimate of the NANOGrav 11 yr continuous wave strain upper limit, $h_{0,\mathrm{11yr}} \approx 10^{-6} \frac{\omega_0}{\pi (1 \ \mathrm{Hz})}$ \citep[see fig. 3 of][]{NG_11yr_cw}, and indicate the 1 kyr coalescence time $\Delta \tau_c$ contour.  These two lines give a sense of what part of parameter space is most interesting.  The lightly shaded region where $\Delta \tau_c < 1$~kyr begins to break the original model assumptions, where frequency evolution becomes very significant (for reference, see assumption xiv of \citetalias{mcgrath2021}).  The NANOGrav upper limit suggests that sources above the line would have already been seen (out to $D_L = 102.35$ Mpc, which is where the distance is fixed in this plot) in the data if they existed.  This leaves us with a portion of parameter space for two sources with $\mathcal{M}_\mathrm{obs} \lesssim 2\times10^9 \ \mathrm{M}_\odot$ and frequencies $\omega_{0,\mathrm{obs}} \gtrsim 20$ nHz where we may be able to find sources that could recover a measurement of $H_0$ in the future.
    
    The right-hand panel of Figure~\ref{fig: 2 Sources - H0 recovery} shows that timing precision makes a significant difference in improving the measurement of $H_0$. For the fiducial PTAs, we gain an order of magnitude improvement in $\mathrm{CV}_{H_0}$ for the same improvement in $\sigma$.  Additionally, we see simply adding more fiducial PTA pulsars (randomly drawn from our distribution equation~\ref{eqn: fiducial PTA}) improves the entire network's ability to recover $H_0$.  Overall, recovery of $H_0$ for the fiducial PTAs with constant timing uncertainties scales as $\mathrm{CV}_{H_0} \sim \sigma/\sqrt{N_p}$.

    Also plotted in Figure~\ref{fig: 2 Sources - H0 recovery} are the $\mathrm{CV}_{H_0}$ recovery uncertainties found from various SKA-era PTAs, with more realistic timing uncertainties (see Section~\ref{S:SKAPTA}).  The upright triangles (optimistic) represent PTAs for which $\sigma \propto \mathrm{SNR}^{-1/2}$ while the upside-down triangles (pessimistic) represent PTAs for which $\sigma \propto \mathrm{SNR}^{-1}$ (equation \ref{eqn:sigmai}). Grey (black) markers represent PTAs for which the best-timed pulsars have $\sigma=1$~ns (10~ns). For small number arrays, the scaling with $N_p$ approaches the $1/\sqrt{N_p}$ scaling of the constant $\sigma$ case. For larger $N_p$ this relation limits towards a constant value. This is due to the way in which we add pulsars to the array, in ranked order of SNR per radial bin, and to the limited number of high SNR pulsars at large distance. As $N_p$ increases, we exhaust the supply of high-SNR pulsars at large distance from Earth so while $N_p$ increases, $N_p$ at large distance and high SNR effectively does not, and these are the pulsars which mostly strongly constrain $D_{\mathrm{par}}$ and hence $H_0$. Given these specific SKA-era PTAs, we find that arrays composed of $\sim 200$ pulsars timed to a best case precision of $\sigma_{\min}=1$~ns (10~ns) could measure the Hubble constant to within $2\%$ ($20\%$) precision using the two-source method described here.

%---------------------------------------------------------------------------------------------------
    \subsection{Prospects for Future SKA-like PTAs}\label{subsec:prospects for future SKA PTAs}
    
\begin{figure}
        \centering
          \begin{subfigure}[b]{0.48\linewidth}
          \centering
            \includegraphics[width=1\linewidth]{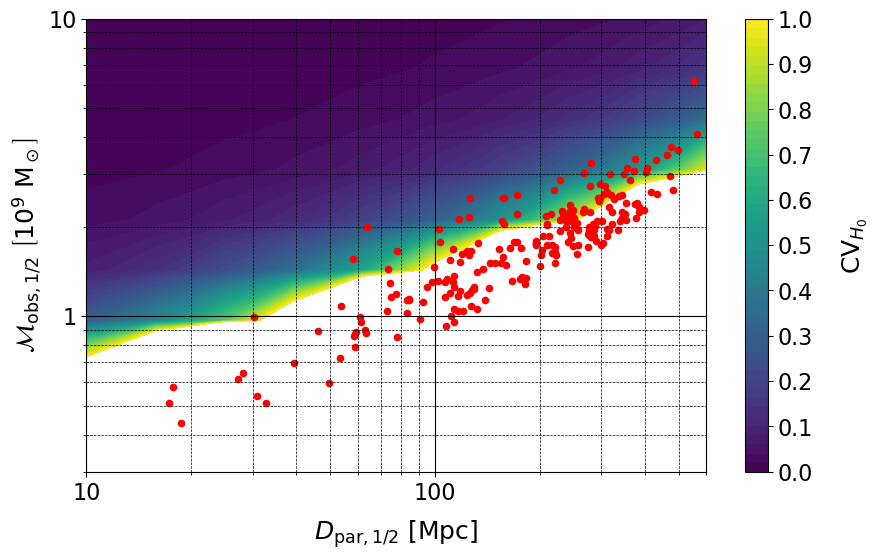}
        % --> located in:  cosmo_study/IIB/Fsurveys_2SH0param/SKA_PTA/variable_SNR/R12-M12-H0_paper/data/figures/NGsourcesSKA_20220809_438_sig1opt.png
          \end{subfigure}
          \hfill
          \begin{subfigure}[b]{0.48\linewidth}
          \centering
          \includegraphics[width=1\linewidth]{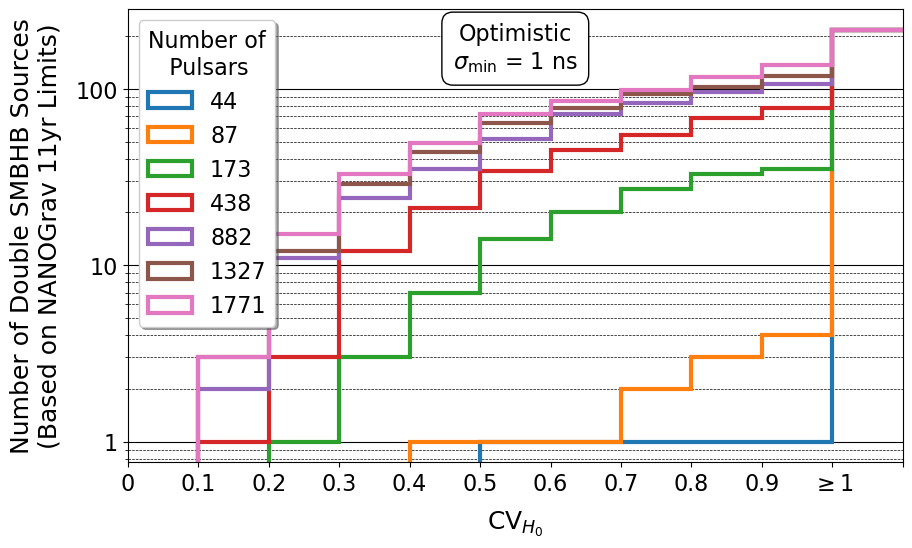}
        % --> located in:  cosmo_study/IIB/Fsurveys_2SH0param/SKA_PTA/variable_SNR/NG_sources_H0/data/figures/NGsources_ska_20220809_H0_opt1ns_cumulative.png
          \end{subfigure}
        \caption{
        \textbf{Left-hand panel:} Fisher analysis predictions of the recovery of $H_0$ (using the IIB model) with two sources using 438 ($5\%$) of our ``optimistic'' SKA-era pulsars with $\sigma_\mathrm{min} = 1$ ns, as a function of the observed chirp mass and parallax distance of the source. In the contour plot, for simplicity, both SMBHB sources are assumed to be identical (hence the ``$1/2$'' notation) \textit{except} for their sky location. Source 1 is at the default location, while source 2 has sky angles $\theta_2 = \frac{\pi}{3}$ rad and $\phi_2 = \frac{\pi}{5}$ rad.  Additionally all sources are simulated with $f_\mathrm{gw} = 10$~nHz (i.e. $\omega_{0,\mathrm{obs}} \approx 31.4$~nHz).  Plotted as red dots on top of the contours, we compare 216 hypothetical SMBHB sources within 600 Mpc published in \citet{NG_11yr_500MpcLimits}.  This gives us a sense of the quality required of an SKA-era PTA in order to make the $H_0$ measurement.  For the same number of pulsars, the pessimistic version of this PTA resulted in a marginal number of sources with $CV_\mathrm{H_0} < 1$. 
        \textbf{Right-hand panel:} For each of the 216 SMBHB sources indicated on the left-hand panel, we simulate \textit{two} identical SMBHBs with those mass and distance values (and with the indicated SKA-era PTAs), and bin them based on their level of recovery of $H_0$ in a cumulative distribution.  Both SMBHB sources are assumed to be identical with the same caveats as the left-hand panel.  As expected from the right-hand panel of Figure~\ref{fig: 2 Sources - H0 recovery}, adding more pulsars into the PTA improves the recovery of $H_0$ in more two-system scenarios.  For the PTAs with $>173$ pulsars, we begin to find double sources where $\mathrm{CV}_{H_0} \leq 0.3$ (comparable to what is shown in Figure~\ref{fig: 2 Sources - MCMC target}).  These plots do not rule out the possibility of a measurement of $H_0$ with a future SKA PTA, and instead suggests that there do exist hypothetical SMBHB candidates wherein this measurement may be viable. 
        }
    \label{fig: NG 11yr 500Mpc limits}
    % ---> Data taken from Table 3 of: The NANOGrav 11 yr Data Set: Limits on Supermassive Black Hole Binaries in Galaxies within 500 Mpc
    % ---> Chirp masses calculated from q values:  95% upper limit on the mass ratio at frequency 10nHz
    \end{figure}

    To further explore the prospects of using future SKA-like PTAs to measure the Hubble constant, we examine the possibility that a detectable SMBHB source exists.  Because closer sources within a few hundred Mpcs will have larger Fresnel effects, and hence are more likely to produce a measurement of $H_0$, we take an observational approach and consider the catalogue of PTA constraints on gravitational waves from massive SMBHBs in galaxies within 600 Mpc generated from the NANOGrav 11 yr data set in \cite{NG_11yr_500MpcLimits}.\footnote{We used the data from Table 3 -- the ``Mass,'' ``Dist,'' and ``$q_{95}$ (10.0)'' columns.} This includes 216 nearby galaxies for which PTA upper limits can constrain the mass ratios of putative SMBHBs at a given gravitational wave frequency (we consider the $f_\mathrm{gw} = 10$~nHz, circular orbit constraints). In the left-hand panel of Figure~\ref{fig: NG 11yr 500Mpc limits} we plot these limiting putative SMBHBs in chirp mass and distance space. Using a best-case, 438 pulsar, SKA-era PTA of Section~\ref{S:SKAPTA} ($5\%$ of all detected MSPs), we overlay contours of the corresponding $H_0$ recoverability $\mathrm{CV}_{H_0}$.  For this optimistic SKA-era PTA, there are 10's of possible nearby sources. 
    
    The right-hand panel of Figure~\ref{fig: NG 11yr 500Mpc limits} explores this further by taking each putative SMBHB (red dots) from the left-hand panel, assuming it has a twin at a different position on the sky, and computing the $H_0$ recovery uncertainty $\mathrm{CV}_{H_0}$.  The cumulative distribution reveals how increasing the size of our PTA increases the number of twin SMBHBs for which $H_0$ recovery at a given precision is possible.  With 438 pulsars or more, we find 10's of systems with $\mathrm{CV}_{H_0} \leq 0.5$, and over $30\%$ of all of these systems procure $\mathrm{CV}_{H_0} \leq 0.9$.  Over $5\%$ (i.e. more than $\sim12$) of these putative systems result in $\mathrm{CV}_{H_0} \leq 0.4$, and there are a few cases where we recover $\mathrm{CV}_{H_0} \sim 0.1$.
    
    Hence, an $\mathcal{O}(10\%)$, purely gravitational wave $H_0$ measurement is possible if two of these optimal putative SMBHBs exist.  Note that already for the `pessimistic,' $\sigma_{\min}=1$~ns SKA-era arrays (see the right-hand panel of Figure~\ref{fig: 2 Sources - H0 recovery}), the overlap between putative sources and $\mathrm{CV}_{H_0}\leq 1$ in Figure~\ref{fig: NG 11yr 500Mpc limits} becomes marginal. Hence these best-case PTAs are likely required.
    
    \section{Conclusions}\label{sec: conclusions}

In this work we have shown that future PTA experiments could make purely gravitational wave-based measurements of the Hubble constant.  This is made possible by accounting for the Fresnel curvature effects in the wavefront across the Earth-pulsar baseline.  By using the fully general Fresnel frequency evolution timing residual model, we can obtain two separate distance measurements to the source: the luminosity distance $D_L$ (from the frequency evolution effects) and the parallax distance $D_\mathrm{par}$ (from the Fresnel effects).

The measurement of the parallax distance is particularly challenging due to the pulsar distance wrapping problem.  Unless the distances to the pulsars in our array can be measured to sub-gravitational wavelength ($\sim$sub-parsec) precision, this measurement cannot be made with a single SMBHB source.  However, we demonstrate that \textit{two} SMBHB sources detected simultaneously break the wrapping cycle degeneracy, allowing a viable scenario for measuring the Hubble constant.  Two or more sources will actively calibrate the PTA by helping to identify the true pulsar distances from the degenerate secondary modes.  The ability to better resolve the pulsar distances using this method will be explored in a follow-up study (\textcolor{blue}{McGrath, D'Orazio, \& Creighton, in preparation}).  Additionally, this method may better improve the recovery of the luminosity distance $D_L$ (to \textit{both} sources), even at higher redshifts where the Fresnel effects may be too small to measure $D_\mathrm{par}$.  It would be interesting for future studies to investigate how significant this improvement is, compared to measuring $D_L$ of just a single source, since the luminosity distance by itself would still be useful for other astrophysical and cosmological studies.

For the two source problem, we explored the measurement of $H_0$ for a wide range of systems and PTAs.  We developed new Fisher matrix-based criteria which can quickly predict the ability of a particular source-PTA set-up to measure $H_0$, and facilitate follow-up with MCMC verification.  While our MCMC techniques are less complicated than other samplers for PTA problems, we find they support our proof-of-principle calculations, and we leave it to future work to improve upon methods for efficiently extracting the model parameters \citep[see for example,][]{samajdar_2022}.

With fiducial PTAs, with constant timing uncertainty $\sigma$ and $N_p$ pulsars, the precision in our simulated $H_0$ measurement scales as $\mathrm{CV}_{H_0} \sim \sigma / \sqrt{N_p}$.  With more realistic SKA-era PTAs, we observe a similar scaling for arrays with less than a few hundred pulsars and a saturation in measurement precision for larger arrays, where $\mathrm{CV}_{H_0} \sim \sigma_\mathrm{min}$.  We also find that hypothetical two-source SMBHB systems, based on the NANOGrav 11 yr data in~\cite{NG_11yr_500MpcLimits}, suggest that SKA-era PTAs may be capable of measuring $H_0$ to as low as $\mathrm{CV}_{H_0} \sim \mathcal{O}\left(0.1\right)$.  This will likely require fortuitous SMBHB sources and ``best-case'' PTAs (such that $\sigma_\mathrm{min} \sim 1$~ns to $10$~ns) containing several hundred pulsars. While we have not considered the possibility of more than two sources here, this will likely improve measurement prospects.

In conclusion, the existence of two individually resolvable gravitational wave signals from inspiraling SMBHBs allows a purely gravitational wave-based measurement of the Hubble constant with PTAs in the SKA-era and beyond.  Current and future $H_0$ measurements can achieve higher precision than what we envision here on that time-scale; for example, the standard sirens approach using binary neutron star mergers with electromagnetic counterparts could reach a 1\% level $H_0$ measurement by the 2030s~\citep{ChenFishHolzH0SS:2018}.  However, as is apparent from the current Hubble tension, precision does not guarantee agreement between measurement methods.  The value of the method discussed here, using only the gravitational messenger and geometry, is that it can provide a novel and independent measurement of $H_0$.

\section*{Acknowledgements}

This material is based upon work supported by the National Aeronautics and Space Administration (NASA) under award number 80GSFC21M0002.  This work was also supported by the National Science Foundation (NSF) PHY-1430284 [through the North American Nanohertz Observatory for Gravitational Waves (NANOGrav's) Physics Frontier Center], PHY-1912649, and PHY-2207728, and by UW-Milwaukee's computational resources PHY-1626190.  DJD received funding from the European Union's Horizon 2020 research and innovation programme under Marie Sklodowska-Curie grant agreement No. 101029157, and from the Danish Independent Research Fund through Sapere Aude Starting Grant No. 121587.  We thank the referee, Neil Cornish, for a constructive report.

\section*{Data Availability}
Calculations in this paper were performed using code developed by the authors, and the data are available on reasonable request to the authors.

%%%%%%%%%%%%%%%%%%%%%%%%%%%%%%%%%%%%%%%%%%%%%%%%%%



\bibliographystyle{mnras}
\bibliography{references}

\begin{thebibliography}{}
\makeatletter
\relax
\def\mn@urlcharsother{\let\do\@makeother \do\$\do\&\do\#\do\^\do\_\do\%\do\~}
\def\mn@doi{\begingroup\mn@urlcharsother \@ifnextchar [ {\mn@doi@}
  {\mn@doi@[]}}
\def\mn@doi@[#1]#2{\def\@tempa{#1}\ifx\@tempa\@empty \href
  {http://dx.doi.org/#2} {doi:#2}\else \href {http://dx.doi.org/#2} {#1}\fi
  \endgroup}
\def\mn@eprint#1#2{\mn@eprint@#1:#2::\@nil}
\def\mn@eprint@arXiv#1{\href {http://arxiv.org/abs/#1} {{\tt arXiv:#1}}}
\def\mn@eprint@dblp#1{\href {http://dblp.uni-trier.de/rec/bibtex/#1.xml}
  {dblp:#1}}
\def\mn@eprint@#1:#2:#3:#4\@nil{\def\@tempa {#1}\def\@tempb {#2}\def\@tempc
  {#3}\ifx \@tempc \@empty \let \@tempc \@tempb \let \@tempb \@tempa \fi \ifx
  \@tempb \@empty \def\@tempb {arXiv}\fi \@ifundefined
  {mn@eprint@\@tempb}{\@tempb:\@tempc}{\expandafter \expandafter \csname
  mn@eprint@\@tempb\endcsname \expandafter{\@tempc}}}

\bibitem[\protect\citeauthoryear{Aggarwal et~al.,}{Aggarwal
  et~al.}{2019}]{NG_11yr_cw}
Aggarwal K.,  et~al., 2019, \mn@doi [\apj] {10.3847/1538-4357/ab2236}, 880, 116

\bibitem[\protect\citeauthoryear{Arzoumanian et~al.,}{Arzoumanian
  et~al.}{2018}]{NG_11yr_data}
Arzoumanian Z.,  et~al., 2018, \mn@doi [\apjs] {10.3847/1538-4365/aab5b0}, 235,
  37

\bibitem[\protect\citeauthoryear{{Arzoumanian} et~al.,}{{Arzoumanian}
  et~al.}{2021}]{NG_11yr_500MpcLimits}
{Arzoumanian} Z.,  et~al., 2021, \mn@doi [\apj] {10.3847/1538-4357/abfcd3},
  \href {https://ui.adsabs.harvard.edu/abs/2021ApJ...914..121A} {914, 121}

\bibitem[\protect\citeauthoryear{Babak \& Sesana}{Babak \&
  Sesana}{2012}]{babak_multipleSMBHBsI}
Babak S.,  Sesana A.,  2012, \mn@doi [Phys. Rev. D]
  {10.1103/PhysRevD.85.044034}, 85, 044034

\bibitem[\protect\citeauthoryear{{Bates}, {Lorimer}, {Rane}  \&
  {Swiggum}}{{Bates} et~al.}{2014}]{PsrPopPy:2014}
{Bates} S.~D.,  {Lorimer} D.~R.,  {Rane} A.,   {Swiggum} J.,  2014, \mn@doi
  [\mnras] {10.1093/mnras/stu157}, \href
  {https://ui.adsabs.harvard.edu/abs/2014MNRAS.439.2893B} {439, 2893}

\bibitem[\protect\citeauthoryear{{Caldwell}}{{Caldwell}}{1993}]{greens_functions_1993}
{Caldwell} R.~R.,  1993, \mn@doi [\prd] {10.1103/PhysRevD.48.4688}, \href
  {https://ui.adsabs.harvard.edu/abs/1993PhRvD..48.4688C} {48, 4688}

\bibitem[\protect\citeauthoryear{{Chen}, {Fishbach}  \& {Holz}}{{Chen}
  et~al.}{2018}]{ChenFishHolzH0SS:2018}
{Chen} H.-Y.,  {Fishbach} M.,   {Holz} D.~E.,  2018, \mn@doi [\nat]
  {10.1038/s41586-018-0606-0}, \href
  {https://ui.adsabs.harvard.edu/abs/2018Natur.562..545C} {562, 545}

\bibitem[\protect\citeauthoryear{{Corbin} \& {Cornish}}{{Corbin} \&
  {Cornish}}{2010}]{CC_main_paper}
{Corbin} V.,  {Cornish} N.~J.,  2010, arXiv e-prints, \href
  {https://ui.adsabs.harvard.edu/abs/2010arXiv1008.1782C} {p. arXiv:1008.1782}

\bibitem[\protect\citeauthoryear{{Cordes} \& {Chernoff}}{{Cordes} \&
  {Chernoff}}{1997}]{Cordes_Chernoff:1997}
{Cordes} J.~M.,  {Chernoff} D.~F.,  1997, \mn@doi [\apj] {10.1086/304179},
  \href {https://ui.adsabs.harvard.edu/abs/1997ApJ...482..971C} {482, 971}

\bibitem[\protect\citeauthoryear{Creighton \& Anderson}{Creighton \&
  Anderson}{2011}]{creighton_anderson_2011}
Creighton J. D.~E.,  Anderson W.~G.,  2011, Gravitational-Wave Physics and
  Astronomy: An Introduction to Theory, Experiment and Data Analysis.
Wiley-VCH

\bibitem[\protect\citeauthoryear{D'Orazio \& Loeb}{D'Orazio \&
  Loeb}{2021}]{DOrazio2021}
D'Orazio D.~J.,  Loeb A.,  2021, \mn@doi [\prd] {10.1103/PhysRevD.104.063015},
  104, 063015

\bibitem[\protect\citeauthoryear{{Deller} et~al.,}{{Deller}
  et~al.}{2019}]{pulsar_parallax2019}
{Deller} A.~T.,  et~al., 2019, \mn@doi [\apj] {10.3847/1538-4357/ab11c7}, \href
  {https://ui.adsabs.harvard.edu/abs/2019ApJ...875..100D} {875, 100}

\bibitem[\protect\citeauthoryear{{Deng} \& {Finn}}{{Deng} \&
  {Finn}}{2011}]{DF_main_paper}
{Deng} X.,  {Finn} L.~S.,  2011, \mn@doi [\mnras]
  {10.1111/j.1365-2966.2010.17913.x}, \href
  {https://ui.adsabs.harvard.edu/abs/2011MNRAS.414...50D} {414, 50}

\bibitem[\protect\citeauthoryear{{Feeney}, {Peiris}, {Williamson}, {Nissanke},
  {Mortlock}, {Alsing}  \& {Scolnic}}{{Feeney}
  et~al.}{2019}]{Feeney_standardsirens}
{Feeney} S.~M.,  {Peiris} H.~V.,  {Williamson} A.~R.,  {Nissanke} S.~M.,
  {Mortlock} D.~J.,  {Alsing} J.,   {Scolnic} D.,  2019, \mn@doi [Physical
  Review Letters] {10.1103/PhysRevLett.122.061105}, \href
  {https://ui.adsabs.harvard.edu/abs/2019PhRvL.122f1105F} {122, 061105}

\bibitem[\protect\citeauthoryear{{Foreman-Mackey}, {Hogg}, {Lang}  \&
  {Goodman}}{{Foreman-Mackey} et~al.}{2013}]{emcee}
{Foreman-Mackey} D.,  {Hogg} D.~W.,  {Lang} D.,   {Goodman} J.,  2013, \mn@doi
  [\pasp] {10.1086/670067}, \href
  {https://ui.adsabs.harvard.edu/abs/2013PASP..125..306F} {125, 306}

\bibitem[\protect\citeauthoryear{{Ghosh}, {Biswas}  \& {Bose}}{{Ghosh}
  et~al.}{2022}]{ghosh_2022}
{Ghosh} T.,  {Biswas} B.,   {Bose} S.,  2022, arXiv e-prints, \href
  {https://ui.adsabs.harvard.edu/abs/2022arXiv220311756G} {p. arXiv:2203.11756}

\bibitem[\protect\citeauthoryear{{Hogg}}{{Hogg}}{1999}]{hogg_distances}
{Hogg} D.~W.,  1999, arXiv e-prints, \href
  {https://ui.adsabs.harvard.edu/abs/1999astro.ph..5116H} {pp
  astro--ph/9905116}

\bibitem[\protect\citeauthoryear{{Holz} \& {Hughes}}{{Holz} \&
  {Hughes}}{2005}]{holz_standardsirens}
{Holz} D.~E.,  {Hughes} S.~A.,  2005, \mn@doi [\apj] {10.1086/431341}, \href
  {https://ui.adsabs.harvard.edu/abs/2005ApJ...629...15H} {629, 15}

\bibitem[\protect\citeauthoryear{{Iacovelli}, {Mancarella}, {Foffa}  \&
  {Maggiore}}{{Iacovelli} et~al.}{2022}]{gwfast_fisher}
{Iacovelli} F.,  {Mancarella} M.,  {Foffa} S.,   {Maggiore} M.,  2022, arXiv
  e-prints, \href {https://ui.adsabs.harvard.edu/abs/2022arXiv220702771I} {p.
  arXiv:2207.02771}

\bibitem[\protect\citeauthoryear{{Kelley}, {Blecha}, {Hernquist}, {Sesana}  \&
  {Taylor}}{{Kelley} et~al.}{2018}]{Kelley+2018}
{Kelley} L.~Z.,  {Blecha} L.,  {Hernquist} L.,  {Sesana} A.,   {Taylor} S.~R.,
  2018, \mn@doi [\mnras] {10.1093/mnras/sty689}, \href
  {https://ui.adsabs.harvard.edu/abs/2018MNRAS.477..964K} {477, 964}

\bibitem[\protect\citeauthoryear{{Lee}, {Wex}, {Kramer}, {Stappers}, {Bassa},
  {Janssen}, {Karuppusamy}  \& {Smits}}{{Lee} et~al.}{2011}]{GWastro_Lee2011}
{Lee} K.~J.,  {Wex} N.,  {Kramer} M.,  {Stappers} B.~W.,  {Bassa} C.~G.,
  {Janssen} G.~H.,  {Karuppusamy} R.,   {Smits} R.,  2011, \mn@doi [\mnras]
  {10.1111/j.1365-2966.2011.18622.x}, \href
  {https://ui.adsabs.harvard.edu/abs/2011MNRAS.414.3251L} {414, 3251}

\bibitem[\protect\citeauthoryear{{Lorimer} et~al.,}{{Lorimer}
  et~al.}{2006}]{Lorimer+2006}
{Lorimer} D.~R.,  et~al., 2006, \mn@doi [\mnras]
  {10.1111/j.1365-2966.2006.10887.x}, \href
  {https://ui.adsabs.harvard.edu/abs/2006MNRAS.372..777L} {372, 777}

\bibitem[\protect\citeauthoryear{Maggiore}{Maggiore}{2008}]{maggiore_2008}
Maggiore M.,  2008, Gravitational Waves: Theory and Experiments.
~ Vol. 1, Oxford University Press

\bibitem[\protect\citeauthoryear{Maggiore}{Maggiore}{2018}]{maggiore_2018}
Maggiore M.,  2018, Gravitational Waves: Astrophysics and Cosmology.
~ Vol. 2, Oxford University Press

\bibitem[\protect\citeauthoryear{{McGrath} \& {Creighton}}{{McGrath} \&
  {Creighton}}{2021}]{mcgrath2021}
{McGrath} C.,  {Creighton} J.,  2021, \mn@doi [\mnras]
  {10.1093/mnras/stab1417}, \href
  {https://ui.adsabs.harvard.edu/abs/2021MNRAS.tmp.1391M} {}

\bibitem[\protect\citeauthoryear{{Messenger} \& {Read}}{{Messenger} \&
  {Read}}{2012}]{messenger_gwH0}
{Messenger} C.,  {Read} J.,  2012, \mn@doi [\prl]
  {10.1103/PhysRevLett.108.091101}, \href
  {https://ui.adsabs.harvard.edu/abs/2012PhRvL.108i1101M} {108, 091101}

\bibitem[\protect\citeauthoryear{Petiteau, Babak, Sesana  \& de
  Ara\'ujo}{Petiteau et~al.}{2013}]{babak_multipleSMBHBsII}
Petiteau A.,  Babak S.,  Sesana A.,   de Ara\'ujo M.,  2013, \mn@doi [Phys.
  Rev. D] {10.1103/PhysRevD.87.064036}, 87, 064036

\bibitem[\protect\citeauthoryear{Qian, Mohanty  \& Wang}{Qian
  et~al.}{2022}]{qian_2022}
Qian Y.-Q.,  Mohanty S.~D.,   Wang Y.,  2022, \mn@doi [Phys. Rev. D]
  {10.1103/PhysRevD.106.023016}, 106, 023016

\bibitem[\protect\citeauthoryear{{Samajdar} et~al.,}{{Samajdar}
  et~al.}{2022}]{samajdar_2022}
{Samajdar} A.,  et~al., 2022, arXiv e-prints, \href
  {https://ui.adsabs.harvard.edu/abs/2022arXiv220504332S} {p. arXiv:2205.04332}

\bibitem[\protect\citeauthoryear{Schneider}{Schneider}{2015}]{schneider_2015}
Schneider P.,  2015, Extragalactic Astronomy and Cosmology: An Introduction,
  2nd edn.
Springer, p. 177–192, \mn@doi{10.1007/978-3-642-54083-7}

\bibitem[\protect\citeauthoryear{Schutz}{Schutz}{1986}]{Schutz_1986}
Schutz B.~F.,  1986, \mn@doi [Nature] {10.1038/323310a0}, 323, 310

\bibitem[\protect\citeauthoryear{{Shiralilou}, {Raaijmakers}, {Duboeuf},
  {Nissanke}, {Foucart}, {Hinderer}  \& {Williamson}}{{Shiralilou}
  et~al.}{2022}]{Shiralilou_2022}
{Shiralilou} B.,  {Raaijmakers} G.,  {Duboeuf} B.,  {Nissanke} S.,  {Foucart}
  F.,  {Hinderer} T.,   {Williamson} A.,  2022, arXiv e-prints, \href
  {https://ui.adsabs.harvard.edu/abs/2022arXiv220711792S} {p. arXiv:2207.11792}

\bibitem[\protect\citeauthoryear{{Smits}, {Kramer}, {Stappers}, {Lorimer},
  {Cordes}  \& {Faulkner}}{{Smits} et~al.}{2009}]{SmitsSKA+2009}
{Smits} R.,  {Kramer} M.,  {Stappers} B.,  {Lorimer} D.~R.,  {Cordes} J.,
  {Faulkner} A.,  2009, \mn@doi [\aap] {10.1051/0004-6361:200810383}, \href
  {https://ui.adsabs.harvard.edu/abs/2009A&A...493.1161S} {493, 1161}

\bibitem[\protect\citeauthoryear{{{The LIGO, Virgo, \& KAGRA
  Collaborations}}}{{{The LIGO, Virgo, \& KAGRA
  Collaborations}}}{2021}]{gw170817_h0_updated}
{{The LIGO, Virgo, \& KAGRA Collaborations}} 2021, arXiv e-prints, \href
  {https://ui.adsabs.harvard.edu/abs/2021arXiv211103604T} {p. arXiv:2111.03604}

\bibitem[\protect\citeauthoryear{{{The LIGO, Virgo, 1M2H, Dark Energy Camera
  GW-EM, \& DES Collaborations}}}{{{The LIGO, Virgo, 1M2H, Dark Energy Camera
  GW-EM, \& DES Collaborations}}}{2017}]{gw170817_h0}
{{The LIGO, Virgo, 1M2H, Dark Energy Camera GW-EM, \& DES Collaborations}}
  2017, \mn@doi [\nat] {10.1038/nature24471}, \href
  {https://ui.adsabs.harvard.edu/abs/2017Natur.551...85A} {551, 85}

\bibitem[\protect\citeauthoryear{{Verbiest} \& {Shaifullah}}{{Verbiest} \&
  {Shaifullah}}{2018}]{Verbiest_timing_noise:2018}
{Verbiest} J. P.~W.,  {Shaifullah} G.~M.,  2018, \mn@doi [Classical and Quantum
  Gravity] {10.1088/1361-6382/aac412}, \href
  {https://ui.adsabs.harvard.edu/abs/2018CQGra..35m3001V} {35, 133001}

\bibitem[\protect\citeauthoryear{{Wang}, {Zhang}, {Shao}  \& {Zhang}}{{Wang}
  et~al.}{2022}]{pta_standardsiren}
{Wang} L.-F.,  {Zhang} G.-P.,  {Shao} Y.,   {Zhang} X.,  2022, arXiv e-prints,
  \href {https://ui.adsabs.harvard.edu/abs/2022arXiv220100607W} {p.
  arXiv:2201.00607}

\bibitem[\protect\citeauthoryear{{Weinberg}}{{Weinberg}}{1972}]{weinberg1972}
{Weinberg} S.,  1972, Gravitation and Cosmology: Principles and Applications of
  the General Theory of Relativity.
New York: Wiley

\makeatother
\end{thebibliography}

%\begin{thebibliography}{99}
%\end{thebibliography}

%%%%%%%%%%%%%%%%%%%%%%%%%%%%%%%%%%%%%%%%%%%%%%%%%%

%%%%%%%%%%%%%%%%% APPENDICES %%%%%%%%%%%%%%%%%%%%%

\appendix

    \section{Cosmological Distances}\label{app: cosmological distances}

For a gravitational wave propagating from our source (at comoving coordinate distance $R$ and redshift $z$) to the Earth (``infalling''), we define the following ``boundary'' conditions:
\begin{align*}
    \textbf{Source:}  & \quad r = R, \quad z' = z, \quad t' = t_\mathrm{ret} , \\
    \textbf{Earth:} & \quad r = 0, \quad z' = 0, \quad t' = t , 
\end{align*}
where our reference frame is centered on the Earth.  From the Friedmann equations and the definition of the redshift in terms of the present day $t_0$ scale factor, $1+z = \frac{a(t_0)}{a(t)}$, we explicity list the following important quantities~\citep{weinberg1972, hogg_distances, schneider_2015}:
\begin{align}
    \textbf{Expansion Rate} &\qquad H &\hspace{-2.1cm}\equiv \ & \left(\frac{\dot{a}(t)}{a(t)}\right) = H_0 E ,  \label{eqn: expansion rate} \\
    \textbf{Dimensionless Hubble Function} &\qquad E &\hspace{-2.1cm}\equiv \ & \left[\frac{\Omega_r}{a^4(t)} + \frac{\Omega_m}{a^3(t)} + \Omega_\Lambda + \frac{\Omega_k}{a^2(t)}\right]^{1/2} , \nonumber \\
    & &\hspace{-2.1cm}= \ & \left[\frac{(1+z)^4}{a^4(t_0)}\Omega_r + \frac{(1+z)^3}{a^3(t_0)}\Omega_m + \Omega_\Lambda + \frac{(1+z)^2}{a^2(t_0)}\Omega_k\right]^{1/2} , \label{eqn: dimensionless Hubble function} \\
    \textbf{Curvature Density} &\qquad \Omega_k &\hspace{-2.1cm}\equiv \ & a^2(t_0) \left[1-\frac{\Omega_r}{a^4(t_0)} - \frac{\Omega_m}{a^3(t_0)} - \Omega_\Lambda\right] = -k D_H^2 , \label{eqn: curvature density parameter} \\
    \textbf{Hubble Distance} &\qquad D_H &\hspace{-2.1cm}\equiv \ & \frac{c}{H_0} , \label{eqn: Hubble distance} \\
    \textbf{Line-of-Sight Comoving Distance} &\qquad D_c &\hspace{-2.1cm}\equiv \ & D_H \int^z_0 \frac{\dd z'}{E} , \nonumber \\
    & & \hspace{-2.1cm}= \ & c a(t_0)\int^t_{t_\mathrm{ret}} \frac{\dd t'}{a},  \label{eqn: line-of-sight comoving distance Dc} \\
    \textbf{Parallax Distance} &\qquad D_\mathrm{par} &\hspace{-2.1cm}\equiv \ & a(t_0) \frac{R}{\sqrt{1-kR^2}} , \label{eqn: parallax distance Dpar}
\end{align}
where $H_0$ is the Hubble constant and $\Omega_r$, $\Omega_m$, and $\Omega_\Lambda$ are the radiation, matter, and vacuum density parameters, respectively.  For completeness here we will continue to write the present day scale factor explicitly, but typical convention is to normalize this to $a(t_0) \equiv 1$.  The two forms of the line-of-sight comoving distance in equation~\ref{eqn: line-of-sight comoving distance Dc} are made through the change of variables $t \rightarrow z$ using the definition of $z$ and equation~\ref{eqn: expansion rate}.

The luminosity distance of our source is defined by considering the energy flux $F_\mathrm{obs}$ (in this case, of our gravitational waves) as measured by the observer \citep{maggiore_2008}.  In a cosmologically expanding universe the observed energy $E_\mathrm{obs}$ is redshifted compared to the energy that was emitted in the source frame $E_\mathrm{s}$, such that $E_\mathrm{s} = (1+z) E_\mathrm{obs}$.  Additionally, time dilation means the time measured by the observer and source clocks are related by $t_\mathrm{s} = \frac{t_\mathrm{obs}}{(1+z)}$.  Finally, from the FLRW metric equation~\ref{eqn: FLRW metric}, the flux at time $t$ will be spread over a total area of $A(t) = 4\pi a^2(t) R^2$.  This means that we can write:
\begin{equation}
    F_\mathrm{obs}(t) \ \ \equiv \ \ \frac{L_\mathrm{obs}}{A(t)} \ \ = \ \ \frac{L_\mathrm{s}}{(1+z)^2 4\pi a^2(t) R^2} \ \ = \ \ \frac{L_\mathrm{s}}{4\pi \Big[(1+z)^2 a^2(t) R^2\Big]} \ \ \equiv \ \ \frac{L_\mathrm{s}}{4\pi D_L^2} ,
\label{eqn: luminosity distance flux relationship}
\end{equation}
where in the final equality we define our luminosity distance $D_L$ such that the observed flux matches our standard notion of flux, but as a function of the source frame luminosity $L_\mathrm{s}$.  Therefore at the present time, we have:
\begin{equation}
    D_L \equiv (1+z) a(t_0) R .
\label{eqn: DL - R relation}
\end{equation}

Next we draw the connection between the comoving coordinate distance $R$, the luminosity distance $D_L$, and the line-of-sight comoving distance $D_c$.  Integrating the metric equation~\ref{eqn: FLRW metric} along the radial ($\dd \theta=\dd \phi=0)$, null ($\dd s=0$), infalling path of the gravitational wave, and using equation~\ref{eqn: line-of-sight comoving distance Dc} gives us:
\begin{align*}
    -\int^0_R \frac{\dd r}{\sqrt{1-kr^2}} &= \int^t_0 \frac{c}{a(t')}\dd t' , \\
    &= \frac{D_c}{a(t_0)},
\end{align*}
The solution of the left-hand integral depends on the sign of the curvature constant $k$ (or alternatively, the density parameter $\Omega_k$).  Solving this integral we now have:
\begin{equation}
    R \ \ \equiv\ \  D_M \ \ =\ \  \left.\begin{cases}
        \frac{1}{\sqrt{k}}\sin\left(\frac{\sqrt{k}}{a(t_0)}D_c\right), & \quad k > 0 \\[4pt]
        \frac{1}{a(t_0)}D_c, & \quad k = 0 \\[4pt]
        \frac{1}{\sqrt{|k|}}\sinh\left(\frac{\sqrt{|k|}}{a(t_0)}D_c\right), & \quad k < 0
    \end{cases}\right\} \ \ =\ \  \left.\begin{cases}
        \frac{D_H}{\sqrt{|\Omega_k|}}\sin\left(\frac{\sqrt{|\Omega_k|}}{a(t_0)D_H}D_c\right), & \quad \Omega_k < 0 \\[4pt]
        \frac{1}{a(t_0)}D_c, & \quad \Omega_k = 0 \\[4pt]
        \frac{D_H}{\sqrt{\Omega_k}}\sinh\left(\frac{\sqrt{\Omega_k}}{a(t_0)D_H}D_c\right), & \quad \Omega_k > 0
    \end{cases}\right\} \ \ =\ \  \frac{D_L}{a(t_0) (1+z)} ,
\label{eqn: R - DM - DC - DL relationship}
\end{equation}
where $D_M$ is introduced as the ``transverse comoving distance'' following~\citeauthor{hogg_distances}, which is just a re-naming of the comoving coordinate distance.  Thus depending on the global universe geometry, the relationship between the various distances $R$, $D_L$, $D_c$, and $D_M$ are given by equation~\ref{eqn: R - DM - DC - DL relationship}.

One final and very important distance for our consideration is the ``parallax distance'' $D_\mathrm{par}$ in equation~\ref{eqn: parallax distance Dpar} \citep{weinberg1972}.  Substituting equation~\ref{eqn: R - DM - DC - DL relationship} into equation~\ref{eqn: parallax distance Dpar} lets us write:
\begin{equation}
    D_\mathrm{par} \ \ \equiv\ \  \frac{a(t_0)R}{\sqrt{1-kR^2}} \ \ =\ \ \frac{a(t_0)D_M}{\sqrt{1-kD_M^2}} \ \ =\ \  \left.\begin{cases}
        \frac{a(t_0)}{\sqrt{k}}\tan\left(\frac{\sqrt{k}}{a(t_0)}D_c\right), & \quad k > 0 \\[4pt]
        D_c, & \quad k = 0 \\[4pt]
        \frac{a(t_0)}{\sqrt{|k|}}\tanh\left(\frac{\sqrt{|k|}}{a(t_0)}D_c\right), & \quad k < 0
    \end{cases}\right\} \ \ =\ \  \frac{D_L}{(1+z)\sqrt{1-\frac{k D_L^2}{a^2(t_0)(1+z)^2}}} .
\label{eqn: Dpar - R - DM - DC - DL relationship}
\end{equation}
Notably in a flat universe (with $a(t_0)=1$ normalization) we have $D_\mathrm{par} = R = D_M = D_c = D_L / (1+z)$.

As a final note, the closed universe case needs further special consideration, as the parallax distance not only diverges, but can take on \textit{negative} values.  For closed geometry, the expression for $D_\mathrm{par}$ in equations~\ref{eqn: parallax distance Dpar} and~\ref{eqn: Dpar - R - DM - DC - DL relationship} has a $\pm$ sign in front of the expression.  Conceptually, consider a point at the north pole of a globe generating some wave.  As the wave propagates away from the source, in the northern hemisphere two points on the wavefront (constant latitude) would produce a positive parallax measurement $D_\mathrm{par} > 0$ back to the source (the wavefront curves \textit{towards} the source and \textit{away} from the direction of propagation).  At the equator, however, all points along the wavefront will experience a plane-wave.  Therefore there is no well-defined parallax distance for points along the equator, and mathematically we see that the value of $D_\mathrm{par} \rightarrow \infty$ (i.e. at $R = R_\mathrm{max}= \frac{1}{\sqrt{k}}$, or equivalently $\chi = \frac{1}{\sqrt{k}}\frac{\pi}{2}$ through equation~\ref{eqn: r(chi) definition}).

Then interestingly, in the southern hemisphere two points on the wavefront (constant latitude) would now produce a \textit{negative} parallax measurement $D_\mathrm{par}$ back to the source (the wavefront curves \textit{away} the source and \textit{towards} the direction of propagation -- towards the source antipode).  Using equation~\ref{eqn: r(chi) definition} to replace $R$ with $\chi$ in equation~\ref{eqn: Dpar - R - DM - DC - DL relationship}, and including the $\pm$ sign out front, we write:
\begin{equation*}
    D_\mathrm{par} = \pm \frac{a(t_0)\sin\left(\sqrt{k}\chi\right)}{\sqrt{k}\sqrt{1-\sin^2\left(\sqrt{k}\chi\right)}} = \pm \frac{a(t_0)\sin\left(\sqrt{k}\chi\right)}{\sqrt{k}\sqrt{\cos^2\left(\sqrt{k}\chi\right)}} = \frac{a(t_0)}{\sqrt{k}}\tan\left(\sqrt{k}\chi\right) . \qquad (\text{for } k>0)
\end{equation*} 
Writing $D_\mathrm{par}$ in terms of $\chi$ takes care of the sign ambiguity in equations~\ref{eqn: parallax distance Dpar} and~\ref{eqn: Dpar - R - DM - DC - DL relationship}.  For values $0\leq \chi < \frac{1}{\sqrt{k}}\frac{\pi}{2}$ (i.e. the northern hemisphere) the parallax distance is positive, for $\chi = \frac{1}{\sqrt{k}}\frac{\pi}{2}$ (i.e. the equator) the parallax distance diverges, and for values $\frac{1}{\sqrt{k}}\frac{\pi}{2} < \chi \leq \frac{1}{\sqrt{k}}\pi$ (i.e. the southern hemisphere) the parallax distance is negative.

    \section{The Retarded Time Calculation}\label{app: tret calculation}

The goal is to calculate $\left|\vec{x}' - \vec{x}\right|$ in equation~\ref{eqn: tret - eta and t versions}, where $\vec{x}'$ is the source location and $\vec{x}$ is the field point.  For our problem, we will primarily be interested in setting our field point at the pulsar.  With the center of our coordinate system at the Earth, let the source be at the coordinate $r=R$ (i.e. $\chi = \chi_S$) with unit vector $\hat{r}$, and let the pulsar be at the coordinate $r=L$ (i.e. $\chi=\chi_P$) with unit vector $\hat{p}$.

Next we write the law of cosines for Euclidean, spherical, and hyperbolic geometries:
\begin{equation}
    \begin{cases}
        \begin{tabular}{r l r}
            $\cos\left(\sqrt{k}\left|\vec{x}' - \vec{x}\right|\right)$ &$= \cos\left(\sqrt{k}\chi_S\right)\cos\left(\sqrt{k}\chi_P\right) + \sin\left(\sqrt{k}\chi_S\right)\sin\left(\sqrt{k}\chi_P\right)\left(\hat{r}\cdot\hat{p}\right) ,$ &$k > 0$ \\[4pt]
            $\left|\vec{x}' - \vec{x}\right|$ &$= \chi_S^2 + \chi_P^2 - 2\chi_S\chi_P\left(\hat{r}\cdot\hat{p}\right) ,$ &$k = 0$ \\[4pt]
            $\cosh\left(\sqrt{|k|}\left|\vec{x}' - \vec{x}\right|\right)$ &$= \cosh\left(\sqrt{|k|}\chi_S\right)\cosh\left(\sqrt{|k|}\chi_P\right) - \sinh\left(\sqrt{|k|}\chi_S\right)\sinh\left(\sqrt{|k|}\chi_P\right)\left(\hat{r}\cdot\hat{p}\right) .$ &$k < 0$
        \end{tabular}
    \end{cases}
\end{equation}
Recall from Section~\ref{subsec: generalizing the flat, static model} that at the present time the Gaussian curvature is $k$.  One approach to solving these equations is to Taylor expand in the small parameter $\sqrt{|k|}\chi_P$ in the closed and open universe cases, and in the small parameter $\frac{\chi_P}{\chi_S}$ the flat universe case.  The result is:
\begin{equation}
    \left|\vec{x}' - \vec{x}\right| \approx \chi_S - \left(\hat{r}\cdot\hat{p}\right)\chi_P + \frac{1}{2}\left( 1 - \left(\hat{r}\cdot\hat{p}\right)^2 \right) \begin{cases}
        \frac{\chi_P^2}{\frac{1}{\sqrt{k}}\tan\left(\sqrt{k}\chi_S\right)} , \quad & k>0 \\[12pt]
        \frac{\chi_P^2}{\chi_S} , \quad & k=0 \\[4pt]
        \frac{\chi_P^2}{{\frac{1}{\sqrt{|k|}}\tanh\left(\sqrt{|k|}\chi_S\right)}} . \quad & k<0
    \end{cases}
\label{eqn: source-field separation}
\end{equation}
Additionally, using equation~\ref{eqn: r(chi) definition} we see that $\chi_P \approx L$ for each universe geometry case (Taylor expanding once again for the closed and open cases in $\sqrt{|k|}\chi_P$).

Consider for a moment a gravitational wave propagating directly from the source to the Earth.  The wave leaves it's position at redshift $z$ and time $t_\mathrm{ret}$, and arrives at Earth at redshift $z=0$ at time $t$ (recall the boundary conditions from Appendix~\ref{app: cosmological distances}).  In this case our field position $\vec{x} = 0$, so $\eta_\mathrm{ret} = \eta - \frac{\chi_S}{c}$ from equation~\ref{eqn: tret - eta and t versions}.  Therefore using equation~\ref{eqn: line-of-sight comoving distance Dc} we can write:
\begin{align*}
    -\frac{\chi_S}{c} = \eta_\mathrm{ret} - \eta &= \int^{t_\mathrm{ret}}_t \frac{\dd t'}{a} \\
    &= -\frac{D_c}{a(t_0)c}.
\end{align*}
This gives us a connection between $\chi_S$ and the comoving coordinate distance $D_c$, namely that $\chi_S = \frac{D_c}{a(t_0)}$, which we can now substitute into equation~\ref{eqn: source-field separation}.  The result is that we can now see that the relevant distance which appears in the Fresnel term in our Taylor expansion (for all three universe cases) is the parallax distance $D_\mathrm{par}$ (substitute in equation~\ref{eqn: Dpar - R - DM - DC - DL relationship}).  Therefore the final form of equation~\ref{eqn: source-field separation} which is then used in equation~\ref{eqn: tret - pulsar term} is: 
\begin{equation}
    \left|\vec{x}' - \vec{x}\right| \approx \frac{D_c}{a(t_0)} - \left(\hat{r}\cdot\hat{p}\right)L + \frac{1}{2}\left( 1 - \left(\hat{r}\cdot\hat{p}\right)^2 \right) a(t_0)\frac{L^2}{D_\mathrm{par}} .
\label{eqn: source-field separation - final}
\end{equation}
Note that in the flat universe case $D_\mathrm{par} = D_c$, which is consistent with the result of \citetalias{DOrazio2021} given that this was their working assumption.

    \section{Assumptions}\label{app: assumptions}

Assumptions (ii)--(xvi) of \citetalias{mcgrath2021} carry over to this work.  Additionally, we work under the following assumptions:

\begin{enumerate}
    \item The universe is described by the FLRW metric with comoving coordinates $(t, r, \theta, \phi)$ and curvature constant $k$.
    \item The FLRW scale factor $a$ does not change appreciably over the time-scale of the travel time of the photon from the pulsar to the Earth: $\dot{a}(t_0) \Delta t_{\mathrm{p}\rightarrow\mathrm{E}}$. \label{as: scale factor evolution}
    \item All ``local'' distances and times between the pulsar and the Earth are on a Minkowski background, $g_{\mu\nu} = \eta_{\mu\nu} = \text{diag}\left(-c^2, 1, 1, 1\right)$.  The transition from the global cosmological FLRW background to the local Minkowski background requires that all points of interest have comoving coordinate distance separations much smaller than the background curvature $\left(r^2 \ll \frac{1}{|k|}\right)$, and that the observation time of our experiment is much smaller than the present day age of the universe $\left(\frac{t_\mathrm{obs}}{t_0} \ll 1\right)$. \label{as: metric transition}
    \item For the results presented in this work we assume a flat universe ($k=0$).
\end{enumerate}

    \section{Pulsar Populations and Example Pulsar Distance Posteriors}\label{app: Pulsar Populations}

\begin{figure}
    \centering
    \includegraphics[width=0.75\linewidth]{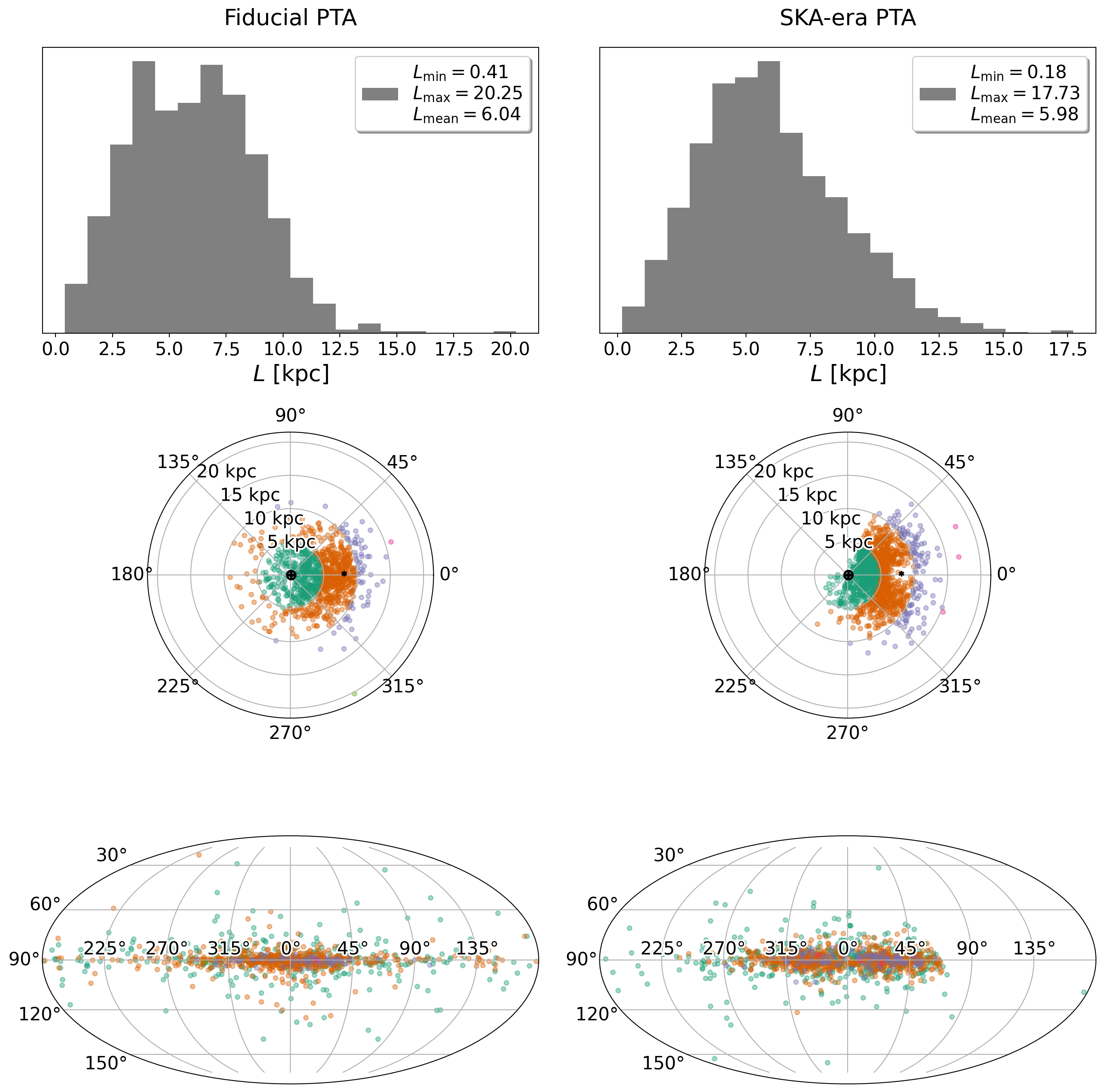}
        % --> located in:  cosmo_study/pulsar_files/PTA_comparison_ska20220809_sigmin1.png
    \caption{A side-by-side comparison of the fiducial PTA (left-hand column) and SKA-era PTA (right-hand column).  These figures show the general distribution properties, in terms of the pulsar distances and locations throughout the Milky Way Galaxy.  In the middle row, Earth is located at the origin and galactic center is located at $8$ kpc (at $\phi = 0^{\circ}$).
    \textbf{Left-hand panel:} This fiducial PTA contains 1000 pulsars generated from the distribution in equation~\ref{eqn: fiducial PTA}.
    \textbf{Right-hand panel:} This SKA-era PTA contains 1194 pulsars.}
    \label{fig: PTAs}
\end{figure}

Two pulsar populations are used for different studies within this paper, a ``fiducial PTA'' and an ``SKA-era PTA''.  The general characteristics of these PTAs are shown in Figure~\ref{fig: PTAs}.

To create the SKA-era PTAs, we use the \psrpoppy package \citep{PsrPopPy:2014} to generate a realistic population of MSPs detectable in the SKA era. To generate the underlying population of pulsars we assume the default model in \psrpoppy (except that we place the galactic center at 8 instead of 8.5 kpc), with a spatial pulsar distribution following \cite{Lorimer+2006}, but with the pulsar period distribution from \cite{Cordes_Chernoff:1997}, chosen specifically for MSPs. We then follow \citep{SmitsSKA+2009} in assuming that there are $3\times10^4$ MSPs in the galaxy to generate the discoverable population.

\begin{figure}
    \centering
    \includegraphics[width=0.5\linewidth]{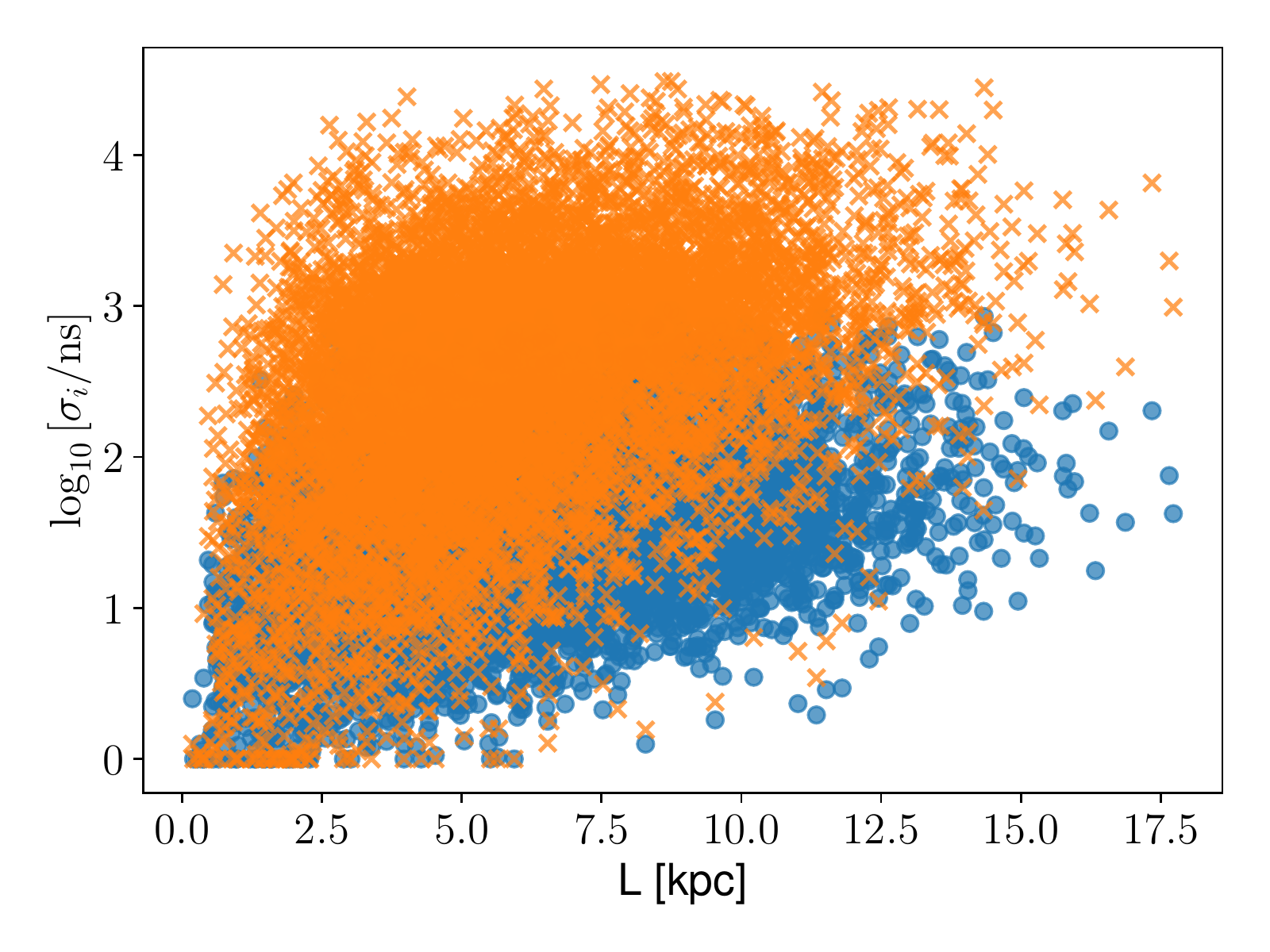}
        % --> located in:  cosmo_study/pulsar_files/8kpc_SKA_PTAs_20220809/sigmares_DGC8_NPTA8893_Pmin1_SNRmax10000SKA_SmitsA_8chk.det.pdf
    \caption{Timing uncertainty distributions of all of the MSPs detected in a mock SKA-era population, with $\sigma_{\min}=1$~ns. The blue circles and orange crosses denote optimistic and pessimistic choices for $\sigma_a$, as written in equation \ref{eqn:sigmai}. Mock SKA-era PTAs are generated by choosing MSPs from this set, as outlined in the text.}
    \label{fig:skaPTA}
\end{figure}

We simulate an SKA-like survey by choosing survey parameters following \cite{SmitsSKA+2009} (model A) and summarized in Table~\ref{table:SKAsurvey}. We find that our SKA survey detects $8893$ of these MSPs with SNR greater than a threshold value of 9, at distances ranging from $0.2$~kpc  out to $18$~kpc.
We assume that only some percentage of these MSPs will be suitable for high precision timing, and enforce this as described by the selection process in Section~\ref{S:SKAPTA}.  We then compute a timing uncertainty for each MSP as per equation \ref{eqn:sigmai}. Timing uncertainty distributions for the different choices of uncertainty scaling with SNR are plotted in Figure \ref{fig:skaPTA}.

% \begin{table}
%     \caption{SKA survey parameters motivated by \citet{SmitsSKA+2009} (their Model A), used to generate mock PTAs via the \psrpoppycaption code \citep{PsrPopPy:2014}.}
%     \label{table:SKAsurvey}
%     \begin{center}
%         \begin{tabular}{ l | c  } 
%         %
%         \hline
%         Name  & \hspace{2cm} Value \\
%         \hline
%          Survey degradation factor  &  \hspace{2cm} 1.0 \\
%          Antenna gain (K $\mathrm{Jy}^{-1}$) &  \hspace{2cm} 130 \\ %(0.6 * 2e4 m^2/K at  Ts=30k) \\
%          Integration time (s)       &  \hspace{2cm} 1800 \\
%          Sampling time (ms)         &  \hspace{2cm} 0.1 \\
%          System temperature (K)     &  \hspace{2cm} 30  \\
%          Centre frequency (MHz)     &  \hspace{2cm} 1400 \\ %(best for high gain, in galactic plane, Smits Figure 1) 
%          Bandwidth (MHz)            &  \hspace{2cm} 500  \\ %500 for dishes 
%          Channel bandwidth (MHz)    &  \hspace{2cm} 0.009   \\
%          No. polarizations          &  \hspace{2cm} 2     \\
%          FWHM (arcmin)              &  \hspace{2cm} 65.5  \\
%          Min RA (deg)               &  \hspace{2cm} 0   \\
%          Max RA (deg)               &  \hspace{2cm} 360 \\
%          Min DEC (deg)              &  \hspace{2cm} -90  \\
%          Max DEC (deg)              &   \hspace{2cm} 30  \\
%         Frac. survey coverage.      &   \hspace{2cm} 1  \\
%         SNR threshold               &   \hspace{2cm} 9 \\
%         \hline
%         %
%         \end{tabular}
%     \end{center}
% \end{table}

\begin{figure}
    \centering
    \includegraphics[width=\linewidth]{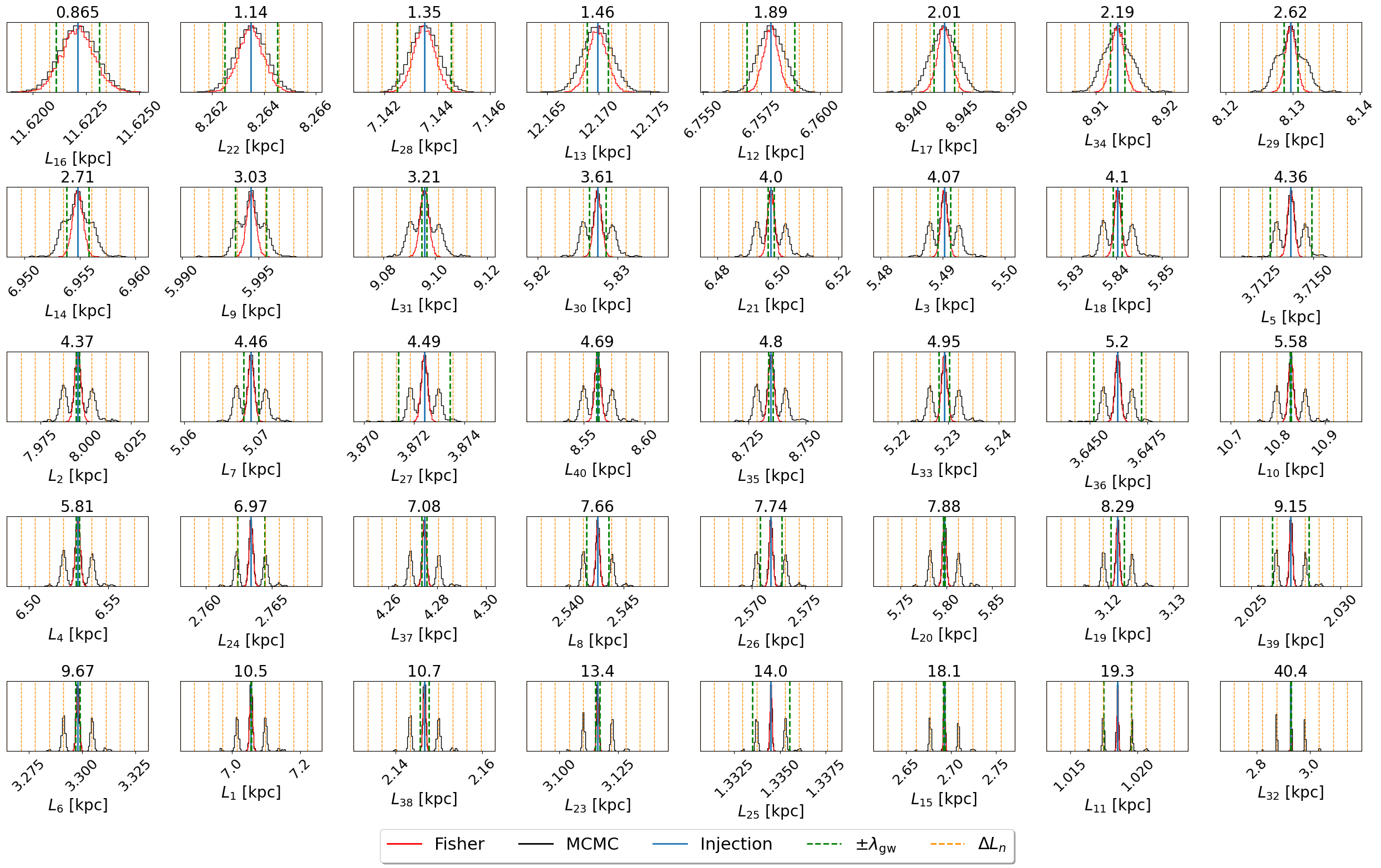}
            % --> located in:  cosmo_study/IB/envelope_runs_allparams/data/figures/run_16_fullPTA_cleaned-100.eps
    \caption{The complete set of 1D posteriors for the same results shown in the right-hand panel of Figure~\ref{fig: error envelopes} using the IB model.  Here we have sorted the pulsars in ascending value by the modal overlap criterion ratio defined in equation~\ref{eqn: error envelope criteria} (given above each figure).  As we can see, this particular criterion successfully shows the progression of the error envelope from ``overlapping'' to ``resolved,'' with a total of 12 overlapping, 18 intermediate, and 10 resolved pulsars.  For ratios less than about $4$, we see that uncertainties about each mode are both sufficiently large and sufficiently close to the primary mode that they blend together around the true value.  However, for ratios between about $4$ and $8$ these secondary modes begin to separate away from the primary mode far enough that they start to become distinguishable from the true mode.  Finally, for ratios greater than about $8$, we see that the secondary modes are distinguishable from the true mode.  In practice this therefore motivates us to search our results to see what the value of this ratio is for each of our pulsars in our desired source-PTA simulation.  If it is greater than $8$, then we would expect that the true distance to that pulsar would be identifiable.  Note the samples in this figure have been cleaned, such that all samples with log-posterior $\leq -100$ have been removed from the data set (see the discussion of cleaning in Section~\ref{subsec:mcmc results}).}
    \label{fig: error envelope progression}
\end{figure}

Finally, Figure~\ref{fig: error envelope progression} shows the complete set of 1D posteriors for the 40 pulsar PTA shown in the right-hand panel of Figure~\ref{fig: error envelopes}, which motivates the modal overlap criterion defined in equation~\ref{eqn: error envelope criteria}.

\begin{table}
    \caption{SKA survey parameters motivated by \citet{SmitsSKA+2009} (their Model A), used to generate mock PTAs via the \psrpoppycaption code \citep{PsrPopPy:2014}.}
    \label{table:SKAsurvey}
    \begin{center}
        \begin{tabular}{ l | c  } 
        \hline
        Name  & \hspace{2cm} Value \\
        \hline
         Survey degradation factor  &  \hspace{2cm} 1.0 \\
         Antenna gain (K $\mathrm{Jy}^{-1}$) &  \hspace{2cm} 130 \\ %(0.6 * 2e4 m^2/K at  Ts=30k) \\
         Integration time (s)       &  \hspace{2cm} 1800 \\
         Sampling time (ms)         &  \hspace{2cm} 0.1 \\
         System temperature (K)     &  \hspace{2cm} 30  \\
         Centre frequency (MHz)     &  \hspace{2cm} 1400 \\ %(best for high gain, in galactic plane, Smits Figure 1) 
         Bandwidth (MHz)            &  \hspace{2cm} 500  \\ %500 for dishes 
         Channel bandwidth (MHz)    &  \hspace{2cm} 0.009   \\
         No. polarizations          &  \hspace{2cm} 2     \\
         FWHM (arcmin)              &  \hspace{2cm} 65.5  \\
         Min RA (deg)               &  \hspace{2cm} 0   \\
         Max RA (deg)               &  \hspace{2cm} 360 \\
         Min DEC (deg)              &  \hspace{2cm} -90  \\
         Max DEC (deg)              &   \hspace{2cm} 30  \\
        Frac. survey coverage.      &   \hspace{2cm} 1  \\
        SNR threshold               &   \hspace{2cm} 9 \\
        \hline
        \end{tabular}
    \end{center}
\end{table}

% If you want to present additional material which would interrupt the flow of the main paper, it can be placed in an Appendix which appears after the list of references.

%%%%%%%%%%%%%%%%%%%%%%%%%%%%%%%%%%%%%%%%%%%%%%%%%%

% Don't change these lines
\bsp	% typesetting comment
\label{lastpage}
\end{document}